\documentclass[%
 reprint,
superscriptaddress,
 bibnotes,
 amsmath,
 amssymb,
 aps,
pra,
]{revtex4-2}

\usepackage{graphicx}
\usepackage{dcolumn}
\usepackage{bm}
\usepackage{physics}
\usepackage[inkscapelatex=false]{svg}
\usepackage{bbm}
\usepackage{mathrsfs}
\usepackage{xcolor}
\usepackage[most]{tcolorbox}
\usepackage[colorlinks = true,]{hyperref}

\usepackage{mathtools}

\DeclarePairedDelimiterX\opbraket[2]{(}{)}{#1\,\delimsize\vert\,\mathopen{}#2}

\newtcolorbox{highlighted}{colback=yellow,coltext=red,breakable}
\begin{document}

\preprint{APS/123-QED}

\title{Phase Transitions in Open Dicke Model: a degenerate perturbation theory approach}

\author{Wenqi Tong}
 \email{tong76@purdue.edu}
 \affiliation{Elmore Family School of Electrical and Computer Engineering, Purdue University, West Lafayette, Indiana 47907, USA}
 
\author{H. Alaeian}%
 \email{halaeian@purdue.edu}
 \affiliation{Elmore Family School of Electrical and Computer Engineering, Purdue University, West Lafayette, Indiana 47907, USA}
 \affiliation{%
 Department of Physics and Astronomy, Purdue University, West Lafayette, Indiana 47907, USA
 }%
 \affiliation{Purdue Quantum Science and Engineering Institute, Purdue University, West Lafayette, Indiana 47907, USA}
 
\author{F. Robicheaux}%
 \email{robichf@purdue.edu}
 \affiliation{%
 Department of Physics and Astronomy, Purdue University, West Lafayette, Indiana 47907, USA
}%
 \affiliation{Purdue Quantum Science and Engineering Institute, Purdue University, West Lafayette, Indiana 47907, USA}




\date{\today}

\begin{abstract}
We study the steady-state behavior of the open Dicke model, which describes the collective interaction of $N$ spin-$1/2$ particles with a lossy, quantized cavity mode and exhibits a superradiant phase transition above a critical light-matter coupling. While the standard model conserves total spin, Kirton and Keeling \cite{PhysRevLett.118.123602} demonstrated that even infinitesimal homogeneous local dephasing destroys this phase transition, and that local atomic decay can restore it. We analyze this interplay using degenerate perturbation theory across subspaces of fixed total spin, $S$. For coupling strengths above the threshold, there exists a critical spin value $S_c$ such that the superradiant phase transition occurs only for $S>S_c$. The perturbative approach captures how weak dephasing and decay induce mixing between different $S$-subspaces, yielding a steady-state spin distribution whose width scales as $1/\sqrt{N}$. This framework requires only the first and second moments and can be implemented via different methods that can yield these two moments (for example, the 2nd-cumulant approach), circumventing the need for full density matrix calculations. These results bridge the quantum Rabi model and Dicke physics, elucidate the roles of dephasing and decay in collective quantum effects, and apply broadly to open quantum systems with degenerate steady states.
\end{abstract}

\maketitle

\section{\label{sec:introduction}Introduction}

Dicke superradiance~\cite{PhysRev.93.99,PhysRevA.3.1735,gross1982superradiance} describes the collective radiation of $N$ inverted atoms confined to zero dimensions. Experiencing the same field, the atoms spontaneously synchronize during the decay. This renders constructive interference between the emitted photons, leading to initially increasing photon emission rate, i.e., a burst of light with peak intensity scaling as $N^2$, instead of $N$ for independent atoms. Referred to as a ``superradiant burst'', this phenomenon has been observed in various physical systems~\cite{PhysRevLett.30.309,PhysRevLett.36.1035,PhysRevLett.49.117,scheibner2007superradiance,rohlsberger2010collective,inouye1999superradiant,PhysRevLett.98.053603,PhysRevLett.76.2049,eschner2001light}. In recent years, the advancements in ordered atomic arrays enables strong control over atomic positions~\cite{kim2016situ,endres2016atom,barredo2016atom,norcia2018microscopic,PhysRevLett.122.143002,PhysRevLett.122.203601,bakr2010probing,sherson2010single,greif2016site,kumar2018sorting}. These extended systems, with inter-atomic separation no longer negligible to the photon emission wavelength, are also reported to show the Dicke superradiance \cite{masson2022universality,PhysRevLett.130.213605,PhysRevResearch.4.023207,PRXQuantum.5.010344} and are reviewed in Ref. ~\cite{PhysRevA.108.030101} for planar atomic arrays and Ref. ~\cite{RevModPhys.95.015002} for 1-D atomic arrays coupled to a waveguide. Such an experimentally accessible phenomenon has been considered as a witness to the atomic entanglement~\cite{PhysRevLett.131.063601}. Recent years, there has been a growing interest in superradiance due to its potential to generate highly entangled photonic states~\cite{PhysRevA.99.043807,Lange2024} and lasers with ultranarrow linewidth~\cite{PhysRevLett.102.163601,maier2014superradiant,bohnet2012steady,PhysRevX.6.011025,norcia2016superradiance,PhysRevX.8.021036,PhysRevLett.123.103601,PhysRevA.101.013819}. 

Instead of the transient superradiance of the fully excited atomic ensemble, Hepp and Lieb~\cite{PhysRevA.8.2517}, later expressed in a more physical language by Wang and Hioe~\cite{PhysRevA.7.831}, focused on the ground state of the system with $N$ atoms coupled to a quantized cavity mode. For coupling strength beyond a critical value, the ground state undergoes a continuous phase transition, characterized by the emergence of macroscopic and coherent occupation of the cavity mode. In the thermodynamic limit where $N \rightarrow \infty$, the average cavity photon number $\langle a^\dagger a  \rangle/N$ vanishes in the normal phase while the superradiant phase is identified by the cavity photon number $\langle a^\dagger a  \rangle$ proportional to $N$. In addition to the co-rotating terms considered by these works, Refs.~\cite{PhysRevA.8.1440,carmichael1973higher,duncan1974effect}, associating the so-called generalized Dicke model, take the counter-rotating terms into account and correct the critical coupling strength by a factor of $1/2$. By Ref.~\cite{PhysRevA.107.043706}, the unbalanced Dicke model, sometimes known as the anisotropic Dicke model, features more degrees of freedom due to different coupling strengths for corotating and counterrotating terms and exhibits exotic phases like the nonergodic~\cite{PhysRevLett.118.080601,hu2021out} and counter-lasing ones~\cite{PhysRevResearch.4.023101}. Despite the intriguing many-body physics, these phases have not been observed experimentally until recent years. The challenge lies in the requirement for an extremely strong coupling strength comparable to the cavity and atomic frequencies. Theoretically, the existence of the phase transition has been under debate due to the so-called $A^2$ term~\cite{PhysRevLett.35.432,nataf2010no,PhysRevLett.107.113602,PhysRevA.86.053807,PhysRevA.90.063825,PhysRevLett.112.073601,PhysRevA.94.033850}. Thanks to the adiabatically-eliminated Raman process through the excited states~\cite{PhysRevA.75.013804,zhiqiang2017nonequilibrium,PhysRevLett.89.253003,PhysRevLett.91.203001,PhysRevLett.104.130401,baumann2010dicke,klinder2015dynamical,PhysRevX.8.011002}, the effective coupling strength can be tuned by the external driving and the $A^2$ term can be circumvented so that the superradiant phase transition is not prohibited by the no-go theorem~\cite{PhysRevLett.35.432}. 

In addition to the equilibrium dynamics above, dissipative processes lead to an open system and the interplay between unitary dynamics and dissipation leads to phase transition in the non-equilibrium steady-states (NESS) of cavity QED systems~\cite{PhysRevA.75.013804,PhysRevLett.104.130401,PhysRevLett.105.043001,oztop2012excitations,piazza2013bose,PhysRevA.85.013817,PhysRevA.87.023831,PhysRevLett.115.043601,PhysRevA.94.061802} and atomic clouds in free space~\cite{PhysRevLett.127.243602,Ferioli2023}. In the standard open Dicke model, the dynamics of $N$ indistinguishable two-level atoms are theoretically described by collective spin operators since the total spin, $S$, is a conserved quantity. However, decoherent processes like local dephasing~\cite{PhysRevA.96.023863,PhysRevA.94.061802} and local decay~\cite{PhysRevA.95.063824} break the spin conservation. As a result, the system is no longer within a fixed total spin subspace of size $O(N)$. Utilizing the permutation symmetry~\cite{PhysRevLett.118.123602,PhysRevA.98.063815}, Ref. \cite{PhysRevLett.118.123602} reduces the complexity to $O(N^3)$ and points out that an infinitesimal dephasing destroys the transition to the superradiant phase, while the introduction of the individual decay restores it. Since the steady states of the open Dicke model exhibit high degeneracy, this phenomenon is reminiscent of degenerate perturbation theory~\cite{PhysRev.33.467,RevModPhys.35.710,klein1974degenerate}, where an infinitesimal perturbation causing the mixing between the degenerate states will qualitatively modify the behavior of the system~\cite{PhysRevA.110.063701}. Despite the discussion on the degeneracy and perturbation theory of the Liouvillian in Refs.~\cite{li2014perturbative,huybrechts2024quantum,thingna2021degenerated,albert2018lindbladians,gomez2018perturbation,li2016nonequilibrium,krishna2023select} as well as the total spin mixing effect in Refs. \cite{PhysRevA.109.032204,ruostekoski2025superradiant,PhysRevResearch.6.023206,PhysRevA.96.023863,PRXQuantum.6.020303, PhysRevA.94.061802,PhysRevA.95.063824,PhysRevA.98.063815}, to the best of our knowledge, the application of degenerate perturbation theory on the phases of the open Dicke model is not fully explored. In fact, one motivation of this study was to show that degenerate perturbation could reproduce results for a system involving a phase transition.

The conventional open Dicke model mainly focuses on the fully symmetric subspace with total spin $S = N/2$. In this work, we explore all the $S$-subspaces and note that the steady-state behavior, especially the existence of a phase transition, depends on the total spin, $S$, as well: given a coupling strength greater than the critical value, there is a critical total spin $S_c$ where the phase transition is available for $S > S_c$ only. We use degenerate perturbation theory to investigate the effects of homogeneous local dephasing and local decay on the superradiant phase transition. Explicitly shown by the coupling matrix obtained from the degenerate perturbation theory, these perturbations mix adjacent $S$-subspaces and yield a new steady state, whose probability distribution in each total spin subspace, $p(S)$, is indicated by the null eigenvector of the coupling matrix. In the thermodynamic limit, this distribution approaches a $\delta$-function with a width scaling as $1 / \sqrt{N}$. In addition, degenerate perturbation theory can give the spectrum Liouvillian eigenvalues which contains information beyond the steady state.

As with all the degenerate perturbation theory, when the perturbation is weak enough, the probability distribution $p(S)$ is determined by the contribution of the two perturbations to the overall perturbation. In particular, in the thermodynamic limit, pure local dephasing yields the $\delta$-function well below $S_c$ while an infinitesimal contribution from local decay renders the $\delta$-function above $S_c$ - that degenerate perturbation theory could reproduce this result was not obvious. Because the coupling matrix only depends on the steady-state values of the first and second order moments, this method can circumvent solving the full density matrix master equations by utilizing mean-field theory or other methods that gives these two moments. This allows the computation for $N \rightarrow \infty$ and is applicable to systems having high degeneracy in steady states (for example, generalized Dicke models~\cite{PhysRevLett.118.080601,hu2021out,PhysRevResearch.4.023101}, two-photon Dicke models~\cite{PhysRevA.109.053702,shah2024dissipative,BANERJEE2022128287,PhysRevA.95.053854,Garbe2020dissipation,PhysRevB.102.245407,PhysRevA.97.053821} or driven Dicke model~\cite{leppenen2024quantumbistabilityinterplaycollective}). 

This paper is organized as follows. In Sec.~\ref{sec:theory}, we present the fundamental equations of motion and outline the perturbative framework, including the key approach for evaluating the coupling matrix. Section~\ref{sec:results} is devoted to benchmarking and validating the degenerate perturbation theory and analyzing critical behavior in terms of total spin. We further examine the spin-mixing effects arising from the two perturbations, explore the Wigner distribution of photons, and discuss the scaling behavior with atom number. Section~\ref{sec:summary} gives the summary and outlook of this framework.

\section{\label{sec:theory}Theory}
\subsection{Open Dicke model, Symmetries, and Equations of Motion\label{sec:eom}}
To investigate the effect of perturbations on the open Dicke model, it is necessary to study the dynamics of the unperturbed system. In this work, we consider $N$ atoms coupled to a single-mode cavity. Each atom is considered as a two-level system, with the ground state denoted by $\ket{0}$ and the excited state represented by $\ket{1}$. The associated spin operators are:
\begin{equation}
    \hat{\sigma}_n^+ \!=\! \ket{1}_n\!\bra{0}_n\!, \  \hat{\sigma}_n^- \!=\! \ket{0}_n\!\bra{1}_n\!,\  \hat{\sigma}_n^z \!=\! \ket{1}_n\!\bra{1}_n \!-\! \ket{0}_n\!\bra{0}_n,
\end{equation}
where $n$ is the index of the atom. Confined in a volume small enough compared to the wavelength, the atoms are indistinguishable due to their identical coupling to the cavity mode. In other words, the system has permutation symmetry~\cite{PhysRevLett.118.123602,PhysRevA.98.063815}, where the density operator elements remains the same after interchanging any pair of atoms:
\begin{equation}
    \hat{P}_{i,j}\hat{\rho}\hat{P}^\dagger_{i, j} = \hat{\rho},
\end{equation}
where $\hat{P}_{i,j}$ is the permutation operator involving atom $i$ and $j$:
\begin{equation}
    \hat{P}_{i,j}\ket{s_1, ..., s_i,..., s_j,..., s_N} = \ket{s_1, ..., s_j,..., s_i,..., s_N},
\end{equation}
where $s_i = \{0, 1\}$. In this case, the atoms can be modeled by a collective spin $\hat{\vec{S}} = \{\hat{S}_x, \hat{S}_y, \hat{S}_z\}$, where:

\begin{equation}
    \hat{S}_\alpha = \frac{1}{2}\sum_n \hat{\sigma}_n^\alpha, \quad \alpha = {x, y, z},
\end{equation}
where $\hat{\sigma}_n^\alpha$ are Pauli matrices of the $n$th atom, and the total spin operator is defined as:
\begin{equation}
    \hat{S}^2 = \sum_\alpha \hat{S}_\alpha^2.
\end{equation}
The elements $\{\hat{S}_x, \hat{S}_y, \hat{S}_z\}$ obey the commutator relations:

\begin{equation}
    [\hat{S}_x, \hat{S}_y] = i\hat{S}_z.
\end{equation}
Throughout this paper, we let $\hbar = 1$. 

The unitary dynamics of the unperturbed system is described by the well-known Dicke Hamiltonian:

\begin{equation}
    H = \omega_c a^\dagger a + 2 \omega_0 \hat{S}_z + \frac{g}{\sqrt{N}}(\hat{S}^+ + \hat{S}^-)(a + a^\dagger),
    \label{eq:hamiltonian}
\end{equation}
where $a$ ($a^\dagger$) is the annihilation (creation) operator of a photon in the cavity and $\omega_c$ is the energy of the corresponding photon. Further, $\omega_0$ is the transition energy of each atom, and $\hat{S}^+$ ($S^-$) is the raising (lowering) operator of the collective spin:

\begin{equation}
    \hat{S}^+ = (\hat{S}_x + i\hat{S}_y), \quad \hat{S}^- = (\hat{S}^+)^\dagger.
\end{equation}
The parameter $g$ describes the coupling strength between the cavity mode and the atomic ensemble. By Hepp and Lieb~\cite{PhysRevA.8.2517}, the coupling strength should scale as $\sim 1/\sqrt{N}$ so that the thermodynamic limit is well-defined. 

In addition to the unitary dynamics, the cavity is lossy, and the dissipative process is described by the Lindblad superoperator:

\begin{equation}
    \mathcal{L}_\kappa[\hat{\rho}] = \kappa (a \rho a^\dagger - \frac{1}{2} a^\dagger a \rho - \frac{1}{2} \rho a^\dagger a),
\end{equation}
where $\kappa$ is the dissipation rate of the cavity photons.

Finally, the density matrix of the \textit{unperturbed} open Dicke model evolves via the following master equation:

\begin{equation}
    \begin{split}
        \frac{d\hat{\rho}}{dt} = & - i [H, \hat{\rho}] + \mathcal{L}_\kappa [\hat{\rho}].
    \end{split}
    \label{eq:eom}
\end{equation}
One can check that the equations of motion for the unperturbed open Dicke model only involve collective spin operators, which only change the projection on the z-axis, $M$, but conserve the total spin, $S$. As a result, the density operators as well as the dynamics with the same $S$ form a series of subspaces labeled by the total spin, $S$. Besides, the similarity between the scaling effect of $g$ and $\hat{S}^\pm$ in Eq.~\eqref{eq:hamiltonian} motivates the exploration through all possible $S$. For the following discussion, we refer to the subspace associated with the total spin $S$ by $S$-subspace and the corresponding steady-state density operators by $\hat{\rho}^S$.

In this work, we explore all the $S$-subspaces so that the critical coupling strength, $g_c$, in Ref.~\cite{PhysRevA.75.013804, PhysRevLett.118.123602} can be generalized to~\cite{kirton2019introduction}:

\begin{equation}
    (g^2 \Tilde{S})_c= \frac{\omega_0}{2\omega_c}\left(\omega_c^2 + \frac{\kappa^2}{4}\right),
    \label{eq:g_c_generalized}
\end{equation}
where $\Tilde{S} = S/(N/2)$, the ratio of a given $S$ to the maximum total spin $S_{max} = N/2$, is called ``normalized total spin'' in this work. Instead of a single critical point $g_c$, Eq. (\ref{eq:g_c_generalized}) defines a critical curve $\Tilde{S}_c(g)$ or $g_c(\Tilde{S})$. In other words, for a fixed coupling $g$, there is a lower bound on the total spin states, corresponding to $\Tilde{S}_c(g)$, that can undergo the superradiant phase transition. Equivalently, for each total spin state, there is a $\Tilde{S}$-dependent lower bound on the coupling for the emergence of the superradiance, associated with $g_c(\Tilde{S})$, as shown in Appendix~\ref{sec:ss_wig}. 

\subsection{Perturbed open Dicke model\label{sec:pert_DDM}}
In the form of Lindblad superoperators, this section describes two $\Delta \mathcal{L}$-type perturbations under investigation in this work: the local dephasing $\Delta\mathcal{L}_\phi[\hat{\rho}]$ and the local decay $\Delta\mathcal{L}_\downarrow[\hat{\rho}]$. The dynamics of local dephasing can be described by~\cite{PhysRevA.98.063815}:
\begin{equation}
    \Delta\mathcal{L}_\phi[\hat{\rho}] = \sum_n \frac{\Gamma_{\phi,n}}{4}(\hat{\sigma}_n^z \hat{\rho} \hat{\sigma}_n^z - \hat{\rho}),
    \label{eq:Delta_L_phi}
\end{equation}
where $\Gamma_{\phi,n}$ is the dephasing rate of the $n$th atom. We note that this perturbation preserves the permutation symmetry only if $\Gamma_{\phi,n} = \Gamma_{\phi}$. We call this process ``homogeneous local dephasing'', where ``homogeneous'' means identical dephasing rate for each atom. 

The local decay dynamics, describing the individual spin losses to the mode outside of the cavity, has the form:
\begin{equation}
    \Delta\mathcal{L}_\downarrow[\hat{\rho}] = \sum_n \Gamma_{\downarrow,n}(\hat{\sigma}_n^- \hat{\rho} \hat{\sigma}_n^+ - \frac{1}{2} \hat{\sigma}_n^+ \hat{\sigma}_n^- \hat{\rho} - \frac{1}{2} \hat{\rho} \hat{\sigma}_n^+ \hat{\sigma}_n^- ),
    \label{eq:Delta_L_down}
\end{equation}
where $\Gamma_{\downarrow, n}$ is the decay rate of the $n$th atom. Similar to the local dephasing, this perturbation preserves the permutation symmetry only when $\Gamma_{\downarrow, n} = \Gamma_{\downarrow}$. This process is named ``homogeneous local decay''.

One can check that even though these two kinds of perturbations preserve the permutation symmetry, they do not conserve the total spin and hence cause the coupling between different $S$-subspaces. One difference between the two perturbations lies in the change in excitations: $\Delta\mathcal{L}_\phi[\hat{\rho}]$ preserves the excitation while $\Delta\mathcal{L}_\downarrow[\hat{\rho}]$ lowers one excitation. To intuitively understand the effects of the two perturbations on $S$, one can imagine a state vector on a generalized Bloch sphere whose radius approximately represents the maximum $S$. The dephasing tends to keep the $z$-component while killing off the coherence (the projection on the $xy$\nobreakdash-plane), shrinking the length of the vector and hence reducing $S$. On the other hand, the local decay push the vector toward lower $z$-component while reducing the $xy$\nobreakdash-component. If the state vector is originally in the half sphere with lower excitation, the projection on the z-axis is enlarged so that the length does not shrink as much as the dephasing case. In other words, local decay tends to preserve $S$ in higher values. By Appendix~\ref{sec:ss_wig}, the superradiant phase requires $S > S_c$ and we can expect that local decay inclines to preserve the phase transition.

\subsection{Degenerate perturbation theory\label{sec:deg_pert_thry}}
Degenerate perturbation theory can explicitly show the coupling between different $S$-subspaces. This section gives a quantitative description of this mixing effect by evaluating the coupling matrix elements; see Ref. \cite{PhysRevA.110.063701,edmonds1996angular} for a more detailed treatment. In this section, the photon degree of freedom is traced out because i)it does not contribute to the degeneracy of the steady states and ii)the perturbations under investigation do not involve photons. In addition to the spin conservation in Sec.~\ref{sec:eom}, the density operators with the same total spin share the same equations of motion and will reach the same steady state. Given the total spin, $S$, and the atom number, $N$, the Dicke state $\ket{S, M}$ with
\begin{equation}
    \hat{S}^2\ket{S, M} = S(S+1)\ket{S, M}, \quad \hat{S}_z\ket{S, M} = M\ket{S, M}
\end{equation}
has the degeneracy of $D_S$~\cite{PhysRev.93.99,PhysRevA.96.023863,PhysRevA.98.063815}:

\begin{equation}
    D_S = (2S + 1)\frac{N!}{(\frac{N}{2} + S + 1)!(\frac{N}{2} - S)!}.
    \label{eq:D_S}
\end{equation} 
By Ref.~\cite{PhysRevA.110.063701}, the steady-state density operators of the unperturbed open Dicke model in each $S$-subspace can be represented by the Dicke states

\begin{equation}
    \hat{\rho}^S = \sum_{M, M^\prime} \rho^S_{M, M^\prime}\ket{S, M}\bra{S, M^\prime},
\end{equation}
and generally have the degeneracy of $D_S^2$. When the permutation symmetry is taken into account, the degree of freedom from the angular momentum coupling can be traced out. As a result, the degeneracy within each $S$-subspace collapses from $D_S^2$ to $1$, and the degeneracy of the overall density operator becomes $N_S$, i.e., the number of possible total spins. These degenerate steady-state density operators can be used as the basis $\{\hat{\rho}^S\}$ for the steady-state density operator of the perturbed model using degenerate perturbation theory. 

Based on the definition of the left-operator:
\begin{equation}
    \hat{\rho}^{S,L} = \sum_{M}\ket{S, M}\bra{S, M},
\end{equation}
the generalized projection of the density operator $\hat{\rho}^{S^\prime}$ on the $S$-subspace resembles the bi-orthogonality of the left- and right-eigenvectors of a non-Hermitian matrix:

\begin{equation}
    Tr(\hat{\rho}^{S^\prime\!,L} \hat{\rho}^S) = \delta_{S S^\prime}.
\end{equation}
The basic idea to evaluate the coupling matrix elements is to apply the perturbation superoperator to the steady-state density operator and evaluate the projection on each $\hat{\rho}^{S^\prime}$:

\begin{equation}
    C_{S^\prime S} = Tr(\hat{\rho}^{S^\prime\!,L} \Delta \mathcal{L}[\hat{\rho}^S]),
    \label{eq:C_SpS}
\end{equation}
where $\Delta \mathcal{L}$, the perturbation operator, can be either the homogeneous dephasing $\Delta \mathcal{L}_\phi$ or the homogeneous individual decay $\Delta \mathcal{L}_\downarrow$ or a mixture of the two. The evaluation process is elaborated in Appendix~\ref{app:evaluation}. One can check that the coupling matrix, under the homogeneous local dephasing and local decay, is a tri-diagonal matrix. This means both the perturbations only couple the adjacent $S$-subspaces.

Despite the tremendous reduction of complexity due to the permutation symmetry, the full density matrix calculation of the steady-state $\{\hat{\rho}^S\}$ becomes challenging for $N \ge 50$. To go beyond this limit, we further derive the expressions of the coupling matrix elements and find that they can be represented just in terms of the expectation values $\langle S_z \rangle$ and $\langle S_z^2 \rangle$ in each $S$-subspaces of the unperturbed system (Appendix~\ref{app:mean_field_calculation}), opening other possibilities to evaluate the coupling matrix elements. These expectation values can be calculated via various methods including the Heisenberg equations of motion (MF2), enabling the calculation for $N \sim 10^5$.

\section{\label{sec:results}Results}

This section shows the effects of the two perturbations on the superradiant phase transition using degenerate perturbation theory. We validate the degenerate perturbation theory via numerical calculations and comparison of the mean-field theory results with the full density matrix calculations. Throughout this paper, all rates are normalized to the cavity decay rate so that $\kappa = 1$ (By the proposal in Ref.~\cite{PhysRevA.75.013804}, the value of $\kappa$ can take $~ 2\pi \times 20$~kHz). We let $\omega_c = 1$, $\omega_0 = 0.5$ so the critical coupling strength for $S = N/2$ is $g_c(\Tilde{S} = 1) \approx 0.56$. Unless otherwise explicitly stated, the coupling strength is $g = 0.9$, leading to $\Tilde{S}_c \approx 0.3858$, Eq.~\eqref{eq:g_c_generalized}. 

The calculation methods in this work are denoted as below: For unperturbed systems, the full density matrix simulation via Eq.~\eqref{eq:eom} is denoted by DM, while the mean-field calculation without/with cumulant expansion up to the second order is denoted by MF1/MF2 (with $\Gamma_\phi = \Gamma_\downarrow = 0$ in Appendix~\ref{app:mean_field_eq}). These methods will be attached with a perturbation strength if any perturbations are directly introduced into the equations of motion (for example, DM at $\Gamma = 10^{-4}$ or MF2 at $\Gamma = 10^{-4}$). Meanwhile, degenerate perturbation theory methods are denoted with a prefix ``DPT''. For example, the degenerate perturbation theory method using $\langle S_z \rangle$ and $\langle S_z^2 \rangle$ from DM/MF2 is denoted by DPT-DM/DPT-MF2.

\subsection{Comparing eigenvalues from degenerate perturbation theory and exact diagonalization of Liouvillian super-operator\label{sec:compare_eigenvalues}}

This section verifies the validity of degenerate perturbation theory by comparing the eigenvalues of the coupling matrix $\hat{C}$ from degenerate perturbation theory and the exact diagonalization of the Liouvillian matrix. Based on the math tool called Fock-Liouville space~\cite{thingna2021degenerated, manzano2020short}, the density operators in the Master equation, $\hat{\rho}$, can be reshaped into vectors, $\vec{\rho}$, and the Lindblad superoperator can be mapped into a matrix, $\hat{\mathscr{L}}$, which is called Liouvillian matrix throughout this work. As a result, the Master equation becomes a matrix-vector product form:

\begin{equation}
    \frac{d\vec{\rho}}{dt} = \hat{\mathscr{L}}\vec{\rho}.
    \label{eq:eom_Liouvillian_vec}
\end{equation}
Since the matrix $\hat{\mathscr{L}}$ is non-Hermitian, the eigenvalues are generally complex, corresponding to left and right eigenvectors subject to biorthogonality. Throughout this work, by ``eigenvector'', we refer to the right eigenvectors unless otherwise stated. The vectorized density operator evolves as:
\begin{equation}
    \vec{\rho}(t) = \sum_n c_n e^{\lambda_n t} \vec{\rho}_n,
    \label{eq:vec_rho_t}
\end{equation}
where $\lambda_n$ is the $n$th eigenvalue corresponding to the $n$th eigenvector $\vec{\rho}_n$:

\begin{equation}
    \lambda_n = -\gamma_n + i\nu_n,
\end{equation}
where $\gamma_n$ indicates the decay rate associated with the $n$th eigenvector while the imaginary part accounts for the oscillatory behavior. In particular, the eigenvalues are either pure real or come in complex conjugate pairs so that the dynamics is hermitian. The parameter $c_n$ is the projection of the initial vector $\vec{\rho}(0)$ on $\vec{\rho}_n$, which can be obtained by

\begin{equation}
    c_n = \vec{\rho}_n^L \cdot \vec{\rho}(0)
\end{equation}
where $\vec{\rho}_n^L$ is the left eigenvector of the $n$th eigenvalue.

For a time-independent Liouvillian, there is at least one eigenvalue with a real part equal to $0$. For the open Dicke model with $\kappa > 0$, these stationary states become the NESS with $\lambda_n = 0$. By Eq.\eqref{eq:vec_rho_t}, if the density operator is prepared in the NESS, it will stay in this state forever. As described in Sec~\ref{sec:deg_pert_thry}, due to the degeneracy, the unperturbed open Dicke model has $N_S$ zero eigenvalues so that the steady state is a linear combination of the corresponding $N_S$ NESS, weighted by their overlaps with the initial state, $c_n$. However, the perturbations of Eqs.~\eqref{eq:Delta_L_phi} and~\eqref{eq:Delta_L_down} lead to only one zero eigenvalue. As a result, the steady state of the perturbed system is unique and not dependent on the initial state. 

\begin{figure}[!htbp]
  \centering
  \includegraphics[width = 0.5\textwidth]{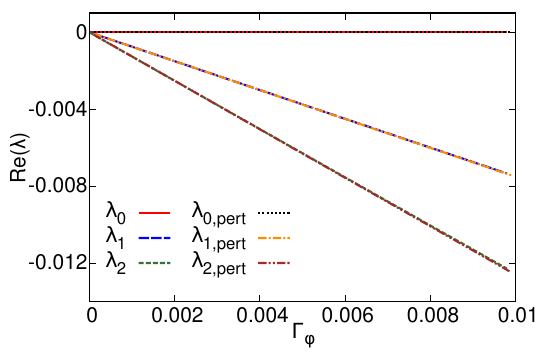} 
  \caption{The cluster of $3$ eigenvalues with
the least negative real parts under the dephasing perturbation, leading to the long-term behavior under investigation. The atom number $N = 4$, with $g = 0.9$, $\omega_c = 1$, $\omega_0 = 0.5$, and $\kappa = 1$. The perturbation strength $\Gamma_\phi$ and the eigenvalues $\lambda$ are scaled to $\kappa$. The results from DPT-DM agree well with the exact diagonalization of the Liouvillian matrix.}
  \label{fig:eig_phi}
\end{figure}

For $N = 4$ atoms, the comparison between the three least negative eigenvalues of the coupling matrix, $\hat{C}$, and the Liouvillian matrix, $\hat{\mathscr{L}}$, is shown in Fig.~\ref{fig:eig_phi} for homogeneous local dephasing. 

The imaginary parts of the eigenvalues are not shown because these eigenvalues are pure real (The imaginary parts from the exact diagonalization of the Liouvillian matrix is $\sim 10^{-14}$, and the degenerate perturbation theory is $\sim 10^{-20}$). When the perturbation strength goes to zero, the largest $3$ eigenvalues approach $0$ and become degenerate. In other words, the degeneracy of the steady states in the unperturbed model is $3$, in agreement with $N_S = 3$ for $N = 4$. Since we are only interested in the long-term behavior of the system, we do not show the fourth and following eigenvalues because they start at $\sim -0.017$ instead of $0$, corresponding to relatively fast decay even without perturbation. When $\Gamma_\phi \neq 0$, the $3$ eigenvalues become proportional to the perturbation strength and split. By Ref.~\cite{PhysRevA.110.063701}, the reason for the linear dependence of the eigenvalues on the perturbation strength is that both homogeneous local dephasing and individual decay only appear in the form of the Lindblad superoperator. By the lowest order degenerate perturbation theory, these perturbations only contribute to the first-order correction to the real parts of the eigenvalues. By Eq.~\eqref{eq:vec_R_dot_C_phi} and \eqref{eq:vec_R_dot_C_down}, the coupling matrix is proportional to the perturbation strength. This not only explains the linear relation between the eigenvalues and the perturbation strength but also implies that the eigenvectors are not dependent on the perturbation strength as $\Gamma_{\phi, \downarrow} \rightarrow 0^+$. As a result, when the perturbation is weak enough, the steady state only depends on the perturbation type but not its strength. 

One can check that the eigenvalues of the coupling matrix agree well with the exact result from the Liouvillian matrix, except for the finite error for the third eigenvalue when the perturbation strength is relatively large ($\Gamma_\phi$ and $\Gamma_\downarrow \sim 0.01$). This is because we are using the lowest-order perturbation theory, and higher-order corrections come into play when the perturbation strength increases. This agreement indicates that degenerate perturbation theory can capture the long-term dynamics of the system, including the steady state behavior. The plot for local decay has similar features and is not shown here. 

\subsection{Effects of homogeneous local dephasing and local decay on NESS\label{sec:frac_phi_down}}

\begin{figure}[!htbp]
  \centering
  \includegraphics[width = 0.5\textwidth]{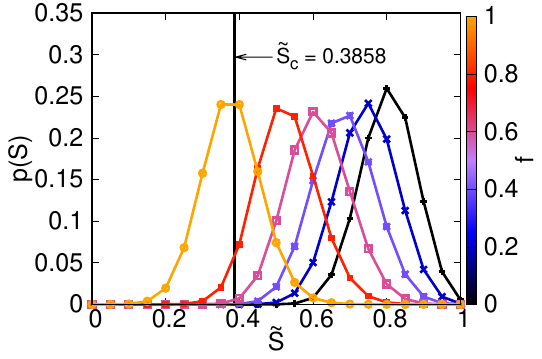}
  \caption{The steady-state probability distribution $p(S)$ under the combined perturbation parameterized by $f = 0, 0.2, 0.4, 0.6, 0.8, 1$ in Eq. (\ref{eq:overall_pert}) obtained from DPT-DM for $N = 40$. The parameters are $g = 0.9$, $\omega_c = 1$, $\omega_0 = 0.5$, and $\kappa = 1$. The superradiant phase transition happens for $\Tilde{S} > 0.3858$ (the region to the right of the vertical black solid line).}
  \label{fig:prob_dist_f}
\end{figure}

\begin{figure}[!htbp]
  \centering
  \includegraphics[width = 0.5\textwidth]{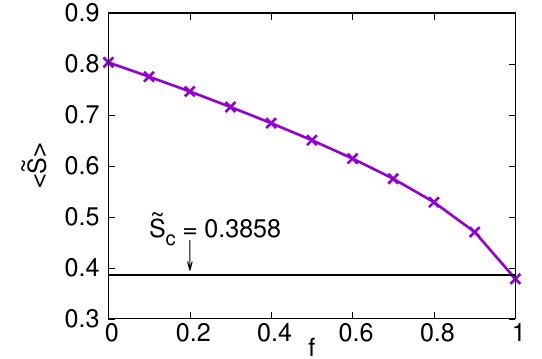}
  \caption{Expectation value of the normalized total spin as a function of $f$. Obtained from DPT-DM for $N = 40$ atoms, the parameters are the same as Fig.~\ref{fig:prob_dist_f}.}
  \label{fig:avg_S}
\end{figure}

Without perturbation, the steady-state behavior of the open Dicke model in different $S$-subspaces is reviewed in Appendix~\ref{sec:ss_wig}, where the phase transition happens near the critical normalized total spin $\Tilde{S}_c$ given a fixed $g$. When the perturbation is present, by Sec.~\ref{sec:deg_pert_thry} and Appendix~\ref{app:evaluation}, the coupling between adjacent total spins leads to a unique null eigenvector of the coupling matrix, $\vec{p}_0 = [p(S_{min}), p(S_{min} +1), ..., p(N/2)]$. By Eq.~\eqref{eq:rho_ss}, the elements of the null eigenvector, $p(S)$, describe the probability of the perturbed system's NESS being in each $S$-subspace and hence is called the ``probability distribution'' in this paper. In this section, we study the behavior of $p(S)$ due to the mixture of perturbations $\Delta \mathcal{L}_\phi$ and $\Delta \mathcal{L}_\downarrow$, parameterized by $f$:

\begin{equation}
    \Gamma_\phi = f\Gamma, \quad \Gamma_\downarrow = (1-f)\Gamma,
\end{equation}
so that:
\begin{equation}
\begin{split}
    \Delta \mathcal{L} &= \Delta \mathcal{L}_\phi + \Delta \mathcal{L}_\downarrow = f\sum_n \frac{\Gamma}{4}(\hat{\sigma}_n^z \hat{\rho} \hat{\sigma}_n^z - \hat{\rho})\\
    &+ (1-f)\sum_n \Gamma(\hat{\sigma}_n^- \hat{\rho} \hat{\sigma}_n^+ - \frac{1}{2} \hat{\sigma}_n^+ \hat{\sigma}_n^- \hat{\rho} - \frac{1}{2} \hat{\rho} \hat{\sigma}_n^+ \hat{\sigma}_n^- ),
\end{split}
\label{eq:overall_pert}
\end{equation}
For $N = 40$, $g = 0.9$, $\omega_c = 1$, $\omega_0 = 0.5$, and $\kappa = 1$, we plot the probability distribution $p(S)$ for various values of $f$ and $\Gamma = 0.01$ in Fig. \ref{fig:prob_dist_f} using DPT-DM, Eq.~\eqref{eq:C_SpS}. To explicitly show the position of the peaks of $p(S)$ under different $f$, the expectation value of the normalized total spin,
\begin{equation}
    \langle \Tilde{S} \rangle = \sum_S p(S) \Tilde{S}, 
    \label{eq:avg_S}
\end{equation}
is shown in Fig.~\ref{fig:avg_S} as a function of $f$, using the $p(S)$ in Fig.~\ref{fig:prob_dist_f}.

The solid black lines in both Figs.~\ref{fig:prob_dist_f} and ~\ref{fig:avg_S} show the critical total spin as a reference. When the homogeneous local dephasing fully dominates the perturbation ($f = 1$, orange curve in Fig.~\ref{fig:prob_dist_f}, the peak of $p(S)$ approaches $\Tilde{S}_c$, leading to $\langle \Tilde{S} \rangle \approx 0.38$, corresponding to the rightmost data point in Fig.~\ref{fig:avg_S}. With increasing contribution from the individual decay ($0 \le f < 1$), the peaks in Fig.~\ref{fig:prob_dist_f} move away from $\Tilde{S}_c$, leading to increasing probability in $\Tilde{S} > \Tilde{S}_c$ and the rising $\langle \Tilde{S} \rangle$ in Fig. \ref{fig:avg_S}. In other words, the position of $p(S)$ is modulated by $f$, leading to different probabilities below/above the threshold of phase transition. Given the steady-state behavior of the open Dicke model in Appendix~\ref{sec:ss_wig}, this probability distribution can be used as weights to calculate the average observables of the perturbed system, such as the Wigner distribution illustrated in Appendix~\ref{sec:wig_perturbed}.

\subsection{The thermodynamic limit\label{sec:gaussian_fit}}
\begin{figure}[!htbp]
  \centering
  \includegraphics[width = 0.5\textwidth]{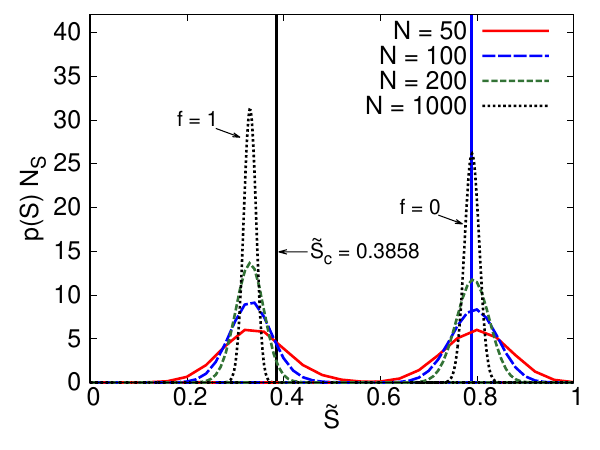}
  \caption{Scaled probability distribution for pure local dephasing ($f = 1$, the peaks on the left) and pure local decay ($f = 0$, the peaks on the right) obtained from DPT-MF2 with $N = 50, 100, 200, 1000$. The parameters are the same as Fig. \ref{fig:prob_dist_f}. The black solid line indicates $\Tilde{S}_c$ and the blue solid line refers to $\langle \Tilde{S} \rangle = 0.7891$ evaluated from MF1 with $f = 0$.}
  \label{fig:pop_dist_scale_N}
\end{figure}

In addition to the perturbations, it is necessary to check the system size effect. To approach the thermodynamic limit, the degenerate perturbation theory is implemented via the MF2 results of the open Dicke model (DPT-MF2, see Appendix~\ref{app:mean_field_calculation}). For $N = 50, 100, 200, 1000$, $g = 0.9$, $\omega_c = 1$, $\omega_0 = 0.5$, and $\kappa = 1$, the probability distribution for $f = 1$ (pure local dephasing) and $f = 0$ (pure local decay) is shown in Fig.~\ref{fig:pop_dist_scale_N}.

\begin{figure}[!htbp]
  \centering
  \includegraphics[width = 0.5\textwidth]{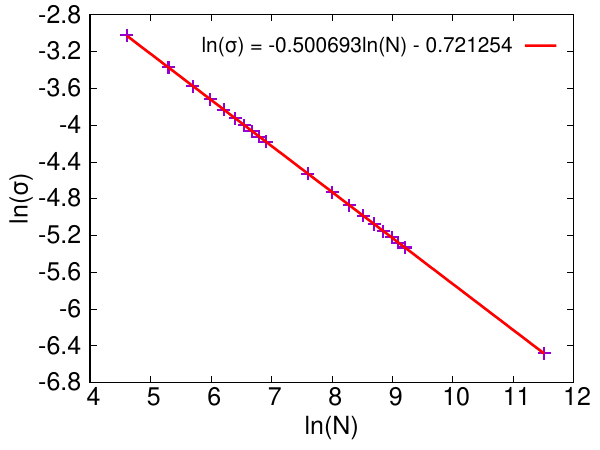}
  \caption{Relation between the logarithm of $\sigma$ and $N$ for the curves with $f = 0$ in Fig.~\ref{fig:pop_dist_scale_N}. The linear relation, indicated by the red straight line from fitting the symbols, infers the power law between them. This relation holds for different 
 ranges of N: $100 \sim 1000$ with increment of $100$, $1000 \sim 10000$ with step of $1000$ and a single point at $N = 100000$.}
  \label{fig:fit_gau_pop_dist}
\end{figure}
Note that here we scale the probability $p(S)$ by the number of possible total spin $N_S$ to avoid $p(S) \rightarrow 0$ when $N \rightarrow \infty$. This is because $\sum_S p(S) = 1$ has $N_S$ terms - greater $N$ leads to more terms and $p(S)$ scales as $\sim 1/N_S$ to ensure total probability conservation. One can notice that the distribution becomes increasingly narrower with increasing $N$. To study the dependence of the width on the number of atoms, the Gaussian function:
\begin{equation}
    p_G(x) = \frac{A}{\sigma\sqrt{2\pi}}e^{-\frac{(x-\Bar{x})^2}{2\sigma^2}}
\end{equation}
is used to fit the curves in Fig.~\ref{fig:pop_dist_scale_N} as well as for higher $N$ up to $10^5$, whose standard deviation $\sigma$ is a measure of the distribution width. As is shown in Fig.~\ref{fig:fit_gau_pop_dist}, the logarithm of $\sigma$ and $N$ shows a linear relation.

This indicates a power law:
\begin{equation}
    \sigma \sim N^\beta.
    \label{eq:N_beta}
\end{equation}
As is shown in the legend, fitting the symbols yields the power $\beta \approx -0.500693$. So the width of the probability distribution is approximately to the order of $1/\sqrt{N}$. The same procedures are performed for other $f$, leading to the same scaling with $N$, except for a narrowing region near $f = 1$ with lower $\beta$, which can be seen in Fig.~\ref{fig:fit_power} in Appendix~\ref{app:other_f}.

\begin{figure}[!htbp]
  \centering
  \includegraphics[width = 0.5\textwidth]{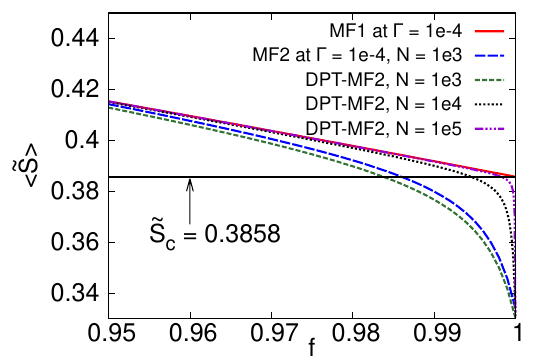}
  \caption{Same as Fig.~\ref{fig:avg_S} except that $\langle \Tilde{S} \rangle$ evaluated using MF2 at $\Gamma = 10^{-4}$, DPT-MF2, and MF1 at $\Gamma = 10^{-4}$ as a reference.}
  \label{fig:avg_S_MF1_MF2_pert_degen}
\end{figure}

In addition to the width, we also investigate the peak position of $p(S)$ by comparing to $\Tilde{S}_c$ (black solid line) and $\langle \Tilde{S} \rangle$ evaluated from MF1 (blue solid line) in Fig.~\ref{fig:pop_dist_scale_N}. 
One can check that for pure local decay ($f = 0$), the peak approaches the MF1 result as $N$ grows, but for the pure local dephasing case ($f = 1$), it is below $\Tilde{S}_c$, regardless of $N$. Here we claim that degenerate perturbation theory is valid while the discrepancy between the peak position of $p(S)$ using MF2 results and the $\langle \hat{S} \rangle$ from MF1 calculations only appears at $f = 1$. To justify degenerate perturbation theory, we compare the $\langle \Tilde{S} \rangle$ obtained: i) directly from MF2 at $\Gamma = 10^{-4}$, and ii) by Eq.~\eqref{eq:avg_S}, with $p(S)$ obtained from DPT-MF2 in Fig. \ref{fig:avg_S_MF1_MF2_pert_degen}.

One can check that for $N = 10^3$ the behavior of DPT-MF2 agrees well with MF2 at $\Gamma = 10^{-4}$ and the relative error $E_{r}$ becomes increasingly small as $N$ becomes larger ($E_r \sim 10^{-4}$ when $N = 10^5$). In other words, degenerate perturbation theory correctly predicts the steady-state behavior of the perturbed equations of motion - this supports the validity of degenerate perturbation theory and we do not distinguish between the $\langle \Tilde{S} \rangle$ obtained from DPT-MF2 and MF2 at $\Gamma = 10^{-4}$ in the following discussion. As a reference corresponding to the thermodynamic limit, with the same parameters, we also plot the $\langle \Tilde{S} \rangle$ obtained from MF1 at $\Gamma = 10^{-4}$. When $f < 0.95$, both DPT-MF2 and MF2 at $\Gamma = 10^{-4}$ agree well with MF1 at $\Gamma = 10^{-4}$, but when $f \rightarrow 1$ the two curves (DPT-MF2 and MF2 at $\Gamma = 10^{-4}$) with $N = 10^3$ gradually deviate from the MF1 result. To understand the dependence of this discrepancy on $N$, the same calculations of DPT-MF2 for $N = 10^4$ and $N = 10^5$ are also plotted in Fig.~\ref{fig:avg_S_MF1_MF2_pert_degen}. One can notice that the DPT-MF2 results are consistently approaching the MF1 result except for the $f = 1$ case, where $\langle \Tilde{S} \rangle \approx 0.3311$ regardless of $N$.

\begin{figure}[!htbp]
  \centering
  \includegraphics[width = 0.5\textwidth]{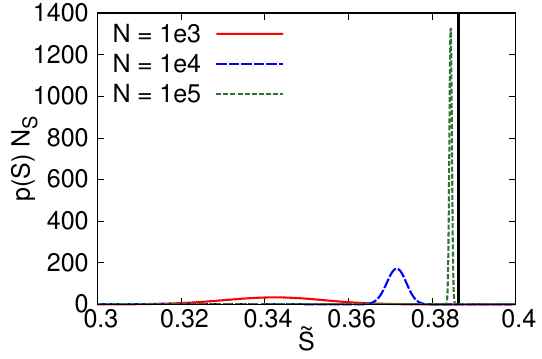}
  \caption{Same as Fig.~\ref{fig:pop_dist_scale_N} except that $f = 0.999$ and $N = 10^3, 10^4, 10^5$. The black solid line shows $\langle \Tilde{S} \rangle = 0.3864$ evaluated by MF1 at $\Gamma = 10^{-4}$.}
  \label{fig:pop_dist_scale_N_f0.999}
\end{figure}

This work focuses on the peak position of $p(S)$ in two cases: $f = 1$ and $f \rightarrow 1^-$, corresponding to the pure local dephasing and the introduction of infinitesimal local decay. The consistency between MF1 at $\Gamma = 10^{-4}$ and DPT-MF2 (or MF2 at $\Gamma = 10^{-4}$) facilitates the prediction of the peak position of $p(S)$ other than $f = 1$ and especially for $f \rightarrow 1^-$. In Fig.~\ref{fig:pop_dist_scale_N_f0.999}, we plot scaled probability distribution the same as Fig.~\ref{fig:pop_dist_scale_N} except that $f = 0.999$ and $N = 10^3, 10^4, 10^5$. One can see that the probability distribution is narrowing and approaching the $\langle \Tilde{S} \rangle$ evaluated by MF1 at $\Gamma = 10^{-4}$ (black solid line). As is shown in Fig.~\ref{fig:avg_S_MF1_MF2_pert_degen} (red curve), this $\langle \Tilde{S} \rangle$ approaches $\Tilde{S}_c$ from above as $f \rightarrow 1^-$ (One can check that the $\langle \Tilde{S} \rangle$ obtained from MF1 at $\Gamma = 10^{-4}$ at $f = 1$ coincides with $\Tilde{S}_c$.).
 
\begin{figure}[!htbp]
  \centering
  \includegraphics[width = 0.5\textwidth]{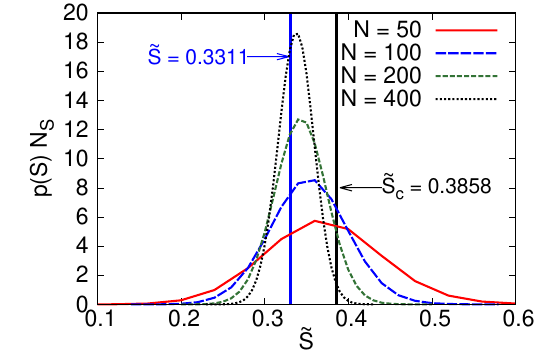}
  \caption{Same as Fig.~\ref{fig:pop_dist_scale_N} except for the $p(S)$ obtained from DPT-DM for $N = 50, 100, 200$ and $400$ at $f = 1$. The black solid line refers to $\Tilde{S}_c$ and the blue solid line indicates $\langle \Tilde{S} \rangle$ evaluated by DPT-MF2 for $N = 10^5$.}
  \label{fig:pop_dist_scale_N_DM}
\end{figure}

Based on the discussion above, we can infer that 
\begin{equation}
    p(S) \rightarrow \delta(S - S_c^+)
    \label{eq:p_S_thermo_limit_f_approach_1}
\end{equation}
in the thermodynamic limit when approaching only dephasing ($f \rightarrow 1^-$). As for the peak position of $p(S)$ exactly at $f = 1$, there are two possible values indicated by the $\langle \Tilde{S} \rangle$ obtained from MF1 at $\Gamma = 10^{-4}$ and DPT-MF2(or MF2 at $\Gamma = 10^{-4}$, equivalently) in Fig.~\ref{fig:avg_S_MF1_MF2_pert_degen}. To confirm which one is accurate,  we plot the $p(S)$ obtained from DPT-DM with $N = 50, 100, 200, 400$ in Fig.~\ref{fig:pop_dist_scale_N_DM}, compared to $\Tilde{S}_c$ (or $\langle \Tilde{S} \rangle$ obtained from MF1 at $\Gamma = 10^{-4}$, black solid line) and the $\langle \Tilde{S} \rangle$ evaluated by DPT-MF2 for $N = 10^5$ (or MF2 at $\Gamma = 10^{-4}$, blue solid line). One can check that the full density matrix $p(S)$ is narrowing and approaching the MF2 result with growing $N$. In other words, at $f = 1$, the MF2 result indicates the correct peak position of $p(S)$, yielding $100\%$ population below $\Tilde{S}_c$ in the thermodynamic limit. This explains why an infinitesimal dephasing destroys the superradiant phase transition. Meanwhile, an infinitesimal individual decay, bringing all the population above $\Tilde{S}_c$ (by Eq.\eqref{eq:p_S_thermo_limit_f_approach_1}), will restore the phase transition.

In addition to the steady state, degenerate perturbation theory can also reveal the information about long-term transient dynamics that is not available from mean-field methods only. For example, Appendix~\ref{app:Liouvillian_spectrum_eig_vec} illustrates the eigenvalues of the coupling matrix (or Liouvillian spectrum) as a function of the mixing ratio $f$ as well as the first $4$ eigenvectors. The properties of these states and the reason for such patterns can be interesting topics for future research.

\section{\label{sec:summary}Summary}

This work investigates the effects of homogeneous local dephasing and individual decay on the superradiant phase transition of the standard open Dicke model from the view of degenerate perturbation theory. Without perturbations, the dynamics and all the states with the same total spin, $S$, form a closed subspace referred to as $S$-subspace. Going beyond the $S = N/2$-subspace considered in the standard open Dicke model, the exploration among different $S$-subspaces of the unperturbed model reveals a critical behavior in terms of the total spin: Given a coupling $g$, there exists a critical total spin $S_c$ so that the phase transition only appears in the regime $S > S_c$. This $g$-$S$ co-determined critical behavior is described by a boundary curve and verified by the phase diagram in the $g$-$S$ plane. 

In addition to the steady-state driven Dicke superradiance in the driven Dicke model, applying degenerate perturbation theory to the open Dicke model yields a different perspective and discusses a different phenomenon - the superradiant phase transition. The coupling matrix $\hat{C}$ explicitly shows that the homogeneous dephasing and individual decay mix the adjacent total spins, rendering the steady state a mixture of different total spins. This mixture can be described by the probability distribution on each $S$-subspace, denoted by $p(S)$, which is the null eigenvector of $\hat{C}$. The width of $p(S)$ narrows as $1/\sqrt{N}$, rendering a $\delta$-function distribution asymptotically in the thermodynamic limit. The peak position of this $\delta$-function can be modulated by the ratio between the two perturbations. In particular, an infinitesimal pure local dephasing renders all the population below $S_c$ and destroys the phase transition, while an infinitesimal local decay, moving all the population above $S_c$, restores the superradiant phase transition. This gives a complementary interpretation to the results in Ref.~\cite{PhysRevLett.118.123602}.

Benefiting from the degeneracy of the steady states of the open Dicke model, the degenerate perturbation theory is useful to investigate the significant modulation of the steady-state behavior due to infinitesimal perturbations. Since this method only requires steady-state $\langle \hat{S}_z \rangle$ and $\langle \hat{S}^2_z \rangle$ in each $S$-subspace, various methods giving these two moments, including mean-field theory, can be used so that the calculation for very large $N$ is feasible. This technique can be used for other systems having high degeneracy in the steady states like generalized Dicke models~\cite{PhysRevLett.118.080601,hu2021out,PhysRevResearch.4.023101,PhysRevA.109.053702}, two-photon Dicke models~\cite{PhysRevA.109.053702,shah2024dissipative,BANERJEE2022128287,PhysRevA.95.053854,Garbe2020dissipation,PhysRevB.102.245407,PhysRevA.97.053821}, and driven Dicke model~\cite{leppenen2024quantumbistabilityinterplaycollective}. In addition to the steady state, degenerate perturbation theory can also reveal the long-term transient dynamics by the Liouvillian spectrum modulated by $f$ and the corresponding eigenvectors (see Appendix~\ref{app:Liouvillian_spectrum_eig_vec}). Investigating the properties of these slowly decaying states is still an open question.

\begin{acknowledgements}

This work was supported by the National Science Foundation under Award No. 2410890-PHY (WT and FR). Research supported as part of QuPIDC, an Energy Frontier Research Center, funded by the US Department of Energy (DOE), Office of Science, Basic Energy Sciences (BES), under award number DE-SC0025620 (HA). This research was supported in part through computational resources provided by Information Technology at Purdue University, West Lafayette, Indiana.

WT mainly wrote the manuscript. WT and FR did calculations, HA and FR supervised the project, and all other aspects (editing the manuscript, generating ideas, deriving equations) were performed by WT, HA, and FR.

\end{acknowledgements}

\section{Data availability}
The data plotted in the figures are available in Ref.~\cite{data_ODM_DPT}.

\appendix

\section{Recap the open Dicke model: Steady-state behavior in different $S$-subspaces\label{sec:ss_wig}}

Although the results in this section can be easily found in or derived from Ref.~\cite{PhysRevLett.118.123602}, for completeness, we include the steady-state behavior of all $S$-subspaces without the perturbations. This is instructive to understand the coupling between different $S$-subspaces due to the homogeneous local dephasing and local decay. For the following discussion, by ``Wigner distribution'', we refer to the Wigner distribution for the photon part. 

\begin{figure}[!htbp]
  \centering
  \includegraphics[width = 0.5\textwidth]{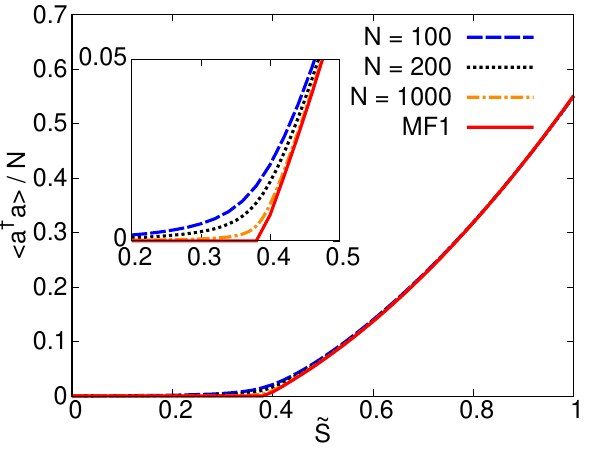}
  \caption{Steady-state average photon number for $N=100, 200, 1000$ from MF2 and, in the thermodynamic limit, obtained from MF1 at $g = 0.9$, $\omega_c = 1$, $\omega_0 = 0.5$, and $\kappa = 1$.}
  \label{fig:scl_ph_num_MF}
\end{figure}

As implied by Eq.~\eqref{eq:g_c_generalized}, the critical total spin, $\Tilde{S}_c(g)$, is a function of the coupling strength, and vice versa. To quantitatively show this relation, we start from a fixed $g$ to find out the corresponding $\Tilde{S}_c$ and then vary $g$ to obtain $\Tilde{S}_c(g)$. For a fixed $g = 0.9$, by Eq.~\eqref{eq:g_c_generalized}, the critical normalized total spin is $\Tilde{S}_c(g = 0.9) \approx 0.3858$. To verify this value, we time integrate the mean-field equations with cumulant expansion up to the second order (MF2) from Ref. \cite{PhysRevLett.118.123602} for $S_{min} \le S \le N/2$, where 

\begin{equation*}
    S_{min} = (N\ mod\ 2) / 2. 
\end{equation*}
The steady-state average photon number obtained from MF2 equations, $\langle a^\dagger a \rangle / N$, is used as the order parameter. In addition to $N = 100, 200, 1000$, we also include the first-order mean-field calculation (MF1), corresponding to the thermodynamic limit, as a reference. The relation between the steady-state average photon number and the normalized total spin, $\Tilde{S}$, is shown in Fig.~\ref{fig:scl_ph_num_MF}.

The behavior of the four curves corresponding to four system sizes is the same when $\Tilde{S} \rightarrow 0$ and $\Tilde{S} \rightarrow 1$, and the difference lies in the region near the critical normalized total spin ($0.2 \le \Tilde{S} \le 0.5$), which is shown in the inset. One can notice that for MF2 calculation, the average photon number increases smoothly. However, the transition becomes increasingly sharp and, in the limit $N \rightarrow \infty$ indicated by MF1, there is a clear critical point at $\Tilde{S} \approx 0.38$, featuring the emergence of non-zero average photon number. This corroborates with the critical total spin $\Tilde{S}_c(g=0.9) \approx 0.3858$ by Eq.~\eqref{eq:g_c_generalized}. 

\begin{figure}[!htbp]
  \centering
  \includegraphics[width = 0.5\textwidth]{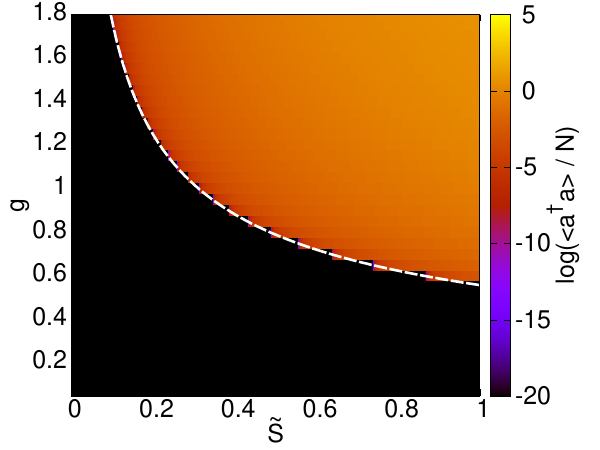}
  \caption{Phase diagram on the  $\Tilde{S}$-$g$ plane for $N = 1000$, $\omega_c = 1$, $\omega_0 = 0.5$, and $\kappa = 1$ using MF1 calculation. The white dashed line is the bondary obtained from Eq. (\ref{eq:g_c_generalized}).}
  \label{fig:scaled_photon_number_Nbit1000}
\end{figure}

In addition to a single critical normalized total spin for a fixed $g$, Eq.~\eqref{eq:g_c_generalized} implies a set of critical points, $(\Tilde{S}, g)$, constituting a boundary in the $\Tilde{S}$-$g$ plane. To verify this, we extend the coupling strength to a range $0.05 \le g \le 1.80$ with the increment of $0.05$ and plot the phase diagram for the open Dicke model in Fig.~\ref{fig:scaled_photon_number_Nbit1000}.

The phase diagram has a clear boundary between the normal phase and the superradiant phase, to the left of which $\langle a^\dagger a\rangle / N \rightarrow 0$ while to the right $\langle a^\dagger a\rangle / N > 0$. Meanwhile, the boundary obtained from Eq.~\eqref{eq:g_c_generalized} (white dashed line) matches the boundary of the phase diagram very well. 

\begin{figure}[!htbp]
  \centering
  \includegraphics[width = 0.5\textwidth]{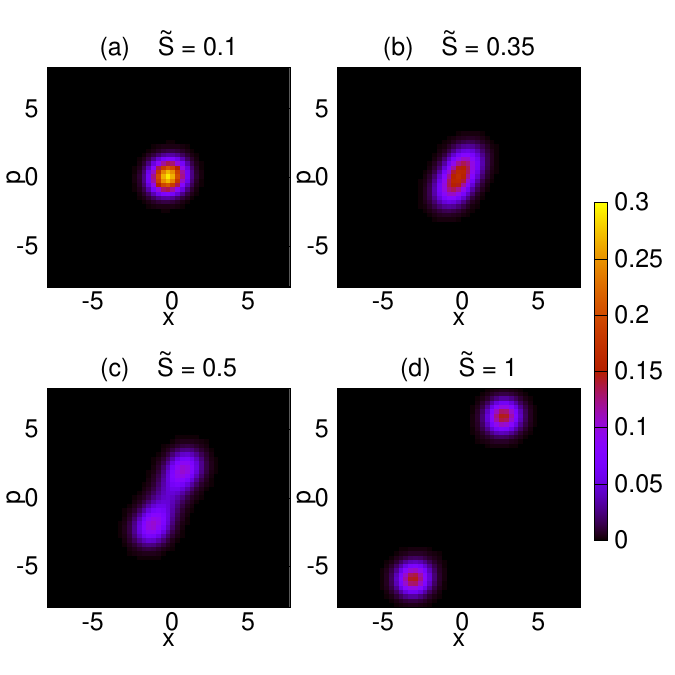}
  \caption{Steady-state Wigner distributions of the photon part for $\Tilde{S} = 0.1, 0.35, 0.5$, and $1$, corresponding to $S = 2, 7, 10$, and $20$. These plots are obtained from DM for $N = 40$ atoms at at $g = 0.9$, $\omega_c = 1$, $\omega_0 = 0.5$, and $\kappa = 1$. }
  \label{fig:wig_pure}
\end{figure}

In addition to the quantitative results using the average photon number, the Wigner distribution is an intuitive tool to understand the behavior of the system for different $S$-subspaces. By Ref.~\cite{PhysRevLett.118.123602}, the superradiant phase, breaking the $\mathbb{Z}_2$ symmetry in thermodynamic limit, is characterized by two well-separated peaks. For $N = 40$ atoms, with $g = 0.9$, $\omega_c = 1$, $\omega_0 = 0.5$, and $\kappa = 1$, we perform full density matrix calculations and trace out the spin part of the steady-state density operator in each $S$-subspace. The consequent reduced density operator for photons can be used to evaluate the steady-state Wigner distribution for each total spin by standard techniques~\cite{PhysRev.40.749,doi:https://doi.org/10.1002/3527602976.ch3}. In Fig.~\ref{fig:wig_pure}, we show the Wigner distributions for a few total spins to demonstrate that the phase transition depends on the total spin.

Note that $g = 0.9 > g_c(\Tilde{S} = 1) = 0.56$ is large enough to see the phase transition in the $S = N/2$-subspace, characterized by two well-separated peaks in $\Tilde{S} = 1$ case in Fig. \ref{fig:wig_pure} (d). These two lobes move progressively closer as $\Tilde{S}$ decreases, until $\Tilde{S} = 0.5$ - as is shown in panel (c), the two lobes are merging. When $\Tilde{S}$ is lower than the critical value, as in panel (b), the two peaks merge into one with finite stretching. Finally, when $\Tilde{S}$ is far below the threshold, as in panel (a), there is only a single lobe at the origin. 

\section{Steady-state Wigner function using degenerate perturbation theory\label{sec:wig_perturbed}}
\begin{figure}[!htbp]
  \centering
  \includegraphics[width = 0.5\textwidth]{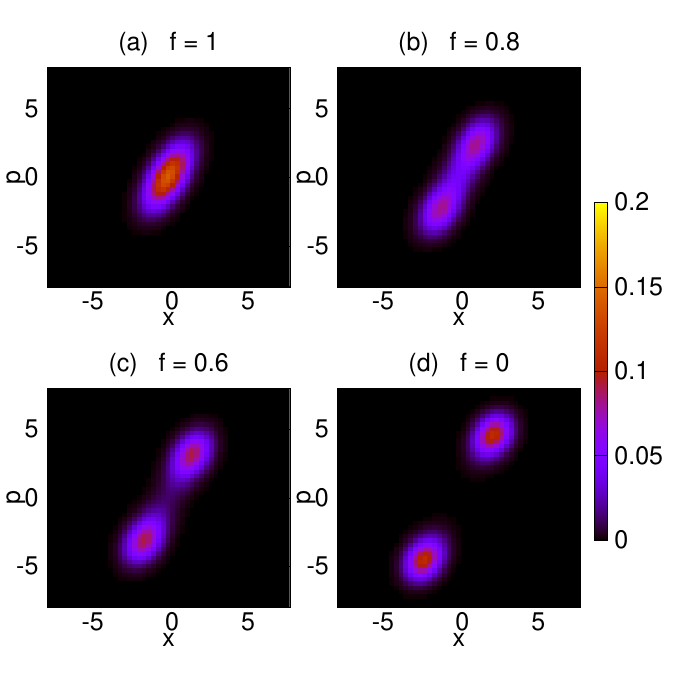}
  \caption{Steady-state Wigner distribution of the perturbed open Dicke model parameterized by Eq.~\eqref{eq:overall_pert} for different fractions $f$ using DPT-DM. The parameters are the same as Fig.~\ref{fig:prob_dist_f}.}
  \label{fig:wig_pert}
\end{figure}

Using the probability distribution $p(S)$ in Fig.~\ref{fig:prob_dist_f}, the Wigner distribution for the perturbed system, $W$, can be obtained by:

\begin{equation}
    W = \sum_S p(S) W_S,
\end{equation}
where $W_S$ refers to the steady-state Wigner distribution of $\hat{\rho}^S$, depicted in Fig.~\ref{fig:wig_pure}. The steady-state Wigner distribution of the perturbed system for $N = 40$ for various $f$ is shown in Fig.~\ref{fig:wig_pert}. 

Corresponding to the orange curve in Fig.~\ref{fig:prob_dist_f}, the full dephasing case ($f = 1$), with the peak almost located at the critical total spin, gives only one lobe at the center with finite stretching (Fig.~\ref{fig:wig_pert} (a)). With increasing contribution from individual decay ($f = 0$), the weights on the higher total spin (and hence more separated lobes in Fig.~\ref{fig:wig_pure}) increase, and the lobe starts to split(Fig.~\ref{fig:wig_pert} (b) and (c)). Finally, for $f = 0$, the pure individual decay perturbation results in two well-separated lobes(Fig.~\ref{fig:wig_pert} (d)). This observation agrees with the trends in Ref.~\cite{PhysRevLett.118.123602}.

\section{Perturbation theory in the degenerate subspace, evaluation of the coupling matrix elements\label{app:evaluation}}

Using the degenerate steady-state density operators $\hat{\rho}^S$ as a basis, the steady-state density operator within the subspace spanned by $\{\hat{\rho}^S\}$ can be written as:

\begin{equation}
    \hat{\rho}_{ss} = \sum_{S} p(S) \hat{\rho}^S.
    \label{eq:rho_ss}
\end{equation}
Since both $\hat{\rho}$ and $\hat{\rho}^S$ are hermitian and have unity trace, $p(S)$ must be real and sum up to $1$. By Eq.~\eqref{eq:pS}, $p(S)$ is non-negative, so $p(S)$ can be considered as a kind of probability. In this work, by ``probability distribution'', we mean the profile of $p(S)$ indicating the probability that the density operator $\hat{\rho}$ can be observed in the density operator $\hat{\rho}^S$.

As an analogy to the left and right eigenvectors for a non-hermitian matrix, we define the left operator:
\begin{equation}
    \hat{\rho}^{S,L} = \sum_{M}\ket{S, M}\bra{S, M}
    \label{eq:rho_eta_L}
\end{equation}
so that the inner product of the left and right operators can be defined as the trace of their matrix product:

\begin{equation}
\begin{split}
    &Tr(\hat{\rho}^{S,L} \hat{\rho}^{S^\prime}\!) \! = \! Tr[(\sum_{\mu} \! \ket{S, \mu} \!\! \bra{S, \mu})\!(\!\!\sum_{M, M^\prime} \!\!\! \rho^{S^\prime}_{M M^\prime} \!\ket{S^\prime\!,\! M} \!\! \bra{S^\prime\!,\! M^\prime})]\\
    & = \delta_{S S^\prime},
    \label{eq:rho_eta_L_orth}
\end{split}
\end{equation}
which implies
\begin{equation}
    p(S) = Tr(\hat{\rho}^{S,L} \hat{\rho}_{ss}) \ge 0.
    \label{eq:pS}
\end{equation}

The essence of the perturbation theory is to evaluate the transition among the basis operators $\hat{\rho}^S$ due to the perturbation. 

For dephasing, the perturbed dynamics can be written as:

\begin{equation}
    \Delta \Dot{\hat{\rho}} = \sum_n \frac{\Gamma_\phi}{4}(\hat{\sigma}_n^z \hat{\rho} \hat{\sigma}_n^z - \hat{\rho}).
    \label{eq:eom_perturbation}
\end{equation}

For first-order perturbation theory, we substitute the steady-state density operator in Eq.~\eqref{eq:rho_ss} into the perturbed dynamics in Eq.~\eqref{eq:eom_perturbation} and take the projection on $\hat{\rho}^S$ by acting the left operator $\hat{\rho}^{S, L}$ on Eq.~\eqref{eq:eom_perturbation} and taking the trace:

\begin{equation}
\begin{split}
    \Dot{p}(S) = Tr(\hat{\rho}^{S,L} \mathcal{L}[\hat{\rho}_{ss}]),
\end{split}
\label{eq:dot_R_S}
\end{equation}

By Ref.~\cite{PhysRevA.98.063815}, thanks to the permutation symmetry, we can trace out the angular momentum coupling degree of freedom, yielding the dynamics of the perturbation:

\begin{equation}
\begin{split}
    \mathcal{L}[\hat{\rho}_{ss}] &= \Gamma_\phi \!\!\!\!\!\! \sum_{S^\prime, M, M^\prime} \!\!\!\!\! \rho_{M, M^\prime}^{S^\prime}[(a_{zz}(S^\prime\!\!, M, M^\prime)\! -\! \frac{N}{4})\!\ket{S^\prime\!\!, M} \! \bra{S^\prime\!\!, M^\prime} \\
    &+ b_{zz}(S^\prime, M, M^\prime)\ket{S^\prime - 1, M} \! \bra{S^\prime - 1, M^\prime} \\
    &+ c_{zz}(S^\prime, M, M^\prime)\ket{S^\prime + 1, M} \! \bra{S^\prime + 1, M^\prime}]
\end{split}
\label{eq:L_rho_ss}
\end{equation}
where the expression of $a_{zz}(S^\prime, M, M^\prime)$, $b_{zz}(S^\prime, M, M^\prime)$, and $c_{zz}(S^\prime, M, M^\prime)$ can be found in Ref.~\cite{PhysRevA.98.063815}. As a result, Eq.~\eqref{eq:dot_R_S} becomes:

\begin{equation}
\begin{split}
    \Dot{p}(S) &= \Gamma_\phi \sum_{M} [p(S)\rho^S_{M, M}(a_{zz}(S, M, M) - \frac{N}{4}) \\
    &+ p(S + 1)\rho^{S+1}_{M, M} b_{zz}(S + 1, M, M) \\
    &+ p(S - 1) \rho^{S - 1}_{M, M} c_{zz}(S - 1, M, M)]
\end{split}
\end{equation}
or in vector form:
\begin{equation}
    \Dot{\vec{p}} = \Gamma_\phi \hat{O}_\phi \vec{p} = \hat{C} \vec{p},
    \label{eq:vec_R_dot_C_phi}
\end{equation}
where $\vec{p} = [p(S_{min}), p(S_{min} + 1), ..., p(N/2)]$ and $\hat{O}_\phi$ is an $N_S \times N_S$ matrix corresponding to the coupling matrix under unit dephasing rate, with matrix element defined as:
\begin{equation}
\begin{split}
    O_\phi(S, S) &= \sum_M [\rho^S_{M, M}(a_{zz}(S, M, M) - \frac{N}{4})],\\
    O_\phi(S, S + 1) &= \sum_M [\rho^{S+1}_{M, M} b_{zz}(S + 1, M, M)],\\
    O_\phi(S, S - 1) &= \sum_M [\rho^{S - 1}_{M, M} c_{zz}(S - 1, M, M)]
\end{split}
\label{eq:cpl_mat_den_phi}
\end{equation}

The matrix elements for individual decay can be obtained using the same strategy, with the form of the dynamics resembling Eq.~\eqref{eq:vec_R_dot_C_phi}:
\begin{equation}
    \Dot{\vec{p}} = \Gamma_\downarrow \hat{O}_\downarrow \vec{p} = \hat{C} \vec{p},
    \label{eq:vec_R_dot_C_down}
\end{equation}
where $\hat{O}_\downarrow$ is an $N_S \times N_S$ matrix corresponding to the coupling matrix under unit individual decay rate. Here, we only give the final results:
\begin{equation}
\begin{split}
    O_\downarrow(S, S) &= \{\sum_M [\rho^S_{M+1, M+1}a_{--}(S, M+1, M+1) \\
    &- M\rho^S_{M, M}]\} - \frac{N}{2}\\
    O_\downarrow(S, S + 1) &= \sum_M [\rho^{S+1}_{M+1, M+1}b_{--}(S + 1, M+1, M+1)]\\
    O_\downarrow(S, S - 1) &= \sum_M [\rho^{S - 1}_{M+1, M+1} c_{--}(S - 1, M+1, M+1)]
\end{split}
\label{eq:cpl_mat_den_down}
\end{equation}

\section{Degenerate perturbation theory using moments up to the second order\label{app:mean_field_calculation}}
One can check that the evaluation of coupling matrix elements in Eq.~\eqref{eq:cpl_mat_den_phi} and \eqref{eq:cpl_mat_den_down} involves the steady-state density operators, which is expensive to obtain even when $N = 50$. Fortunately, the forms of the expressions resemble the expectation values of two observables. Further derivation reveals the expressions of the matrix elements in terms of the expectation values of $\hat{S}_z$ and $\hat{S}_z^2$:
\begin{equation}
\begin{split}
    O_\phi(S, S) &= \frac{\frac{N}{2} + 1}{2S(S+1)}\langle\hat{S}_z^2\rangle_S - \frac{N}{4},\\
    O_\phi(S, S + 1) &= \frac{\frac{N}{2} + S + 2}{2(S+1)(2S+3)}[(S+1)^2 - \langle\hat{S}_z^2\rangle_{S+1}],\\
    O_\phi(S, S - 1) &= \frac{\frac{N}{2} - S + 1}{2S(2S-1)}[S^2 - \langle\hat{S}_z^2\rangle_{S-1}]
\end{split}
\label{eq:cpl_mat_exp_phi}
\end{equation}

\begin{equation}
\begin{split}
    O_\downarrow(S, S) & = \frac{\frac{N}{2} + 1}{2S(S+1)}[S(S+1) - \langle\hat{S}_z^2\rangle_S + \langle\hat{S}_z\rangle_S]\\
    &- \langle \hat{S}_z \rangle_S - \frac{N}{2},\\
    O_\downarrow(S, S + 1) & = \frac{\frac{N}{2} + S + 2}{2(S+1)(2S+3)}[S(S+1) + \langle\hat{S}_z^2\rangle_{S+1} \\
    &+ (2S+1)\langle\hat{S}_z\rangle_{S+1}],\\
    O_\downarrow(S, S - 1) & = \frac{\frac{N}{2} - S + 1}{2S(2S-1)}[S(S+1) + \langle\hat{S}_z^2\rangle_{S-1} \\
    &- (2S+1)\langle\hat{S}_z\rangle_{S-1}]
\end{split}
\label{eq:cpl_mat_exp_down}
\end{equation}
where $\langle \cdot \rangle_S$ means the steady-state expectation value within the $S$-subspace. Involving only expectation values instead of density matrix elements, the mean-field calculation, instead of the full density matrix simulation, can be used to evaluate the coupling matrix elements, facilitating calculation for atom number $N\sim10^5$ or even higher. The mean-field equations for both MF1 and MF2 can be found in Appendix~\ref{app:mean_field_eq}~\cite{PhysRevLett.118.123602}.

\section{Mean-field equations with cumulant expansion upto the second order \label{app:mean_field_eq}}

For completeness, this section includes the mean-field equations in Ref.~\cite{PhysRevLett.118.123602}. For an arbitrary operator $\hat{A}$, the dynamics of its expectation value can be obtained by $\partial_t\langle \hat{A} \rangle = Tr(\hat{A}\Dot{\hat{\rho}})$ and calculating the commutators between $\hat{A}$ and operators in the Master equation. For simplicity, here we define $g^\prime = g/\sqrt{N}$ and $\Tilde{\Gamma} = (\Gamma_\phi + \Gamma_\downarrow)/2$. The mean-field calculation to the first order (MF1) includes the following $4$ equations:

\begin{equation}
    \partial_t\langle \hat{a} \rangle = -(i\omega_c + \frac{\kappa}{2}) \langle \hat{a} \rangle - i 2 g^\prime \langle \hat{S}_x \rangle
\end{equation}

\begin{equation}
    \partial_t\langle \hat{S}_x \rangle = -2\omega_0 \langle \hat{S}_y \rangle - \Tilde{\Gamma} \langle \hat{S}_x \rangle
\end{equation}

\begin{equation}
    \partial_t\langle \hat{S}_y \rangle = 2 \omega_0 \langle \hat{S}_x \rangle - 4 g^\prime Re[\langle \hat{a} \hat{S}_z \rangle] - \Tilde{\Gamma} \langle \hat{S}_y \rangle
\end{equation}

\begin{equation}
    \partial_t\langle \hat{S}_z \rangle = 4 g^\prime Re[\langle \hat{a} \hat{S}_y \rangle] - \Gamma_\downarrow (\langle \hat{S}_z \rangle + \frac{N}{2})
\end{equation}
To calculate the mean-field equations up to the second order (MF2), in addition to the equations above, the following equations should be taken into account:

\begin{equation}
    \partial_t\langle \hat{a}^\dagger \hat{a}\rangle = - \kappa \langle \hat{a}^\dagger \hat{a} \rangle - 4 g^\prime Im[\langle \hat{a} \hat{S}_x \rangle]
\end{equation}

\begin{equation}
    \partial_t\langle \hat{a} \hat{a}\rangle = -(i 2 \omega_c + \kappa) \langle \hat{a} \hat{a} \rangle - i 4 g^\prime \langle a \hat{S}_x \rangle
\end{equation}

\begin{equation}
    \partial_t\langle \hat{a} \hat{S}_x \rangle = -(i\omega_c + \frac{\kappa}{2} + \Tilde{\Gamma}) \langle \hat{a} \hat{S}_x \rangle - 2 \omega_0 \langle \hat{a} \hat{S}_y \rangle - i 2 g^\prime \langle \hat{S}_x^2 \rangle
\end{equation}

\begin{equation}
\begin{split}
    \partial_t\langle \hat{a} \hat{S}_y \rangle &= -(i\omega_c + \frac{\kappa}{2} + \Tilde{\Gamma}) \langle \hat{a} \hat{S}_y \rangle  - 2 g^\prime \langle \hat{S}_z \rangle (\langle \hat{a} \hat{a} \rangle + \langle \hat{a}^\dagger \hat{a} \rangle)\\
    &+ 2 \omega_0 \langle \hat{a} \hat{S}_x \rangle - i 2 g^\prime [\langle \hat{S}_x \hat{S}_y \rangle - i \hat{S}_z \rangle]
\end{split}
\end{equation}

\begin{equation}
    \partial_t\langle\hat{S}_x^2\rangle = - 2 \omega_0 (2 \langle \hat{S}_x \hat{S}_y \rangle - i \hat{S}_z \rangle) - \Tilde{\Gamma}(2\langle \hat{S}_x^2 \rangle - \frac{N}{2})
\end{equation}

\begin{equation}
\begin{split}
    \partial_t\langle \hat{S}_y^2 \rangle &= 2 \omega_0 (2 \langle \hat{S}_x \hat{S}_y \rangle - i \hat{S}_z \rangle) - 8 g^\prime \frac{N - 1}{N}\langle \hat{S}_z \rangle Re[\langle \hat{a} \hat{S}_y \rangle]\\
    &- \Tilde{\Gamma}(2\langle \hat{S}_y^2 \rangle - \frac{N}{2})
\end{split}
\end{equation}

\begin{equation}
\begin{split}
    \partial_t\langle \hat{S}_z^2 \rangle &=  g^\prime \frac{N - 1}{N}\langle \hat{S}_z \rangle Re[\langle \hat{a} \hat{S}_y \rangle]\\
    &- \Gamma_\downarrow [2\langle \hat{S}_z^2 \rangle - \frac{N}{2} + (N - 1)\langle \hat{S}_z \rangle]
\end{split}
\end{equation}

\begin{equation}
\begin{split}
    \partial_t\langle \hat{S}_x \hat{S}_y \rangle &= 2 \omega_0 (\langle \hat{S}_x^2 \rangle - \langle \hat{S}_y^2 \rangle) - 4 g^\prime \frac{N - 1}{N}\langle \hat{S}_z \rangle Re[\langle \hat{a} \hat{S}_x \rangle]\\
    &+i 2 g^\prime Re[\langle \hat{a} \hat{S}_y \rangle] - \Tilde{\Gamma}(2\langle \hat{S}_z^2 \rangle - i\langle \hat{S}_z \rangle)\\
    &-i\frac{\Gamma_\downarrow}{2}(\langle \hat{S}_z \rangle + \frac{N}{2})
\end{split}
\end{equation}

To find the steady state within each $S$-subspace using MF2, the equations are time-integrated for the starting condition of $\ket{S, M = -S, N_{ph} = 0}$. In the MF1 calculation, the initial state remains the same for the spin part, except that the photon part is a coherent state.

\section{Fitted powers for different $f$ \label{app:other_f}}

\begin{figure}[!htbp]
  \centering
  \includegraphics[width = 0.5\textwidth]{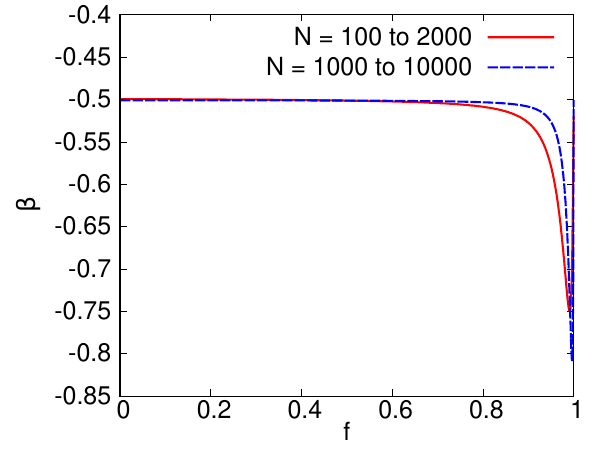}
  \caption{Powers $\beta$ in Eq.~\eqref{eq:N_beta} as a function of $f$ obtained from the same parameters and procedures as Fig.~\ref{fig:fit_gau_pop_dist}. The red curve corresponds to $N$ ranging from $100$ to $2000$ with increment of $100$ while the blue line represents the range of $N$ from $1000$ to $10000$ with step size $1000$.}
  \label{fig:fit_power}
\end{figure}

With the same parameters and the same procedures as Fig.~\ref{fig:fit_gau_pop_dist}, we fit the $ln(\sigma)$-$ln(N)$ curves for $f$ ranging from $0$ to $1$ with increment of $0.001$, yielding the powers as a function of $f$ illustrated in Fig.~\ref{fig:fit_power}, labeled by ``N = 100 to 2000''.
\nocite{*}
One can check that $\beta \approx -0.5$ holds for $f < 0.8$ and gradually drops down with increasing $f$ but approaches back to $\beta = -0.5$ at $f = 1$. By the discussion about Fig.~\ref{fig:avg_S_MF1_MF2_pert_degen}, when $f \rightarrow 1^-$, much greater $N$ is required so that the MF2 calculation results can approach the MF1 results. To understand whether the decreasing of $\beta$ results from the limited range of $N$, we repeat the fitting procedure except that $N$ ranges from $1000$ to $10000$ with step size $1000$ and plot the corresponding $\beta - f$ relation in Fig.~\ref{fig:fit_power}, labeled by ``N = 1000 to 10000''. Compared to the case ``N = 100 to 2000'', this curve has more points ($0.8 < f < 0.95$) approaching $\beta = -0.5$, corresponding to better consistency of MF2 calculation with MF1 in Fig.~\ref{fig:avg_S_MF1_MF2_pert_degen}. By this trend we infer that the power $\beta$ should approach $-0.5$  for $N \rightarrow \infty$, except for a narrowing window near $f = 1$ with lower $\beta$.

\section{Liouvillian spectrum modulated by $f$ and the corresponding eigenvectors\label{app:Liouvillian_spectrum_eig_vec} at $f = 1$}

\begin{figure}[!htbp]
  \centering
  \includegraphics[width = 0.5\textwidth]{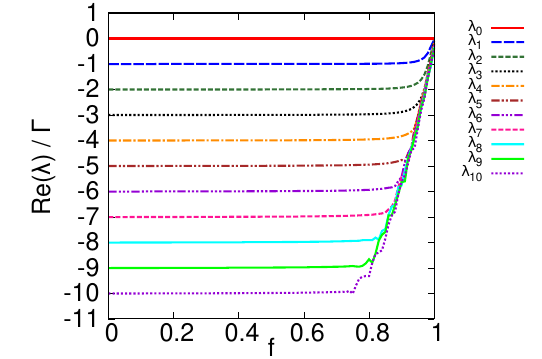}
  \caption{Eigenvalues of the coupling matrix (or Liouvillian spectrum) with the least negative real parts modulated by $f$. The parameters are the same as Fig.~\ref{fig:eig_phi} except for $N = 1000$ and infinitesimal $\Gamma$.}
  \label{fig:eig_val_f}
\end{figure}

In addition to the steady state, degenerate perturbation theory can also reveal the long-term transient dynamics via the eigenvalues of the coupling matrix (or Liouvillian spectrum) and the eigenvectors. The Liouvillian spectrum modulated by $f$ can not be determined from low order mean field equations. With the same parameters as Fig.~\ref{fig:eig_phi} except for $N = 1000$ and the infinitesimal perturbation strength $\Gamma$, the Liouvillian spectrum is shown in Fig.~\ref{fig:eig_val_f}.

\begin{figure}[!htbp]
  \centering
  \includegraphics[width = 0.5\textwidth]{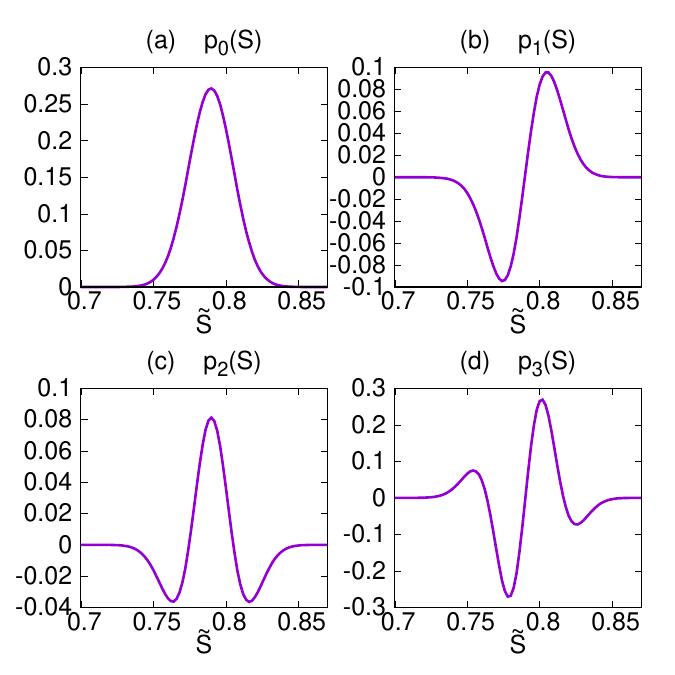}
  \caption{First $4$ eigenvectors of the coupling matrix with the least negative real parts. The parameters are the same as Fig.~\ref{fig:eig_val_f} except for $f = 0$.}
  \label{fig:eig_vec_f1}
\end{figure}

Here we only plot the first $11$ eigenvalues with the smallest real parts because we are mainly interested in the long-term behavior. One can notice a large difference in the orders of magnitude between $f = 0$ and $f = 1$. With $f$ decreasing from $f = 1$, the eigenvalue $\lambda_n$ becomes increasingly negative at first and then saturates to $\lambda_n \approx n\Gamma$ at $f = 0$, where $n$ is an integer. Similar to $p(S)$, we also plot the eigenvectors $p_n(S)$ for the first $4$ eigenvalues with the smallest real parts in Fig.~\ref{fig:eig_vec_f1}.

Here we scale the eigenvectors so that the norm is $1$. One can check that $p_n(S)$ is approximately enveloped by a Gaussian function and the number of nodes is $n$. The properties of these eigenstates and the reason for the eigenvalues being integers of $\Gamma$ are open questions. 

\newpage

\bibliography{ref}

\begin{thebibliography}{107}%
\makeatletter
\providecommand \@ifxundefined [1]{%
 \@ifx{#1\undefined}
}%
\providecommand \@ifnum [1]{%
 \ifnum #1\expandafter \@firstoftwo
 \else \expandafter \@secondoftwo
 \fi
}%
\providecommand \@ifx [1]{%
 \ifx #1\expandafter \@firstoftwo
 \else \expandafter \@secondoftwo
 \fi
}%
\providecommand \natexlab [1]{#1}%
\providecommand \enquote  [1]{``#1''}%
\providecommand \bibnamefont  [1]{#1}%
\providecommand \bibfnamefont [1]{#1}%
\providecommand \citenamefont [1]{#1}%
\providecommand \href@noop [0]{\@secondoftwo}%
\providecommand \href [0]{\begingroup \@sanitize@url \@href}%
\providecommand \@href[1]{\@@startlink{#1}\@@href}%
\providecommand \@@href[1]{\endgroup#1\@@endlink}%
\providecommand \@sanitize@url [0]{\catcode `\\12\catcode `\$12\catcode `\&12\catcode `\#12\catcode `\^12\catcode `\_12\catcode `\%12\relax}%
\providecommand \@@startlink[1]{}%
\providecommand \@@endlink[0]{}%
\providecommand \url  [0]{\begingroup\@sanitize@url \@url }%
\providecommand \@url [1]{\endgroup\@href {#1}{\urlprefix }}%
\providecommand \urlprefix  [0]{URL }%
\providecommand \Eprint [0]{\href }%
\providecommand \doibase [0]{https://doi.org/}%
\providecommand \selectlanguage [0]{\@gobble}%
\providecommand \bibinfo  [0]{\@secondoftwo}%
\providecommand \bibfield  [0]{\@secondoftwo}%
\providecommand \translation [1]{[#1]}%
\providecommand \BibitemOpen [0]{}%
\providecommand \bibitemStop [0]{}%
\providecommand \bibitemNoStop [0]{.\EOS\space}%
\providecommand \EOS [0]{\spacefactor3000\relax}%
\providecommand \BibitemShut  [1]{\csname bibitem#1\endcsname}%
\let\auto@bib@innerbib\@empty
\bibitem [{\citenamefont {Kirton}\ and\ \citenamefont {Keeling}(2017)}]{PhysRevLett.118.123602}%
  \BibitemOpen
  \bibfield  {author} {\bibinfo {author} {\bibfnamefont {P.}~\bibnamefont {Kirton}}\ and\ \bibinfo {author} {\bibfnamefont {J.}~\bibnamefont {Keeling}},\ }\bibfield  {title} {\bibinfo {title} {Suppressing and restoring the {D}icke superradiance transition by dephasing and decay},\ }\href {https://doi.org/10.1103/PhysRevLett.118.123602} {\bibfield  {journal} {\bibinfo  {journal} {Phys. Rev. Lett.}\ }\textbf {\bibinfo {volume} {118}},\ \bibinfo {pages} {123602} (\bibinfo {year} {2017})}\BibitemShut {NoStop}%
\bibitem [{\citenamefont {Dicke}(1954)}]{PhysRev.93.99}%
  \BibitemOpen
  \bibfield  {author} {\bibinfo {author} {\bibfnamefont {R.~H.}\ \bibnamefont {Dicke}},\ }\bibfield  {title} {\bibinfo {title} {Coherence in spontaneous radiation processes},\ }\href {https://doi.org/10.1103/PhysRev.93.99} {\bibfield  {journal} {\bibinfo  {journal} {Phys. Rev.}\ }\textbf {\bibinfo {volume} {93}},\ \bibinfo {pages} {99} (\bibinfo {year} {1954})}\BibitemShut {NoStop}%
\bibitem [{\citenamefont {Rehler}\ and\ \citenamefont {Eberly}(1971)}]{PhysRevA.3.1735}%
  \BibitemOpen
  \bibfield  {author} {\bibinfo {author} {\bibfnamefont {N.~E.}\ \bibnamefont {Rehler}}\ and\ \bibinfo {author} {\bibfnamefont {J.~H.}\ \bibnamefont {Eberly}},\ }\bibfield  {title} {\bibinfo {title} {Superradiance},\ }\href {https://doi.org/10.1103/PhysRevA.3.1735} {\bibfield  {journal} {\bibinfo  {journal} {Phys. Rev. A}\ }\textbf {\bibinfo {volume} {3}},\ \bibinfo {pages} {1735} (\bibinfo {year} {1971})}\BibitemShut {NoStop}%
\bibitem [{\citenamefont {Gross}\ and\ \citenamefont {Haroche}(1982)}]{gross1982superradiance}%
  \BibitemOpen
  \bibfield  {author} {\bibinfo {author} {\bibfnamefont {M.}~\bibnamefont {Gross}}\ and\ \bibinfo {author} {\bibfnamefont {S.}~\bibnamefont {Haroche}},\ }\bibfield  {title} {\bibinfo {title} {Superradiance: An essay on the theory of collective spontaneous emission},\ }\href {https://doi.org/https://doi.org/10.1016/0370-1573(82)90102-8} {\bibfield  {journal} {\bibinfo  {journal} {Physics reports}\ }\textbf {\bibinfo {volume} {93}},\ \bibinfo {pages} {301} (\bibinfo {year} {1982})}\BibitemShut {NoStop}%
\bibitem [{\citenamefont {Skribanowitz}\ \emph {et~al.}(1973)\citenamefont {Skribanowitz}, \citenamefont {Herman}, \citenamefont {MacGillivray},\ and\ \citenamefont {Feld}}]{PhysRevLett.30.309}%
  \BibitemOpen
  \bibfield  {author} {\bibinfo {author} {\bibfnamefont {N.}~\bibnamefont {Skribanowitz}}, \bibinfo {author} {\bibfnamefont {I.~P.}\ \bibnamefont {Herman}}, \bibinfo {author} {\bibfnamefont {J.~C.}\ \bibnamefont {MacGillivray}},\ and\ \bibinfo {author} {\bibfnamefont {M.~S.}\ \bibnamefont {Feld}},\ }\bibfield  {title} {\bibinfo {title} {Observation of {D}icke superradiance in optically pumped hf gas},\ }\href {https://doi.org/10.1103/PhysRevLett.30.309} {\bibfield  {journal} {\bibinfo  {journal} {Phys. Rev. Lett.}\ }\textbf {\bibinfo {volume} {30}},\ \bibinfo {pages} {309} (\bibinfo {year} {1973})}\BibitemShut {NoStop}%
\bibitem [{\citenamefont {Gross}\ \emph {et~al.}(1976)\citenamefont {Gross}, \citenamefont {Fabre}, \citenamefont {Pillet},\ and\ \citenamefont {Haroche}}]{PhysRevLett.36.1035}%
  \BibitemOpen
  \bibfield  {author} {\bibinfo {author} {\bibfnamefont {M.}~\bibnamefont {Gross}}, \bibinfo {author} {\bibfnamefont {C.}~\bibnamefont {Fabre}}, \bibinfo {author} {\bibfnamefont {P.}~\bibnamefont {Pillet}},\ and\ \bibinfo {author} {\bibfnamefont {S.}~\bibnamefont {Haroche}},\ }\bibfield  {title} {\bibinfo {title} {Observation of near-infrared {D}icke superradiance on cascading transitions in atomic sodium},\ }\href {https://doi.org/10.1103/PhysRevLett.36.1035} {\bibfield  {journal} {\bibinfo  {journal} {Phys. Rev. Lett.}\ }\textbf {\bibinfo {volume} {36}},\ \bibinfo {pages} {1035} (\bibinfo {year} {1976})}\BibitemShut {NoStop}%
\bibitem [{\citenamefont {Raimond}\ \emph {et~al.}(1982)\citenamefont {Raimond}, \citenamefont {Goy}, \citenamefont {Gross}, \citenamefont {Fabre},\ and\ \citenamefont {Haroche}}]{PhysRevLett.49.117}%
  \BibitemOpen
  \bibfield  {author} {\bibinfo {author} {\bibfnamefont {J.~M.}\ \bibnamefont {Raimond}}, \bibinfo {author} {\bibfnamefont {P.}~\bibnamefont {Goy}}, \bibinfo {author} {\bibfnamefont {M.}~\bibnamefont {Gross}}, \bibinfo {author} {\bibfnamefont {C.}~\bibnamefont {Fabre}},\ and\ \bibinfo {author} {\bibfnamefont {S.}~\bibnamefont {Haroche}},\ }\bibfield  {title} {\bibinfo {title} {Collective absorption of blackbody radiation by {R}ydberg atoms in a cavity: An experiment on {B}ose statistics and brownian motion},\ }\href {https://doi.org/10.1103/PhysRevLett.49.117} {\bibfield  {journal} {\bibinfo  {journal} {Phys. Rev. Lett.}\ }\textbf {\bibinfo {volume} {49}},\ \bibinfo {pages} {117} (\bibinfo {year} {1982})}\BibitemShut {NoStop}%
\bibitem [{\citenamefont {Scheibner}\ \emph {et~al.}(2007)\citenamefont {Scheibner}, \citenamefont {Schmidt}, \citenamefont {Worschech}, \citenamefont {Forchel}, \citenamefont {Bacher}, \citenamefont {Passow},\ and\ \citenamefont {Hommel}}]{scheibner2007superradiance}%
  \BibitemOpen
  \bibfield  {author} {\bibinfo {author} {\bibfnamefont {M.}~\bibnamefont {Scheibner}}, \bibinfo {author} {\bibfnamefont {T.}~\bibnamefont {Schmidt}}, \bibinfo {author} {\bibfnamefont {L.}~\bibnamefont {Worschech}}, \bibinfo {author} {\bibfnamefont {A.}~\bibnamefont {Forchel}}, \bibinfo {author} {\bibfnamefont {G.}~\bibnamefont {Bacher}}, \bibinfo {author} {\bibfnamefont {T.}~\bibnamefont {Passow}},\ and\ \bibinfo {author} {\bibfnamefont {D.}~\bibnamefont {Hommel}},\ }\bibfield  {title} {\bibinfo {title} {Superradiance of quantum dots},\ }\href {https://doi.org/10.1038/nphys494} {\bibfield  {journal} {\bibinfo  {journal} {Nature Physics}\ }\textbf {\bibinfo {volume} {3}},\ \bibinfo {pages} {106} (\bibinfo {year} {2007})}\BibitemShut {NoStop}%
\bibitem [{\citenamefont {R{\"o}hlsberger}\ \emph {et~al.}(2010)\citenamefont {R{\"o}hlsberger}, \citenamefont {Schlage}, \citenamefont {Sahoo}, \citenamefont {Couet},\ and\ \citenamefont {R{\"u}ffer}}]{rohlsberger2010collective}%
  \BibitemOpen
  \bibfield  {author} {\bibinfo {author} {\bibfnamefont {R.}~\bibnamefont {R{\"o}hlsberger}}, \bibinfo {author} {\bibfnamefont {K.}~\bibnamefont {Schlage}}, \bibinfo {author} {\bibfnamefont {B.}~\bibnamefont {Sahoo}}, \bibinfo {author} {\bibfnamefont {S.}~\bibnamefont {Couet}},\ and\ \bibinfo {author} {\bibfnamefont {R.}~\bibnamefont {R{\"u}ffer}},\ }\bibfield  {title} {\bibinfo {title} {Collective {L}amb shift in single-photon superradiance},\ }\href {https://doi.org/10.1126/science.1187770} {\bibfield  {journal} {\bibinfo  {journal} {Science}\ }\textbf {\bibinfo {volume} {328}},\ \bibinfo {pages} {1248} (\bibinfo {year} {2010})}\BibitemShut {NoStop}%
\bibitem [{\citenamefont {Inouye}\ \emph {et~al.}(1999)\citenamefont {Inouye}, \citenamefont {Chikkatur}, \citenamefont {Stamper-Kurn}, \citenamefont {Stenger}, \citenamefont {Pritchard},\ and\ \citenamefont {Ketterle}}]{inouye1999superradiant}%
  \BibitemOpen
  \bibfield  {author} {\bibinfo {author} {\bibfnamefont {S.}~\bibnamefont {Inouye}}, \bibinfo {author} {\bibfnamefont {A.}~\bibnamefont {Chikkatur}}, \bibinfo {author} {\bibfnamefont {D.}~\bibnamefont {Stamper-Kurn}}, \bibinfo {author} {\bibfnamefont {J.}~\bibnamefont {Stenger}}, \bibinfo {author} {\bibfnamefont {D.}~\bibnamefont {Pritchard}},\ and\ \bibinfo {author} {\bibfnamefont {W.}~\bibnamefont {Ketterle}},\ }\bibfield  {title} {\bibinfo {title} {Superradiant {R}ayleigh scattering from a {B}ose-{E}instein condensate},\ }\href {https://doi.org/10.1126/science.285.5427.571} {\bibfield  {journal} {\bibinfo  {journal} {Science}\ }\textbf {\bibinfo {volume} {285}},\ \bibinfo {pages} {571} (\bibinfo {year} {1999})}\BibitemShut {NoStop}%
\bibitem [{\citenamefont {Slama}\ \emph {et~al.}(2007)\citenamefont {Slama}, \citenamefont {Bux}, \citenamefont {Krenz}, \citenamefont {Zimmermann},\ and\ \citenamefont {Courteille}}]{PhysRevLett.98.053603}%
  \BibitemOpen
  \bibfield  {author} {\bibinfo {author} {\bibfnamefont {S.}~\bibnamefont {Slama}}, \bibinfo {author} {\bibfnamefont {S.}~\bibnamefont {Bux}}, \bibinfo {author} {\bibfnamefont {G.}~\bibnamefont {Krenz}}, \bibinfo {author} {\bibfnamefont {C.}~\bibnamefont {Zimmermann}},\ and\ \bibinfo {author} {\bibfnamefont {P.~W.}\ \bibnamefont {Courteille}},\ }\bibfield  {title} {\bibinfo {title} {Superradiant {R}ayleigh scattering and collective atomic recoil lasing in a ring cavity},\ }\href {https://doi.org/10.1103/PhysRevLett.98.053603} {\bibfield  {journal} {\bibinfo  {journal} {Phys. Rev. Lett.}\ }\textbf {\bibinfo {volume} {98}},\ \bibinfo {pages} {053603} (\bibinfo {year} {2007})}\BibitemShut {NoStop}%
\bibitem [{\citenamefont {DeVoe}\ and\ \citenamefont {Brewer}(1996)}]{PhysRevLett.76.2049}%
  \BibitemOpen
  \bibfield  {author} {\bibinfo {author} {\bibfnamefont {R.~G.}\ \bibnamefont {DeVoe}}\ and\ \bibinfo {author} {\bibfnamefont {R.~G.}\ \bibnamefont {Brewer}},\ }\bibfield  {title} {\bibinfo {title} {Observation of superradiant and subradiant spontaneous emission of two trapped ions},\ }\href {https://doi.org/10.1103/PhysRevLett.76.2049} {\bibfield  {journal} {\bibinfo  {journal} {Phys. Rev. Lett.}\ }\textbf {\bibinfo {volume} {76}},\ \bibinfo {pages} {2049} (\bibinfo {year} {1996})}\BibitemShut {NoStop}%
\bibitem [{\citenamefont {Eschner}\ \emph {et~al.}(2001)\citenamefont {Eschner}, \citenamefont {Raab}, \citenamefont {Schmidt-Kaler},\ and\ \citenamefont {Blatt}}]{eschner2001light}%
  \BibitemOpen
  \bibfield  {author} {\bibinfo {author} {\bibfnamefont {J.}~\bibnamefont {Eschner}}, \bibinfo {author} {\bibfnamefont {C.}~\bibnamefont {Raab}}, \bibinfo {author} {\bibfnamefont {F.}~\bibnamefont {Schmidt-Kaler}},\ and\ \bibinfo {author} {\bibfnamefont {R.}~\bibnamefont {Blatt}},\ }\bibfield  {title} {\bibinfo {title} {Light interference from single atoms and their mirror images},\ }\href {https://doi.org/10.1038/35097017} {\bibfield  {journal} {\bibinfo  {journal} {Nature}\ }\textbf {\bibinfo {volume} {413}},\ \bibinfo {pages} {495} (\bibinfo {year} {2001})}\BibitemShut {NoStop}%
\bibitem [{\citenamefont {Kim}\ \emph {et~al.}(2016)\citenamefont {Kim}, \citenamefont {Lee}, \citenamefont {Lee}, \citenamefont {Jo}, \citenamefont {Song},\ and\ \citenamefont {Ahn}}]{kim2016situ}%
  \BibitemOpen
  \bibfield  {author} {\bibinfo {author} {\bibfnamefont {H.}~\bibnamefont {Kim}}, \bibinfo {author} {\bibfnamefont {W.}~\bibnamefont {Lee}}, \bibinfo {author} {\bibfnamefont {H.-g.}\ \bibnamefont {Lee}}, \bibinfo {author} {\bibfnamefont {H.}~\bibnamefont {Jo}}, \bibinfo {author} {\bibfnamefont {Y.}~\bibnamefont {Song}},\ and\ \bibinfo {author} {\bibfnamefont {J.}~\bibnamefont {Ahn}},\ }\bibfield  {title} {\bibinfo {title} {In situ single-atom array synthesis using dynamic holographic optical tweezers},\ }\href {https://doi.org/10.1038/ncomms13317} {\bibfield  {journal} {\bibinfo  {journal} {Nature communications}\ }\textbf {\bibinfo {volume} {7}},\ \bibinfo {pages} {13317} (\bibinfo {year} {2016})}\BibitemShut {NoStop}%
\bibitem [{\citenamefont {Endres}\ \emph {et~al.}(2016)\citenamefont {Endres}, \citenamefont {Bernien}, \citenamefont {Keesling}, \citenamefont {Levine}, \citenamefont {Anschuetz}, \citenamefont {Krajenbrink}, \citenamefont {Senko}, \citenamefont {Vuletic}, \citenamefont {Greiner},\ and\ \citenamefont {Lukin}}]{endres2016atom}%
  \BibitemOpen
  \bibfield  {author} {\bibinfo {author} {\bibfnamefont {M.}~\bibnamefont {Endres}}, \bibinfo {author} {\bibfnamefont {H.}~\bibnamefont {Bernien}}, \bibinfo {author} {\bibfnamefont {A.}~\bibnamefont {Keesling}}, \bibinfo {author} {\bibfnamefont {H.}~\bibnamefont {Levine}}, \bibinfo {author} {\bibfnamefont {E.~R.}\ \bibnamefont {Anschuetz}}, \bibinfo {author} {\bibfnamefont {A.}~\bibnamefont {Krajenbrink}}, \bibinfo {author} {\bibfnamefont {C.}~\bibnamefont {Senko}}, \bibinfo {author} {\bibfnamefont {V.}~\bibnamefont {Vuletic}}, \bibinfo {author} {\bibfnamefont {M.}~\bibnamefont {Greiner}},\ and\ \bibinfo {author} {\bibfnamefont {M.~D.}\ \bibnamefont {Lukin}},\ }\bibfield  {title} {\bibinfo {title} {Atom-by-atom assembly of defect-free one-dimensional cold atom arrays},\ }\href {https://doi.org/10.1126/science.aah3752} {\bibfield  {journal} {\bibinfo  {journal} {Science}\ }\textbf {\bibinfo {volume} {354}},\ \bibinfo {pages} {1024} (\bibinfo {year} {2016})}\BibitemShut {NoStop}%
\bibitem [{\citenamefont {Barredo}\ \emph {et~al.}(2016)\citenamefont {Barredo}, \citenamefont {De~L{\'e}s{\'e}leuc}, \citenamefont {Lienhard}, \citenamefont {Lahaye},\ and\ \citenamefont {Browaeys}}]{barredo2016atom}%
  \BibitemOpen
  \bibfield  {author} {\bibinfo {author} {\bibfnamefont {D.}~\bibnamefont {Barredo}}, \bibinfo {author} {\bibfnamefont {S.}~\bibnamefont {De~L{\'e}s{\'e}leuc}}, \bibinfo {author} {\bibfnamefont {V.}~\bibnamefont {Lienhard}}, \bibinfo {author} {\bibfnamefont {T.}~\bibnamefont {Lahaye}},\ and\ \bibinfo {author} {\bibfnamefont {A.}~\bibnamefont {Browaeys}},\ }\bibfield  {title} {\bibinfo {title} {An atom-by-atom assembler of defect-free arbitrary two-dimensional atomic arrays},\ }\href {https://doi.org/10.1126/science.aah3778} {\bibfield  {journal} {\bibinfo  {journal} {Science}\ }\textbf {\bibinfo {volume} {354}},\ \bibinfo {pages} {1021} (\bibinfo {year} {2016})}\BibitemShut {NoStop}%
\bibitem [{\citenamefont {Norcia}\ \emph {et~al.}(2018{\natexlab{a}})\citenamefont {Norcia}, \citenamefont {Young},\ and\ \citenamefont {Kaufman}}]{norcia2018microscopic}%
  \BibitemOpen
  \bibfield  {author} {\bibinfo {author} {\bibfnamefont {M.~A.}\ \bibnamefont {Norcia}}, \bibinfo {author} {\bibfnamefont {A.~W.}\ \bibnamefont {Young}},\ and\ \bibinfo {author} {\bibfnamefont {A.~M.}\ \bibnamefont {Kaufman}},\ }\bibfield  {title} {\bibinfo {title} {Microscopic control and detection of ultracold strontium in optical-tweezer arrays},\ }\href {https://doi.org/10.1103/PhysRevX.8.041054} {\bibfield  {journal} {\bibinfo  {journal} {Phys. Rev. X}\ }\textbf {\bibinfo {volume} {8}},\ \bibinfo {pages} {041054} (\bibinfo {year} {2018}{\natexlab{a}})}\BibitemShut {NoStop}%
\bibitem [{\citenamefont {Saskin}\ \emph {et~al.}(2019)\citenamefont {Saskin}, \citenamefont {Wilson}, \citenamefont {Grinkemeyer},\ and\ \citenamefont {Thompson}}]{PhysRevLett.122.143002}%
  \BibitemOpen
  \bibfield  {author} {\bibinfo {author} {\bibfnamefont {S.}~\bibnamefont {Saskin}}, \bibinfo {author} {\bibfnamefont {J.~T.}\ \bibnamefont {Wilson}}, \bibinfo {author} {\bibfnamefont {B.}~\bibnamefont {Grinkemeyer}},\ and\ \bibinfo {author} {\bibfnamefont {J.~D.}\ \bibnamefont {Thompson}},\ }\bibfield  {title} {\bibinfo {title} {Narrow-line cooling and imaging of ytterbium atoms in an optical tweezer array},\ }\href {https://doi.org/10.1103/PhysRevLett.122.143002} {\bibfield  {journal} {\bibinfo  {journal} {Phys. Rev. Lett.}\ }\textbf {\bibinfo {volume} {122}},\ \bibinfo {pages} {143002} (\bibinfo {year} {2019})}\BibitemShut {NoStop}%
\bibitem [{\citenamefont {Ohl~de Mello}\ \emph {et~al.}(2019)\citenamefont {Ohl~de Mello}, \citenamefont {Sch\"affner}, \citenamefont {Werkmann}, \citenamefont {Preuschoff}, \citenamefont {Kohfahl}, \citenamefont {Schlosser},\ and\ \citenamefont {Birkl}}]{PhysRevLett.122.203601}%
  \BibitemOpen
  \bibfield  {author} {\bibinfo {author} {\bibfnamefont {D.}~\bibnamefont {Ohl~de Mello}}, \bibinfo {author} {\bibfnamefont {D.}~\bibnamefont {Sch\"affner}}, \bibinfo {author} {\bibfnamefont {J.}~\bibnamefont {Werkmann}}, \bibinfo {author} {\bibfnamefont {T.}~\bibnamefont {Preuschoff}}, \bibinfo {author} {\bibfnamefont {L.}~\bibnamefont {Kohfahl}}, \bibinfo {author} {\bibfnamefont {M.}~\bibnamefont {Schlosser}},\ and\ \bibinfo {author} {\bibfnamefont {G.}~\bibnamefont {Birkl}},\ }\bibfield  {title} {\bibinfo {title} {Defect-free assembly of 2d clusters of more than 100 single-atom quantum systems},\ }\href {https://doi.org/10.1103/PhysRevLett.122.203601} {\bibfield  {journal} {\bibinfo  {journal} {Phys. Rev. Lett.}\ }\textbf {\bibinfo {volume} {122}},\ \bibinfo {pages} {203601} (\bibinfo {year} {2019})}\BibitemShut {NoStop}%
\bibitem [{\citenamefont {Bakr}\ \emph {et~al.}(2010)\citenamefont {Bakr}, \citenamefont {Peng}, \citenamefont {Tai}, \citenamefont {Ma}, \citenamefont {Simon}, \citenamefont {Gillen}, \citenamefont {Foelling}, \citenamefont {Pollet},\ and\ \citenamefont {Greiner}}]{bakr2010probing}%
  \BibitemOpen
  \bibfield  {author} {\bibinfo {author} {\bibfnamefont {W.~S.}\ \bibnamefont {Bakr}}, \bibinfo {author} {\bibfnamefont {A.}~\bibnamefont {Peng}}, \bibinfo {author} {\bibfnamefont {M.~E.}\ \bibnamefont {Tai}}, \bibinfo {author} {\bibfnamefont {R.}~\bibnamefont {Ma}}, \bibinfo {author} {\bibfnamefont {J.}~\bibnamefont {Simon}}, \bibinfo {author} {\bibfnamefont {J.~I.}\ \bibnamefont {Gillen}}, \bibinfo {author} {\bibfnamefont {S.}~\bibnamefont {Foelling}}, \bibinfo {author} {\bibfnamefont {L.}~\bibnamefont {Pollet}},\ and\ \bibinfo {author} {\bibfnamefont {M.}~\bibnamefont {Greiner}},\ }\bibfield  {title} {\bibinfo {title} {Probing the superfluid--to--{M}ott insulator transition at the single-atom level},\ }\href {https://doi.org/10.1126/science.1192368} {\bibfield  {journal} {\bibinfo  {journal} {Science}\ }\textbf {\bibinfo {volume} {329}},\ \bibinfo {pages} {547} (\bibinfo {year} {2010})}\BibitemShut {NoStop}%
\bibitem [{\citenamefont {Sherson}\ \emph {et~al.}(2010)\citenamefont {Sherson}, \citenamefont {Weitenberg}, \citenamefont {Endres}, \citenamefont {Cheneau}, \citenamefont {Bloch},\ and\ \citenamefont {Kuhr}}]{sherson2010single}%
  \BibitemOpen
  \bibfield  {author} {\bibinfo {author} {\bibfnamefont {J.~F.}\ \bibnamefont {Sherson}}, \bibinfo {author} {\bibfnamefont {C.}~\bibnamefont {Weitenberg}}, \bibinfo {author} {\bibfnamefont {M.}~\bibnamefont {Endres}}, \bibinfo {author} {\bibfnamefont {M.}~\bibnamefont {Cheneau}}, \bibinfo {author} {\bibfnamefont {I.}~\bibnamefont {Bloch}},\ and\ \bibinfo {author} {\bibfnamefont {S.}~\bibnamefont {Kuhr}},\ }\bibfield  {title} {\bibinfo {title} {Single-atom-resolved fluorescence imaging of an atomic {M}ott insulator},\ }\href {https://doi.org/10.1038/nature09378} {\bibfield  {journal} {\bibinfo  {journal} {Nature}\ }\textbf {\bibinfo {volume} {467}},\ \bibinfo {pages} {68} (\bibinfo {year} {2010})}\BibitemShut {NoStop}%
\bibitem [{\citenamefont {Greif}\ \emph {et~al.}(2016)\citenamefont {Greif}, \citenamefont {Parsons}, \citenamefont {Mazurenko}, \citenamefont {Chiu}, \citenamefont {Blatt}, \citenamefont {Huber}, \citenamefont {Ji},\ and\ \citenamefont {Greiner}}]{greif2016site}%
  \BibitemOpen
  \bibfield  {author} {\bibinfo {author} {\bibfnamefont {D.}~\bibnamefont {Greif}}, \bibinfo {author} {\bibfnamefont {M.~F.}\ \bibnamefont {Parsons}}, \bibinfo {author} {\bibfnamefont {A.}~\bibnamefont {Mazurenko}}, \bibinfo {author} {\bibfnamefont {C.~S.}\ \bibnamefont {Chiu}}, \bibinfo {author} {\bibfnamefont {S.}~\bibnamefont {Blatt}}, \bibinfo {author} {\bibfnamefont {F.}~\bibnamefont {Huber}}, \bibinfo {author} {\bibfnamefont {G.}~\bibnamefont {Ji}},\ and\ \bibinfo {author} {\bibfnamefont {M.}~\bibnamefont {Greiner}},\ }\bibfield  {title} {\bibinfo {title} {Site-resolved imaging of a fermionic {M}ott insulator},\ }\href {https://doi.org/10.1126/science.aad9041} {\bibfield  {journal} {\bibinfo  {journal} {Science}\ }\textbf {\bibinfo {volume} {351}},\ \bibinfo {pages} {953} (\bibinfo {year} {2016})}\BibitemShut {NoStop}%
\bibitem [{\citenamefont {Kumar}\ \emph {et~al.}(2018)\citenamefont {Kumar}, \citenamefont {Wu}, \citenamefont {Giraldo},\ and\ \citenamefont {Weiss}}]{kumar2018sorting}%
  \BibitemOpen
  \bibfield  {author} {\bibinfo {author} {\bibfnamefont {A.}~\bibnamefont {Kumar}}, \bibinfo {author} {\bibfnamefont {T.-Y.}\ \bibnamefont {Wu}}, \bibinfo {author} {\bibfnamefont {F.}~\bibnamefont {Giraldo}},\ and\ \bibinfo {author} {\bibfnamefont {D.~S.}\ \bibnamefont {Weiss}},\ }\bibfield  {title} {\bibinfo {title} {Sorting ultracold atoms in a three-dimensional optical lattice in a realization of maxwell’s demon},\ }\href {https://doi.org/10.1038/s41586-018-0458-7} {\bibfield  {journal} {\bibinfo  {journal} {Nature}\ }\textbf {\bibinfo {volume} {561}},\ \bibinfo {pages} {83} (\bibinfo {year} {2018})}\BibitemShut {NoStop}%
\bibitem [{\citenamefont {Masson}\ and\ \citenamefont {Asenjo-Garcia}(2022)}]{masson2022universality}%
  \BibitemOpen
  \bibfield  {author} {\bibinfo {author} {\bibfnamefont {S.~J.}\ \bibnamefont {Masson}}\ and\ \bibinfo {author} {\bibfnamefont {A.}~\bibnamefont {Asenjo-Garcia}},\ }\bibfield  {title} {\bibinfo {title} {Universality of {D}icke superradiance in arrays of quantum emitters},\ }\href {https://www.nature.com/articles/s41467-022-29805-4} {\bibfield  {journal} {\bibinfo  {journal} {Nature Communications}\ }\textbf {\bibinfo {volume} {13}},\ \bibinfo {pages} {2285} (\bibinfo {year} {2022})}\BibitemShut {NoStop}%
\bibitem [{\citenamefont {Mok}\ \emph {et~al.}(2023)\citenamefont {Mok}, \citenamefont {Asenjo-Garcia}, \citenamefont {Sum},\ and\ \citenamefont {Kwek}}]{PhysRevLett.130.213605}%
  \BibitemOpen
  \bibfield  {author} {\bibinfo {author} {\bibfnamefont {W.-K.}\ \bibnamefont {Mok}}, \bibinfo {author} {\bibfnamefont {A.}~\bibnamefont {Asenjo-Garcia}}, \bibinfo {author} {\bibfnamefont {T.~C.}\ \bibnamefont {Sum}},\ and\ \bibinfo {author} {\bibfnamefont {L.-C.}\ \bibnamefont {Kwek}},\ }\bibfield  {title} {\bibinfo {title} {{D}icke superradiance requires interactions beyond nearest neighbors},\ }\href {https://doi.org/10.1103/PhysRevLett.130.213605} {\bibfield  {journal} {\bibinfo  {journal} {Phys. Rev. Lett.}\ }\textbf {\bibinfo {volume} {130}},\ \bibinfo {pages} {213605} (\bibinfo {year} {2023})}\BibitemShut {NoStop}%
\bibitem [{\citenamefont {Sierra}\ \emph {et~al.}(2022)\citenamefont {Sierra}, \citenamefont {Masson},\ and\ \citenamefont {Asenjo-Garcia}}]{PhysRevResearch.4.023207}%
  \BibitemOpen
  \bibfield  {author} {\bibinfo {author} {\bibfnamefont {E.}~\bibnamefont {Sierra}}, \bibinfo {author} {\bibfnamefont {S.~J.}\ \bibnamefont {Masson}},\ and\ \bibinfo {author} {\bibfnamefont {A.}~\bibnamefont {Asenjo-Garcia}},\ }\bibfield  {title} {\bibinfo {title} {{D}icke superradiance in ordered lattices: Dimensionality matters},\ }\href {https://doi.org/10.1103/PhysRevResearch.4.023207} {\bibfield  {journal} {\bibinfo  {journal} {Phys. Rev. Res.}\ }\textbf {\bibinfo {volume} {4}},\ \bibinfo {pages} {023207} (\bibinfo {year} {2022})}\BibitemShut {NoStop}%
\bibitem [{\citenamefont {Masson}\ \emph {et~al.}(2024)\citenamefont {Masson}, \citenamefont {Covey}, \citenamefont {Will},\ and\ \citenamefont {Asenjo-Garcia}}]{PRXQuantum.5.010344}%
  \BibitemOpen
  \bibfield  {author} {\bibinfo {author} {\bibfnamefont {S.~J.}\ \bibnamefont {Masson}}, \bibinfo {author} {\bibfnamefont {J.~P.}\ \bibnamefont {Covey}}, \bibinfo {author} {\bibfnamefont {S.}~\bibnamefont {Will}},\ and\ \bibinfo {author} {\bibfnamefont {A.}~\bibnamefont {Asenjo-Garcia}},\ }\bibfield  {title} {\bibinfo {title} {{D}icke superradiance in ordered arrays of multilevel atoms},\ }\href {https://doi.org/10.1103/PRXQuantum.5.010344} {\bibfield  {journal} {\bibinfo  {journal} {PRX Quantum}\ }\textbf {\bibinfo {volume} {5}},\ \bibinfo {pages} {010344} (\bibinfo {year} {2024})}\BibitemShut {NoStop}%
\bibitem [{\citenamefont {Ruostekoski}(2023)}]{PhysRevA.108.030101}%
  \BibitemOpen
  \bibfield  {author} {\bibinfo {author} {\bibfnamefont {J.}~\bibnamefont {Ruostekoski}},\ }\bibfield  {title} {\bibinfo {title} {Cooperative quantum-optical planar arrays of atoms},\ }\href {https://doi.org/10.1103/PhysRevA.108.030101} {\bibfield  {journal} {\bibinfo  {journal} {Phys. Rev. A}\ }\textbf {\bibinfo {volume} {108}},\ \bibinfo {pages} {030101} (\bibinfo {year} {2023})}\BibitemShut {NoStop}%
\bibitem [{\citenamefont {Sheremet}\ \emph {et~al.}(2023)\citenamefont {Sheremet}, \citenamefont {Petrov}, \citenamefont {Iorsh}, \citenamefont {Poshakinskiy},\ and\ \citenamefont {Poddubny}}]{RevModPhys.95.015002}%
  \BibitemOpen
  \bibfield  {author} {\bibinfo {author} {\bibfnamefont {A.~S.}\ \bibnamefont {Sheremet}}, \bibinfo {author} {\bibfnamefont {M.~I.}\ \bibnamefont {Petrov}}, \bibinfo {author} {\bibfnamefont {I.~V.}\ \bibnamefont {Iorsh}}, \bibinfo {author} {\bibfnamefont {A.~V.}\ \bibnamefont {Poshakinskiy}},\ and\ \bibinfo {author} {\bibfnamefont {A.~N.}\ \bibnamefont {Poddubny}},\ }\bibfield  {title} {\bibinfo {title} {Waveguide quantum electrodynamics: Collective radiance and photon-photon correlations},\ }\href {https://doi.org/10.1103/RevModPhys.95.015002} {\bibfield  {journal} {\bibinfo  {journal} {Rev. Mod. Phys.}\ }\textbf {\bibinfo {volume} {95}},\ \bibinfo {pages} {015002} (\bibinfo {year} {2023})}\BibitemShut {NoStop}%
\bibitem [{\citenamefont {Lohof}\ \emph {et~al.}(2023)\citenamefont {Lohof}, \citenamefont {Schumayer}, \citenamefont {Hutchinson},\ and\ \citenamefont {Gies}}]{PhysRevLett.131.063601}%
  \BibitemOpen
  \bibfield  {author} {\bibinfo {author} {\bibfnamefont {F.}~\bibnamefont {Lohof}}, \bibinfo {author} {\bibfnamefont {D.}~\bibnamefont {Schumayer}}, \bibinfo {author} {\bibfnamefont {D.~A.~W.}\ \bibnamefont {Hutchinson}},\ and\ \bibinfo {author} {\bibfnamefont {C.}~\bibnamefont {Gies}},\ }\bibfield  {title} {\bibinfo {title} {Signatures of superradiance as a witness to multipartite entanglement},\ }\href {https://doi.org/10.1103/PhysRevLett.131.063601} {\bibfield  {journal} {\bibinfo  {journal} {Phys. Rev. Lett.}\ }\textbf {\bibinfo {volume} {131}},\ \bibinfo {pages} {063601} (\bibinfo {year} {2023})}\BibitemShut {NoStop}%
\bibitem [{\citenamefont {Paulisch}\ \emph {et~al.}(2019)\citenamefont {Paulisch}, \citenamefont {Perarnau-Llobet}, \citenamefont {Gonz\'alez-Tudela},\ and\ \citenamefont {Cirac}}]{PhysRevA.99.043807}%
  \BibitemOpen
  \bibfield  {author} {\bibinfo {author} {\bibfnamefont {V.}~\bibnamefont {Paulisch}}, \bibinfo {author} {\bibfnamefont {M.}~\bibnamefont {Perarnau-Llobet}}, \bibinfo {author} {\bibfnamefont {A.}~\bibnamefont {Gonz\'alez-Tudela}},\ and\ \bibinfo {author} {\bibfnamefont {J.~I.}\ \bibnamefont {Cirac}},\ }\bibfield  {title} {\bibinfo {title} {Quantum metrology with one-dimensional superradiant photonic states},\ }\href {https://doi.org/10.1103/PhysRevA.99.043807} {\bibfield  {journal} {\bibinfo  {journal} {Phys. Rev. A}\ }\textbf {\bibinfo {volume} {99}},\ \bibinfo {pages} {043807} (\bibinfo {year} {2019})}\BibitemShut {NoStop}%
\bibitem [{\citenamefont {Lange}\ \emph {et~al.}(2024)\citenamefont {Lange}, \citenamefont {Daggett}, \citenamefont {Walther}, \citenamefont {Huang},\ and\ \citenamefont {Hood}}]{Lange2024}%
  \BibitemOpen
  \bibfield  {author} {\bibinfo {author} {\bibfnamefont {C.~M.}\ \bibnamefont {Lange}}, \bibinfo {author} {\bibfnamefont {E.}~\bibnamefont {Daggett}}, \bibinfo {author} {\bibfnamefont {V.}~\bibnamefont {Walther}}, \bibinfo {author} {\bibfnamefont {L.}~\bibnamefont {Huang}},\ and\ \bibinfo {author} {\bibfnamefont {J.~D.}\ \bibnamefont {Hood}},\ }\bibfield  {title} {\bibinfo {title} {Superradiant and subradiant states in lifetime-limited organic molecules through laser-induced tuning},\ }\href {https://doi.org/10.1038/s41567-024-02404-4} {\bibfield  {journal} {\bibinfo  {journal} {Nature Physics}\ }\textbf {\bibinfo {volume} {20}},\ \bibinfo {pages} {836} (\bibinfo {year} {2024})}\BibitemShut {NoStop}%
\bibitem [{\citenamefont {Meiser}\ \emph {et~al.}(2009)\citenamefont {Meiser}, \citenamefont {Ye}, \citenamefont {Carlson},\ and\ \citenamefont {Holland}}]{PhysRevLett.102.163601}%
  \BibitemOpen
  \bibfield  {author} {\bibinfo {author} {\bibfnamefont {D.}~\bibnamefont {Meiser}}, \bibinfo {author} {\bibfnamefont {J.}~\bibnamefont {Ye}}, \bibinfo {author} {\bibfnamefont {D.~R.}\ \bibnamefont {Carlson}},\ and\ \bibinfo {author} {\bibfnamefont {M.~J.}\ \bibnamefont {Holland}},\ }\bibfield  {title} {\bibinfo {title} {Prospects for a millihertz-linewidth laser},\ }\href {https://doi.org/10.1103/PhysRevLett.102.163601} {\bibfield  {journal} {\bibinfo  {journal} {Phys. Rev. Lett.}\ }\textbf {\bibinfo {volume} {102}},\ \bibinfo {pages} {163601} (\bibinfo {year} {2009})}\BibitemShut {NoStop}%
\bibitem [{\citenamefont {Maier}\ \emph {et~al.}(2014)\citenamefont {Maier}, \citenamefont {Kraemer}, \citenamefont {Ostermann},\ and\ \citenamefont {Ritsch}}]{maier2014superradiant}%
  \BibitemOpen
  \bibfield  {author} {\bibinfo {author} {\bibfnamefont {T.}~\bibnamefont {Maier}}, \bibinfo {author} {\bibfnamefont {S.}~\bibnamefont {Kraemer}}, \bibinfo {author} {\bibfnamefont {L.}~\bibnamefont {Ostermann}},\ and\ \bibinfo {author} {\bibfnamefont {H.}~\bibnamefont {Ritsch}},\ }\bibfield  {title} {\bibinfo {title} {A superradiant clock laser on a magic wavelength optical lattice},\ }\href {https://doi.org/10.1364/OE.22.013269} {\bibfield  {journal} {\bibinfo  {journal} {Optics express}\ }\textbf {\bibinfo {volume} {22}},\ \bibinfo {pages} {13269} (\bibinfo {year} {2014})}\BibitemShut {NoStop}%
\bibitem [{\citenamefont {Bohnet}\ \emph {et~al.}(2012)\citenamefont {Bohnet}, \citenamefont {Chen}, \citenamefont {Weiner}, \citenamefont {Meiser}, \citenamefont {Holland},\ and\ \citenamefont {Thompson}}]{bohnet2012steady}%
  \BibitemOpen
  \bibfield  {author} {\bibinfo {author} {\bibfnamefont {J.~G.}\ \bibnamefont {Bohnet}}, \bibinfo {author} {\bibfnamefont {Z.}~\bibnamefont {Chen}}, \bibinfo {author} {\bibfnamefont {J.~M.}\ \bibnamefont {Weiner}}, \bibinfo {author} {\bibfnamefont {D.}~\bibnamefont {Meiser}}, \bibinfo {author} {\bibfnamefont {M.~J.}\ \bibnamefont {Holland}},\ and\ \bibinfo {author} {\bibfnamefont {J.~K.}\ \bibnamefont {Thompson}},\ }\bibfield  {title} {\bibinfo {title} {A steady-state superradiant laser with less than one intracavity photon},\ }\href {https://doi.org/10.1038/nature10920} {\bibfield  {journal} {\bibinfo  {journal} {Nature}\ }\textbf {\bibinfo {volume} {484}},\ \bibinfo {pages} {78} (\bibinfo {year} {2012})}\BibitemShut {NoStop}%
\bibitem [{\citenamefont {Norcia}\ and\ \citenamefont {Thompson}(2016)}]{PhysRevX.6.011025}%
  \BibitemOpen
  \bibfield  {author} {\bibinfo {author} {\bibfnamefont {M.~A.}\ \bibnamefont {Norcia}}\ and\ \bibinfo {author} {\bibfnamefont {J.~K.}\ \bibnamefont {Thompson}},\ }\bibfield  {title} {\bibinfo {title} {Cold-strontium laser in the superradiant crossover regime},\ }\href {https://doi.org/10.1103/PhysRevX.6.011025} {\bibfield  {journal} {\bibinfo  {journal} {Phys. Rev. X}\ }\textbf {\bibinfo {volume} {6}},\ \bibinfo {pages} {011025} (\bibinfo {year} {2016})}\BibitemShut {NoStop}%
\bibitem [{\citenamefont {Norcia}\ \emph {et~al.}(2016)\citenamefont {Norcia}, \citenamefont {Winchester}, \citenamefont {Cline},\ and\ \citenamefont {Thompson}}]{norcia2016superradiance}%
  \BibitemOpen
  \bibfield  {author} {\bibinfo {author} {\bibfnamefont {M.~A.}\ \bibnamefont {Norcia}}, \bibinfo {author} {\bibfnamefont {M.~N.}\ \bibnamefont {Winchester}}, \bibinfo {author} {\bibfnamefont {J.~R.}\ \bibnamefont {Cline}},\ and\ \bibinfo {author} {\bibfnamefont {J.~K.}\ \bibnamefont {Thompson}},\ }\bibfield  {title} {\bibinfo {title} {Superradiance on the millihertz linewidth strontium clock transition},\ }\href {https://doi.org/10.1126/sciadv.1601231} {\bibfield  {journal} {\bibinfo  {journal} {Science advances}\ }\textbf {\bibinfo {volume} {2}},\ \bibinfo {pages} {e1601231} (\bibinfo {year} {2016})}\BibitemShut {NoStop}%
\bibitem [{\citenamefont {Norcia}\ \emph {et~al.}(2018{\natexlab{b}})\citenamefont {Norcia}, \citenamefont {Cline}, \citenamefont {Muniz}, \citenamefont {Robinson}, \citenamefont {Hutson}, \citenamefont {Goban}, \citenamefont {Marti}, \citenamefont {Ye},\ and\ \citenamefont {Thompson}}]{PhysRevX.8.021036}%
  \BibitemOpen
  \bibfield  {author} {\bibinfo {author} {\bibfnamefont {M.~A.}\ \bibnamefont {Norcia}}, \bibinfo {author} {\bibfnamefont {J.~R.~K.}\ \bibnamefont {Cline}}, \bibinfo {author} {\bibfnamefont {J.~A.}\ \bibnamefont {Muniz}}, \bibinfo {author} {\bibfnamefont {J.~M.}\ \bibnamefont {Robinson}}, \bibinfo {author} {\bibfnamefont {R.~B.}\ \bibnamefont {Hutson}}, \bibinfo {author} {\bibfnamefont {A.}~\bibnamefont {Goban}}, \bibinfo {author} {\bibfnamefont {G.~E.}\ \bibnamefont {Marti}}, \bibinfo {author} {\bibfnamefont {J.}~\bibnamefont {Ye}},\ and\ \bibinfo {author} {\bibfnamefont {J.~K.}\ \bibnamefont {Thompson}},\ }\bibfield  {title} {\bibinfo {title} {Frequency measurements of superradiance from the strontium clock transition},\ }\href {https://doi.org/10.1103/PhysRevX.8.021036} {\bibfield  {journal} {\bibinfo  {journal} {Phys. Rev. X}\ }\textbf {\bibinfo {volume} {8}},\ \bibinfo {pages} {021036} (\bibinfo {year} {2018}{\natexlab{b}})}\BibitemShut {NoStop}%
\bibitem [{\citenamefont {Laske}\ \emph {et~al.}(2019)\citenamefont {Laske}, \citenamefont {Winter},\ and\ \citenamefont {Hemmerich}}]{PhysRevLett.123.103601}%
  \BibitemOpen
  \bibfield  {author} {\bibinfo {author} {\bibfnamefont {T.}~\bibnamefont {Laske}}, \bibinfo {author} {\bibfnamefont {H.}~\bibnamefont {Winter}},\ and\ \bibinfo {author} {\bibfnamefont {A.}~\bibnamefont {Hemmerich}},\ }\bibfield  {title} {\bibinfo {title} {Pulse delay time statistics in a superradiant laser with calcium atoms},\ }\href {https://doi.org/10.1103/PhysRevLett.123.103601} {\bibfield  {journal} {\bibinfo  {journal} {Phys. Rev. Lett.}\ }\textbf {\bibinfo {volume} {123}},\ \bibinfo {pages} {103601} (\bibinfo {year} {2019})}\BibitemShut {NoStop}%
\bibitem [{\citenamefont {Sch\"affer}\ \emph {et~al.}(2020)\citenamefont {Sch\"affer}, \citenamefont {Tang}, \citenamefont {Henriksen}, \citenamefont {J\o{}rgensen}, \citenamefont {Christensen},\ and\ \citenamefont {Thomsen}}]{PhysRevA.101.013819}%
  \BibitemOpen
  \bibfield  {author} {\bibinfo {author} {\bibfnamefont {S.~A.}\ \bibnamefont {Sch\"affer}}, \bibinfo {author} {\bibfnamefont {M.}~\bibnamefont {Tang}}, \bibinfo {author} {\bibfnamefont {M.~R.}\ \bibnamefont {Henriksen}}, \bibinfo {author} {\bibfnamefont {A.~A.}\ \bibnamefont {J\o{}rgensen}}, \bibinfo {author} {\bibfnamefont {B.~T.~R.}\ \bibnamefont {Christensen}},\ and\ \bibinfo {author} {\bibfnamefont {J.~W.}\ \bibnamefont {Thomsen}},\ }\bibfield  {title} {\bibinfo {title} {Lasing on a narrow transition in a cold thermal strontium ensemble},\ }\href {https://doi.org/10.1103/PhysRevA.101.013819} {\bibfield  {journal} {\bibinfo  {journal} {Phys. Rev. A}\ }\textbf {\bibinfo {volume} {101}},\ \bibinfo {pages} {013819} (\bibinfo {year} {2020})}\BibitemShut {NoStop}%
\bibitem [{\citenamefont {Hepp}\ and\ \citenamefont {Lieb}(1973)}]{PhysRevA.8.2517}%
  \BibitemOpen
  \bibfield  {author} {\bibinfo {author} {\bibfnamefont {K.}~\bibnamefont {Hepp}}\ and\ \bibinfo {author} {\bibfnamefont {E.~H.}\ \bibnamefont {Lieb}},\ }\bibfield  {title} {\bibinfo {title} {Equilibrium statistical mechanics of matter interacting with the quantized radiation field},\ }\href {https://doi.org/10.1103/PhysRevA.8.2517} {\bibfield  {journal} {\bibinfo  {journal} {Phys. Rev. A}\ }\textbf {\bibinfo {volume} {8}},\ \bibinfo {pages} {2517} (\bibinfo {year} {1973})}\BibitemShut {NoStop}%
\bibitem [{\citenamefont {Wang}\ and\ \citenamefont {Hioe}(1973)}]{PhysRevA.7.831}%
  \BibitemOpen
  \bibfield  {author} {\bibinfo {author} {\bibfnamefont {Y.~K.}\ \bibnamefont {Wang}}\ and\ \bibinfo {author} {\bibfnamefont {F.~T.}\ \bibnamefont {Hioe}},\ }\bibfield  {title} {\bibinfo {title} {Phase transition in the {D}icke model of superradiance},\ }\href {https://doi.org/10.1103/PhysRevA.7.831} {\bibfield  {journal} {\bibinfo  {journal} {Phys. Rev. A}\ }\textbf {\bibinfo {volume} {7}},\ \bibinfo {pages} {831} (\bibinfo {year} {1973})}\BibitemShut {NoStop}%
\bibitem [{\citenamefont {Hioe}(1973)}]{PhysRevA.8.1440}%
  \BibitemOpen
  \bibfield  {author} {\bibinfo {author} {\bibfnamefont {F.~T.}\ \bibnamefont {Hioe}},\ }\bibfield  {title} {\bibinfo {title} {Phase transitions in some generalized {D}icke models of superradiance},\ }\href {https://doi.org/10.1103/PhysRevA.8.1440} {\bibfield  {journal} {\bibinfo  {journal} {Phys. Rev. A}\ }\textbf {\bibinfo {volume} {8}},\ \bibinfo {pages} {1440} (\bibinfo {year} {1973})}\BibitemShut {NoStop}%
\bibitem [{\citenamefont {Carmichael}\ \emph {et~al.}(1973)\citenamefont {Carmichael}, \citenamefont {Gardiner},\ and\ \citenamefont {Walls}}]{carmichael1973higher}%
  \BibitemOpen
  \bibfield  {author} {\bibinfo {author} {\bibfnamefont {H.}~\bibnamefont {Carmichael}}, \bibinfo {author} {\bibfnamefont {C.}~\bibnamefont {Gardiner}},\ and\ \bibinfo {author} {\bibfnamefont {D.}~\bibnamefont {Walls}},\ }\bibfield  {title} {\bibinfo {title} {Higher order corrections to the {D}icke superradiant phase transition},\ }\href {https://doi.org/https://doi.org/10.1016/0375-9601(73)90679-8} {\bibfield  {journal} {\bibinfo  {journal} {Physics Letters A}\ }\textbf {\bibinfo {volume} {46}},\ \bibinfo {pages} {47} (\bibinfo {year} {1973})}\BibitemShut {NoStop}%
\bibitem [{\citenamefont {Duncan}(1974)}]{duncan1974effect}%
  \BibitemOpen
  \bibfield  {author} {\bibinfo {author} {\bibfnamefont {G.~C.}\ \bibnamefont {Duncan}},\ }\bibfield  {title} {\bibinfo {title} {Effect of antiresonant atom-field interactions on phase transitions in the {D}icke model},\ }\href {https://doi.org/10.1103/PhysRevA.9.418} {\bibfield  {journal} {\bibinfo  {journal} {Phys. Rev. A}\ }\textbf {\bibinfo {volume} {9}},\ \bibinfo {pages} {418} (\bibinfo {year} {1974})}\BibitemShut {NoStop}%
\bibitem [{\citenamefont {Das}\ \emph {et~al.}(2023)\citenamefont {Das}, \citenamefont {Bhakuni},\ and\ \citenamefont {Sharma}}]{PhysRevA.107.043706}%
  \BibitemOpen
  \bibfield  {author} {\bibinfo {author} {\bibfnamefont {P.}~\bibnamefont {Das}}, \bibinfo {author} {\bibfnamefont {D.~S.}\ \bibnamefont {Bhakuni}},\ and\ \bibinfo {author} {\bibfnamefont {A.}~\bibnamefont {Sharma}},\ }\bibfield  {title} {\bibinfo {title} {Phase transitions of the anisotropic {D}icke model},\ }\href {https://doi.org/10.1103/PhysRevA.107.043706} {\bibfield  {journal} {\bibinfo  {journal} {Phys. Rev. A}\ }\textbf {\bibinfo {volume} {107}},\ \bibinfo {pages} {043706} (\bibinfo {year} {2023})}\BibitemShut {NoStop}%
\bibitem [{\citenamefont {Buijsman}\ \emph {et~al.}(2017)\citenamefont {Buijsman}, \citenamefont {Gritsev},\ and\ \citenamefont {Sprik}}]{PhysRevLett.118.080601}%
  \BibitemOpen
  \bibfield  {author} {\bibinfo {author} {\bibfnamefont {W.}~\bibnamefont {Buijsman}}, \bibinfo {author} {\bibfnamefont {V.}~\bibnamefont {Gritsev}},\ and\ \bibinfo {author} {\bibfnamefont {R.}~\bibnamefont {Sprik}},\ }\bibfield  {title} {\bibinfo {title} {Nonergodicity in the anisotropic {D}icke model},\ }\href {https://doi.org/10.1103/PhysRevLett.118.080601} {\bibfield  {journal} {\bibinfo  {journal} {Phys. Rev. Lett.}\ }\textbf {\bibinfo {volume} {118}},\ \bibinfo {pages} {080601} (\bibinfo {year} {2017})}\BibitemShut {NoStop}%
\bibitem [{\citenamefont {Hu}\ and\ \citenamefont {Wan}(2021)}]{hu2021out}%
  \BibitemOpen
  \bibfield  {author} {\bibinfo {author} {\bibfnamefont {J.}~\bibnamefont {Hu}}\ and\ \bibinfo {author} {\bibfnamefont {S.}~\bibnamefont {Wan}},\ }\bibfield  {title} {\bibinfo {title} {Out-of-time-ordered correlation in the anisotropic {D}icke model},\ }\href {https://doi.org/10.1088/1572-9494/ac256d} {\bibfield  {journal} {\bibinfo  {journal} {Communications in Theoretical Physics}\ }\textbf {\bibinfo {volume} {73}},\ \bibinfo {pages} {125703} (\bibinfo {year} {2021})}\BibitemShut {NoStop}%
\bibitem [{\citenamefont {Stitely}\ \emph {et~al.}(2022)\citenamefont {Stitely}, \citenamefont {Giraldo}, \citenamefont {Krauskopf},\ and\ \citenamefont {Parkins}}]{PhysRevResearch.4.023101}%
  \BibitemOpen
  \bibfield  {author} {\bibinfo {author} {\bibfnamefont {K.~C.}\ \bibnamefont {Stitely}}, \bibinfo {author} {\bibfnamefont {A.}~\bibnamefont {Giraldo}}, \bibinfo {author} {\bibfnamefont {B.}~\bibnamefont {Krauskopf}},\ and\ \bibinfo {author} {\bibfnamefont {S.}~\bibnamefont {Parkins}},\ }\bibfield  {title} {\bibinfo {title} {Lasing and counter-lasing phase transitions in a cavity-{QED} system},\ }\href {https://doi.org/10.1103/PhysRevResearch.4.023101} {\bibfield  {journal} {\bibinfo  {journal} {Phys. Rev. Res.}\ }\textbf {\bibinfo {volume} {4}},\ \bibinfo {pages} {023101} (\bibinfo {year} {2022})}\BibitemShut {NoStop}%
\bibitem [{\citenamefont {Rza\ifmmode~\dot{z}\else \.{z}\fi{}ewski}\ \emph {et~al.}(1975)\citenamefont {Rza\ifmmode~\dot{z}\else \.{z}\fi{}ewski}, \citenamefont {W\'odkiewicz},\ and\ \citenamefont {\ifmmode~\dot{Z}\else \.{Z}\fi{}akowicz}}]{PhysRevLett.35.432}%
  \BibitemOpen
  \bibfield  {author} {\bibinfo {author} {\bibfnamefont {K.}~\bibnamefont {Rza\ifmmode~\dot{z}\else \.{z}\fi{}ewski}}, \bibinfo {author} {\bibfnamefont {K.}~\bibnamefont {W\'odkiewicz}},\ and\ \bibinfo {author} {\bibfnamefont {W.}~\bibnamefont {\ifmmode~\dot{Z}\else \.{Z}\fi{}akowicz}},\ }\bibfield  {title} {\bibinfo {title} {Phase transitions, two-level atoms, and the ${A}^{2}$ term},\ }\href {https://doi.org/10.1103/PhysRevLett.35.432} {\bibfield  {journal} {\bibinfo  {journal} {Phys. Rev. Lett.}\ }\textbf {\bibinfo {volume} {35}},\ \bibinfo {pages} {432} (\bibinfo {year} {1975})}\BibitemShut {NoStop}%
\bibitem [{\citenamefont {Nataf}\ and\ \citenamefont {Ciuti}(2010)}]{nataf2010no}%
  \BibitemOpen
  \bibfield  {author} {\bibinfo {author} {\bibfnamefont {P.}~\bibnamefont {Nataf}}\ and\ \bibinfo {author} {\bibfnamefont {C.}~\bibnamefont {Ciuti}},\ }\bibfield  {title} {\bibinfo {title} {No-go theorem for superradiant quantum phase transitions in cavity {QED} and counter-example in circuit {QED}},\ }\href {https://doi.org/10.1038/ncomms1069} {\bibfield  {journal} {\bibinfo  {journal} {Nature communications}\ }\textbf {\bibinfo {volume} {1}},\ \bibinfo {pages} {72} (\bibinfo {year} {2010})}\BibitemShut {NoStop}%
\bibitem [{\citenamefont {Viehmann}\ \emph {et~al.}(2011)\citenamefont {Viehmann}, \citenamefont {von Delft},\ and\ \citenamefont {Marquardt}}]{PhysRevLett.107.113602}%
  \BibitemOpen
  \bibfield  {author} {\bibinfo {author} {\bibfnamefont {O.}~\bibnamefont {Viehmann}}, \bibinfo {author} {\bibfnamefont {J.}~\bibnamefont {von Delft}},\ and\ \bibinfo {author} {\bibfnamefont {F.}~\bibnamefont {Marquardt}},\ }\bibfield  {title} {\bibinfo {title} {Superradiant phase transitions and the standard description of circuit {QED}},\ }\href {https://doi.org/10.1103/PhysRevLett.107.113602} {\bibfield  {journal} {\bibinfo  {journal} {Phys. Rev. Lett.}\ }\textbf {\bibinfo {volume} {107}},\ \bibinfo {pages} {113602} (\bibinfo {year} {2011})}\BibitemShut {NoStop}%
\bibitem [{\citenamefont {Vukics}\ and\ \citenamefont {Domokos}(2012)}]{PhysRevA.86.053807}%
  \BibitemOpen
  \bibfield  {author} {\bibinfo {author} {\bibfnamefont {A.}~\bibnamefont {Vukics}}\ and\ \bibinfo {author} {\bibfnamefont {P.}~\bibnamefont {Domokos}},\ }\bibfield  {title} {\bibinfo {title} {Adequacy of the {D}icke model in cavity {QED}: A counter-no-go statement},\ }\href {https://doi.org/10.1103/PhysRevA.86.053807} {\bibfield  {journal} {\bibinfo  {journal} {Phys. Rev. A}\ }\textbf {\bibinfo {volume} {86}},\ \bibinfo {pages} {053807} (\bibinfo {year} {2012})}\BibitemShut {NoStop}%
\bibitem [{\citenamefont {Bamba}\ and\ \citenamefont {Ogawa}(2014)}]{PhysRevA.90.063825}%
  \BibitemOpen
  \bibfield  {author} {\bibinfo {author} {\bibfnamefont {M.}~\bibnamefont {Bamba}}\ and\ \bibinfo {author} {\bibfnamefont {T.}~\bibnamefont {Ogawa}},\ }\bibfield  {title} {\bibinfo {title} {Stability of polarizable materials against superradiant phase transition},\ }\href {https://doi.org/10.1103/PhysRevA.90.063825} {\bibfield  {journal} {\bibinfo  {journal} {Phys. Rev. A}\ }\textbf {\bibinfo {volume} {90}},\ \bibinfo {pages} {063825} (\bibinfo {year} {2014})}\BibitemShut {NoStop}%
\bibitem [{\citenamefont {Vukics}\ \emph {et~al.}(2014)\citenamefont {Vukics}, \citenamefont {Grie\ss{}er},\ and\ \citenamefont {Domokos}}]{PhysRevLett.112.073601}%
  \BibitemOpen
  \bibfield  {author} {\bibinfo {author} {\bibfnamefont {A.}~\bibnamefont {Vukics}}, \bibinfo {author} {\bibfnamefont {T.}~\bibnamefont {Grie\ss{}er}},\ and\ \bibinfo {author} {\bibfnamefont {P.}~\bibnamefont {Domokos}},\ }\bibfield  {title} {\bibinfo {title} {Elimination of the $a$-square problem from cavity {QED}},\ }\href {https://doi.org/10.1103/PhysRevLett.112.073601} {\bibfield  {journal} {\bibinfo  {journal} {Phys. Rev. Lett.}\ }\textbf {\bibinfo {volume} {112}},\ \bibinfo {pages} {073601} (\bibinfo {year} {2014})}\BibitemShut {NoStop}%
\bibitem [{\citenamefont {Jaako}\ \emph {et~al.}(2016)\citenamefont {Jaako}, \citenamefont {Xiang}, \citenamefont {Garcia-Ripoll},\ and\ \citenamefont {Rabl}}]{PhysRevA.94.033850}%
  \BibitemOpen
  \bibfield  {author} {\bibinfo {author} {\bibfnamefont {T.}~\bibnamefont {Jaako}}, \bibinfo {author} {\bibfnamefont {Z.-L.}\ \bibnamefont {Xiang}}, \bibinfo {author} {\bibfnamefont {J.~J.}\ \bibnamefont {Garcia-Ripoll}},\ and\ \bibinfo {author} {\bibfnamefont {P.}~\bibnamefont {Rabl}},\ }\bibfield  {title} {\bibinfo {title} {Ultrastrong-coupling phenomena beyond the {D}icke model},\ }\href {https://doi.org/10.1103/PhysRevA.94.033850} {\bibfield  {journal} {\bibinfo  {journal} {Phys. Rev. A}\ }\textbf {\bibinfo {volume} {94}},\ \bibinfo {pages} {033850} (\bibinfo {year} {2016})}\BibitemShut {NoStop}%
\bibitem [{\citenamefont {Dimer}\ \emph {et~al.}(2007)\citenamefont {Dimer}, \citenamefont {Estienne}, \citenamefont {Parkins},\ and\ \citenamefont {Carmichael}}]{PhysRevA.75.013804}%
  \BibitemOpen
  \bibfield  {author} {\bibinfo {author} {\bibfnamefont {F.}~\bibnamefont {Dimer}}, \bibinfo {author} {\bibfnamefont {B.}~\bibnamefont {Estienne}}, \bibinfo {author} {\bibfnamefont {A.~S.}\ \bibnamefont {Parkins}},\ and\ \bibinfo {author} {\bibfnamefont {H.~J.}\ \bibnamefont {Carmichael}},\ }\bibfield  {title} {\bibinfo {title} {Proposed realization of the {D}icke-model quantum phase transition in an optical cavity {QED} system},\ }\href {https://doi.org/10.1103/PhysRevA.75.013804} {\bibfield  {journal} {\bibinfo  {journal} {Phys. Rev. A}\ }\textbf {\bibinfo {volume} {75}},\ \bibinfo {pages} {013804} (\bibinfo {year} {2007})}\BibitemShut {NoStop}%
\bibitem [{\citenamefont {Zhiqiang}\ \emph {et~al.}(2017)\citenamefont {Zhiqiang}, \citenamefont {Lee}, \citenamefont {Kumar}, \citenamefont {Arnold}, \citenamefont {Masson}, \citenamefont {Parkins},\ and\ \citenamefont {Barrett}}]{zhiqiang2017nonequilibrium}%
  \BibitemOpen
  \bibfield  {author} {\bibinfo {author} {\bibfnamefont {Z.}~\bibnamefont {Zhiqiang}}, \bibinfo {author} {\bibfnamefont {C.~H.}\ \bibnamefont {Lee}}, \bibinfo {author} {\bibfnamefont {R.}~\bibnamefont {Kumar}}, \bibinfo {author} {\bibfnamefont {K.}~\bibnamefont {Arnold}}, \bibinfo {author} {\bibfnamefont {S.~J.}\ \bibnamefont {Masson}}, \bibinfo {author} {\bibfnamefont {A.}~\bibnamefont {Parkins}},\ and\ \bibinfo {author} {\bibfnamefont {M.}~\bibnamefont {Barrett}},\ }\bibfield  {title} {\bibinfo {title} {Nonequilibrium phase transition in a spin-1 {D}icke model},\ }\href {https://doi.org/10.1364/OPTICA.4.000424} {\bibfield  {journal} {\bibinfo  {journal} {Optica}\ }\textbf {\bibinfo {volume} {4}},\ \bibinfo {pages} {424} (\bibinfo {year} {2017})}\BibitemShut {NoStop}%
\bibitem [{\citenamefont {Domokos}\ and\ \citenamefont {Ritsch}(2002)}]{PhysRevLett.89.253003}%
  \BibitemOpen
  \bibfield  {author} {\bibinfo {author} {\bibfnamefont {P.}~\bibnamefont {Domokos}}\ and\ \bibinfo {author} {\bibfnamefont {H.}~\bibnamefont {Ritsch}},\ }\bibfield  {title} {\bibinfo {title} {Collective cooling and self-organization of atoms in a cavity},\ }\href {https://doi.org/10.1103/PhysRevLett.89.253003} {\bibfield  {journal} {\bibinfo  {journal} {Phys. Rev. Lett.}\ }\textbf {\bibinfo {volume} {89}},\ \bibinfo {pages} {253003} (\bibinfo {year} {2002})}\BibitemShut {NoStop}%
\bibitem [{\citenamefont {Black}\ \emph {et~al.}(2003)\citenamefont {Black}, \citenamefont {Chan},\ and\ \citenamefont {Vuleti\ifmmode~\acute{c}\else \'{c}\fi{}}}]{PhysRevLett.91.203001}%
  \BibitemOpen
  \bibfield  {author} {\bibinfo {author} {\bibfnamefont {A.~T.}\ \bibnamefont {Black}}, \bibinfo {author} {\bibfnamefont {H.~W.}\ \bibnamefont {Chan}},\ and\ \bibinfo {author} {\bibfnamefont {V.}~\bibnamefont {Vuleti\ifmmode~\acute{c}\else \'{c}\fi{}}},\ }\bibfield  {title} {\bibinfo {title} {Observation of collective friction forces due to spatial self-organization of atoms: From {R}ayleigh to {B}ragg scattering},\ }\href {https://doi.org/10.1103/PhysRevLett.91.203001} {\bibfield  {journal} {\bibinfo  {journal} {Phys. Rev. Lett.}\ }\textbf {\bibinfo {volume} {91}},\ \bibinfo {pages} {203001} (\bibinfo {year} {2003})}\BibitemShut {NoStop}%
\bibitem [{\citenamefont {Nagy}\ \emph {et~al.}(2010)\citenamefont {Nagy}, \citenamefont {K\'onya}, \citenamefont {Szirmai},\ and\ \citenamefont {Domokos}}]{PhysRevLett.104.130401}%
  \BibitemOpen
  \bibfield  {author} {\bibinfo {author} {\bibfnamefont {D.}~\bibnamefont {Nagy}}, \bibinfo {author} {\bibfnamefont {G.}~\bibnamefont {K\'onya}}, \bibinfo {author} {\bibfnamefont {G.}~\bibnamefont {Szirmai}},\ and\ \bibinfo {author} {\bibfnamefont {P.}~\bibnamefont {Domokos}},\ }\bibfield  {title} {\bibinfo {title} {{D}icke-model phase transition in the quantum motion of a {B}ose-{E}instein condensate in an optical cavity},\ }\href {https://doi.org/10.1103/PhysRevLett.104.130401} {\bibfield  {journal} {\bibinfo  {journal} {Phys. Rev. Lett.}\ }\textbf {\bibinfo {volume} {104}},\ \bibinfo {pages} {130401} (\bibinfo {year} {2010})}\BibitemShut {NoStop}%
\bibitem [{\citenamefont {Baumann}\ \emph {et~al.}(2010)\citenamefont {Baumann}, \citenamefont {Guerlin}, \citenamefont {Brennecke},\ and\ \citenamefont {Esslinger}}]{baumann2010dicke}%
  \BibitemOpen
  \bibfield  {author} {\bibinfo {author} {\bibfnamefont {K.}~\bibnamefont {Baumann}}, \bibinfo {author} {\bibfnamefont {C.}~\bibnamefont {Guerlin}}, \bibinfo {author} {\bibfnamefont {F.}~\bibnamefont {Brennecke}},\ and\ \bibinfo {author} {\bibfnamefont {T.}~\bibnamefont {Esslinger}},\ }\bibfield  {title} {\bibinfo {title} {{D}icke quantum phase transition with a superfluid gas in an optical cavity},\ }\href {https://doi.org/10.1038/nature09009} {\bibfield  {journal} {\bibinfo  {journal} {Nature}\ }\textbf {\bibinfo {volume} {464}},\ \bibinfo {pages} {1301} (\bibinfo {year} {2010})}\BibitemShut {NoStop}%
\bibitem [{\citenamefont {Klinder}\ \emph {et~al.}(2015)\citenamefont {Klinder}, \citenamefont {Ke{\ss}ler}, \citenamefont {Wolke}, \citenamefont {Mathey},\ and\ \citenamefont {Hemmerich}}]{klinder2015dynamical}%
  \BibitemOpen
  \bibfield  {author} {\bibinfo {author} {\bibfnamefont {J.}~\bibnamefont {Klinder}}, \bibinfo {author} {\bibfnamefont {H.}~\bibnamefont {Ke{\ss}ler}}, \bibinfo {author} {\bibfnamefont {M.}~\bibnamefont {Wolke}}, \bibinfo {author} {\bibfnamefont {L.}~\bibnamefont {Mathey}},\ and\ \bibinfo {author} {\bibfnamefont {A.}~\bibnamefont {Hemmerich}},\ }\bibfield  {title} {\bibinfo {title} {Dynamical phase transition in the open {D}icke model},\ }\href {https://doi.org/10.1073/pnas.1417132112} {\bibfield  {journal} {\bibinfo  {journal} {Proceedings of the National Academy of Sciences}\ }\textbf {\bibinfo {volume} {112}},\ \bibinfo {pages} {3290} (\bibinfo {year} {2015})}\BibitemShut {NoStop}%
\bibitem [{\citenamefont {Vaidya}\ \emph {et~al.}(2018)\citenamefont {Vaidya}, \citenamefont {Guo}, \citenamefont {Kroeze}, \citenamefont {Ballantine}, \citenamefont {Koll\'ar}, \citenamefont {Keeling},\ and\ \citenamefont {Lev}}]{PhysRevX.8.011002}%
  \BibitemOpen
  \bibfield  {author} {\bibinfo {author} {\bibfnamefont {V.~D.}\ \bibnamefont {Vaidya}}, \bibinfo {author} {\bibfnamefont {Y.}~\bibnamefont {Guo}}, \bibinfo {author} {\bibfnamefont {R.~M.}\ \bibnamefont {Kroeze}}, \bibinfo {author} {\bibfnamefont {K.~E.}\ \bibnamefont {Ballantine}}, \bibinfo {author} {\bibfnamefont {A.~J.}\ \bibnamefont {Koll\'ar}}, \bibinfo {author} {\bibfnamefont {J.}~\bibnamefont {Keeling}},\ and\ \bibinfo {author} {\bibfnamefont {B.~L.}\ \bibnamefont {Lev}},\ }\bibfield  {title} {\bibinfo {title} {Tunable-range, photon-mediated atomic interactions in multimode cavity {QED}},\ }\href {https://doi.org/10.1103/PhysRevX.8.011002} {\bibfield  {journal} {\bibinfo  {journal} {Phys. Rev. X}\ }\textbf {\bibinfo {volume} {8}},\ \bibinfo {pages} {011002} (\bibinfo {year} {2018})}\BibitemShut {NoStop}%
\bibitem [{\citenamefont {Keeling}\ \emph {et~al.}(2010)\citenamefont {Keeling}, \citenamefont {Bhaseen},\ and\ \citenamefont {Simons}}]{PhysRevLett.105.043001}%
  \BibitemOpen
  \bibfield  {author} {\bibinfo {author} {\bibfnamefont {J.}~\bibnamefont {Keeling}}, \bibinfo {author} {\bibfnamefont {M.~J.}\ \bibnamefont {Bhaseen}},\ and\ \bibinfo {author} {\bibfnamefont {B.~D.}\ \bibnamefont {Simons}},\ }\bibfield  {title} {\bibinfo {title} {Collective dynamics of {B}ose-{E}instein condensates in optical cavities},\ }\href {https://doi.org/10.1103/PhysRevLett.105.043001} {\bibfield  {journal} {\bibinfo  {journal} {Phys. Rev. Lett.}\ }\textbf {\bibinfo {volume} {105}},\ \bibinfo {pages} {043001} (\bibinfo {year} {2010})}\BibitemShut {NoStop}%
\bibitem [{\citenamefont {{\"O}ztop}\ \emph {et~al.}(2012)\citenamefont {{\"O}ztop}, \citenamefont {Bordyuh}, \citenamefont {M{\"u}stecapl{\i}o{\u{g}}lu},\ and\ \citenamefont {T{\"u}reci}}]{oztop2012excitations}%
  \BibitemOpen
  \bibfield  {author} {\bibinfo {author} {\bibfnamefont {B.}~\bibnamefont {{\"O}ztop}}, \bibinfo {author} {\bibfnamefont {M.}~\bibnamefont {Bordyuh}}, \bibinfo {author} {\bibfnamefont {{\"O}.~E.}\ \bibnamefont {M{\"u}stecapl{\i}o{\u{g}}lu}},\ and\ \bibinfo {author} {\bibfnamefont {H.~E.}\ \bibnamefont {T{\"u}reci}},\ }\bibfield  {title} {\bibinfo {title} {Excitations of optically driven atomic condensate in a cavity: theory of photodetection measurements},\ }\href {https://doi.org/10.1088/1367-2630/14/8/085011} {\bibfield  {journal} {\bibinfo  {journal} {New Journal of Physics}\ }\textbf {\bibinfo {volume} {14}},\ \bibinfo {pages} {085011} (\bibinfo {year} {2012})}\BibitemShut {NoStop}%
\bibitem [{\citenamefont {Piazza}\ \emph {et~al.}(2013)\citenamefont {Piazza}, \citenamefont {Strack},\ and\ \citenamefont {Zwerger}}]{piazza2013bose}%
  \BibitemOpen
  \bibfield  {author} {\bibinfo {author} {\bibfnamefont {F.}~\bibnamefont {Piazza}}, \bibinfo {author} {\bibfnamefont {P.}~\bibnamefont {Strack}},\ and\ \bibinfo {author} {\bibfnamefont {W.}~\bibnamefont {Zwerger}},\ }\bibfield  {title} {\bibinfo {title} {{B}ose--{E}instein condensation versus {D}icke--{H}epp--{L}ieb transition in an optical cavity},\ }\href {https://doi.org/https://doi.org/10.1016/j.aop.2013.08.015} {\bibfield  {journal} {\bibinfo  {journal} {Annals of Physics}\ }\textbf {\bibinfo {volume} {339}},\ \bibinfo {pages} {135} (\bibinfo {year} {2013})}\BibitemShut {NoStop}%
\bibitem [{\citenamefont {Bhaseen}\ \emph {et~al.}(2012)\citenamefont {Bhaseen}, \citenamefont {Mayoh}, \citenamefont {Simons},\ and\ \citenamefont {Keeling}}]{PhysRevA.85.013817}%
  \BibitemOpen
  \bibfield  {author} {\bibinfo {author} {\bibfnamefont {M.~J.}\ \bibnamefont {Bhaseen}}, \bibinfo {author} {\bibfnamefont {J.}~\bibnamefont {Mayoh}}, \bibinfo {author} {\bibfnamefont {B.~D.}\ \bibnamefont {Simons}},\ and\ \bibinfo {author} {\bibfnamefont {J.}~\bibnamefont {Keeling}},\ }\bibfield  {title} {\bibinfo {title} {Dynamics of nonequilibrium {D}icke models},\ }\href {https://doi.org/10.1103/PhysRevA.85.013817} {\bibfield  {journal} {\bibinfo  {journal} {Phys. Rev. A}\ }\textbf {\bibinfo {volume} {85}},\ \bibinfo {pages} {013817} (\bibinfo {year} {2012})}\BibitemShut {NoStop}%
\bibitem [{\citenamefont {Torre}\ \emph {et~al.}(2013)\citenamefont {Torre}, \citenamefont {Diehl}, \citenamefont {Lukin}, \citenamefont {Sachdev},\ and\ \citenamefont {Strack}}]{PhysRevA.87.023831}%
  \BibitemOpen
  \bibfield  {author} {\bibinfo {author} {\bibfnamefont {E.~G.~D.}\ \bibnamefont {Torre}}, \bibinfo {author} {\bibfnamefont {S.}~\bibnamefont {Diehl}}, \bibinfo {author} {\bibfnamefont {M.~D.}\ \bibnamefont {Lukin}}, \bibinfo {author} {\bibfnamefont {S.}~\bibnamefont {Sachdev}},\ and\ \bibinfo {author} {\bibfnamefont {P.}~\bibnamefont {Strack}},\ }\bibfield  {title} {\bibinfo {title} {Keldysh approach for nonequilibrium phase transitions in quantum optics: Beyond the {D}icke model in optical cavities},\ }\href {https://doi.org/10.1103/PhysRevA.87.023831} {\bibfield  {journal} {\bibinfo  {journal} {Phys. Rev. A}\ }\textbf {\bibinfo {volume} {87}},\ \bibinfo {pages} {023831} (\bibinfo {year} {2013})}\BibitemShut {NoStop}%
\bibitem [{\citenamefont {Nagy}\ and\ \citenamefont {Domokos}(2015)}]{PhysRevLett.115.043601}%
  \BibitemOpen
  \bibfield  {author} {\bibinfo {author} {\bibfnamefont {D.}~\bibnamefont {Nagy}}\ and\ \bibinfo {author} {\bibfnamefont {P.}~\bibnamefont {Domokos}},\ }\bibfield  {title} {\bibinfo {title} {Nonequilibrium quantum criticality and non-markovian environment: Critical exponent of a quantum phase transition},\ }\href {https://doi.org/10.1103/PhysRevLett.115.043601} {\bibfield  {journal} {\bibinfo  {journal} {Phys. Rev. Lett.}\ }\textbf {\bibinfo {volume} {115}},\ \bibinfo {pages} {043601} (\bibinfo {year} {2015})}\BibitemShut {NoStop}%
\bibitem [{\citenamefont {Dalla~Torre}\ \emph {et~al.}(2016)\citenamefont {Dalla~Torre}, \citenamefont {Shchadilova}, \citenamefont {Wilner}, \citenamefont {Lukin},\ and\ \citenamefont {Demler}}]{PhysRevA.94.061802}%
  \BibitemOpen
  \bibfield  {author} {\bibinfo {author} {\bibfnamefont {E.~G.}\ \bibnamefont {Dalla~Torre}}, \bibinfo {author} {\bibfnamefont {Y.}~\bibnamefont {Shchadilova}}, \bibinfo {author} {\bibfnamefont {E.~Y.}\ \bibnamefont {Wilner}}, \bibinfo {author} {\bibfnamefont {M.~D.}\ \bibnamefont {Lukin}},\ and\ \bibinfo {author} {\bibfnamefont {E.}~\bibnamefont {Demler}},\ }\bibfield  {title} {\bibinfo {title} {{D}icke phase transition without total spin conservation},\ }\href {https://doi.org/10.1103/PhysRevA.94.061802} {\bibfield  {journal} {\bibinfo  {journal} {Phys. Rev. A}\ }\textbf {\bibinfo {volume} {94}},\ \bibinfo {pages} {061802} (\bibinfo {year} {2016})}\BibitemShut {NoStop}%
\bibitem [{\citenamefont {Ferioli}\ \emph {et~al.}(2021)\citenamefont {Ferioli}, \citenamefont {Glicenstein}, \citenamefont {Robicheaux}, \citenamefont {Sutherland}, \citenamefont {Browaeys},\ and\ \citenamefont {Ferrier-Barbut}}]{PhysRevLett.127.243602}%
  \BibitemOpen
  \bibfield  {author} {\bibinfo {author} {\bibfnamefont {G.}~\bibnamefont {Ferioli}}, \bibinfo {author} {\bibfnamefont {A.}~\bibnamefont {Glicenstein}}, \bibinfo {author} {\bibfnamefont {F.}~\bibnamefont {Robicheaux}}, \bibinfo {author} {\bibfnamefont {R.~T.}\ \bibnamefont {Sutherland}}, \bibinfo {author} {\bibfnamefont {A.}~\bibnamefont {Browaeys}},\ and\ \bibinfo {author} {\bibfnamefont {I.}~\bibnamefont {Ferrier-Barbut}},\ }\bibfield  {title} {\bibinfo {title} {Laser-driven superradiant ensembles of two-level atoms near dicke regime},\ }\href {https://doi.org/10.1103/PhysRevLett.127.243602} {\bibfield  {journal} {\bibinfo  {journal} {Phys. Rev. Lett.}\ }\textbf {\bibinfo {volume} {127}},\ \bibinfo {pages} {243602} (\bibinfo {year} {2021})}\BibitemShut {NoStop}%
\bibitem [{\citenamefont {Ferioli}\ \emph {et~al.}(2023)\citenamefont {Ferioli}, \citenamefont {Glicenstein}, \citenamefont {Ferrier-Barbut},\ and\ \citenamefont {Browaeys}}]{Ferioli2023}%
  \BibitemOpen
  \bibfield  {author} {\bibinfo {author} {\bibfnamefont {G.}~\bibnamefont {Ferioli}}, \bibinfo {author} {\bibfnamefont {A.}~\bibnamefont {Glicenstein}}, \bibinfo {author} {\bibfnamefont {I.}~\bibnamefont {Ferrier-Barbut}},\ and\ \bibinfo {author} {\bibfnamefont {A.}~\bibnamefont {Browaeys}},\ }\bibfield  {title} {\bibinfo {title} {A non-equilibrium superradiant phase transition in free space},\ }\href {https://doi.org/10.1038/s41567-023-02064-w} {\bibfield  {journal} {\bibinfo  {journal} {Nature Physics}\ }\textbf {\bibinfo {volume} {19}},\ \bibinfo {pages} {1345} (\bibinfo {year} {2023})}\BibitemShut {NoStop}%
\bibitem [{\citenamefont {Shammah}\ \emph {et~al.}(2017)\citenamefont {Shammah}, \citenamefont {Lambert}, \citenamefont {Nori},\ and\ \citenamefont {De~Liberato}}]{PhysRevA.96.023863}%
  \BibitemOpen
  \bibfield  {author} {\bibinfo {author} {\bibfnamefont {N.}~\bibnamefont {Shammah}}, \bibinfo {author} {\bibfnamefont {N.}~\bibnamefont {Lambert}}, \bibinfo {author} {\bibfnamefont {F.}~\bibnamefont {Nori}},\ and\ \bibinfo {author} {\bibfnamefont {S.}~\bibnamefont {De~Liberato}},\ }\bibfield  {title} {\bibinfo {title} {Superradiance with local phase-breaking effects},\ }\href {https://doi.org/10.1103/PhysRevA.96.023863} {\bibfield  {journal} {\bibinfo  {journal} {Phys. Rev. A}\ }\textbf {\bibinfo {volume} {96}},\ \bibinfo {pages} {023863} (\bibinfo {year} {2017})}\BibitemShut {NoStop}%
\bibitem [{\citenamefont {Gelhausen}\ \emph {et~al.}(2017)\citenamefont {Gelhausen}, \citenamefont {Buchhold},\ and\ \citenamefont {Strack}}]{PhysRevA.95.063824}%
  \BibitemOpen
  \bibfield  {author} {\bibinfo {author} {\bibfnamefont {J.}~\bibnamefont {Gelhausen}}, \bibinfo {author} {\bibfnamefont {M.}~\bibnamefont {Buchhold}},\ and\ \bibinfo {author} {\bibfnamefont {P.}~\bibnamefont {Strack}},\ }\bibfield  {title} {\bibinfo {title} {Many-body quantum optics with decaying atomic spin states: ($\ensuremath{\gamma}, \ensuremath{\kappa}$) {D}icke model},\ }\href {https://doi.org/10.1103/PhysRevA.95.063824} {\bibfield  {journal} {\bibinfo  {journal} {Phys. Rev. A}\ }\textbf {\bibinfo {volume} {95}},\ \bibinfo {pages} {063824} (\bibinfo {year} {2017})}\BibitemShut {NoStop}%
\bibitem [{\citenamefont {Shammah}\ \emph {et~al.}(2018)\citenamefont {Shammah}, \citenamefont {Ahmed}, \citenamefont {Lambert}, \citenamefont {De~Liberato},\ and\ \citenamefont {Nori}}]{PhysRevA.98.063815}%
  \BibitemOpen
  \bibfield  {author} {\bibinfo {author} {\bibfnamefont {N.}~\bibnamefont {Shammah}}, \bibinfo {author} {\bibfnamefont {S.}~\bibnamefont {Ahmed}}, \bibinfo {author} {\bibfnamefont {N.}~\bibnamefont {Lambert}}, \bibinfo {author} {\bibfnamefont {S.}~\bibnamefont {De~Liberato}},\ and\ \bibinfo {author} {\bibfnamefont {F.}~\bibnamefont {Nori}},\ }\bibfield  {title} {\bibinfo {title} {Open quantum systems with local and collective incoherent processes: Efficient numerical simulations using permutational invariance},\ }\href {https://doi.org/10.1103/PhysRevA.98.063815} {\bibfield  {journal} {\bibinfo  {journal} {Phys. Rev. A}\ }\textbf {\bibinfo {volume} {98}},\ \bibinfo {pages} {063815} (\bibinfo {year} {2018})}\BibitemShut {NoStop}%
\bibitem [{\citenamefont {Van~Vleck}(1929)}]{PhysRev.33.467}%
  \BibitemOpen
  \bibfield  {author} {\bibinfo {author} {\bibfnamefont {J.~H.}\ \bibnamefont {Van~Vleck}},\ }\bibfield  {title} {\bibinfo {title} {On $\ensuremath{\sigma}$-type doubling and electron spin in the spectra of diatomic molecules},\ }\href {https://doi.org/10.1103/PhysRev.33.467} {\bibfield  {journal} {\bibinfo  {journal} {Phys. Rev.}\ }\textbf {\bibinfo {volume} {33}},\ \bibinfo {pages} {467} (\bibinfo {year} {1929})}\BibitemShut {NoStop}%
\bibitem [{\citenamefont {Primas}(1963)}]{RevModPhys.35.710}%
  \BibitemOpen
  \bibfield  {author} {\bibinfo {author} {\bibfnamefont {H.}~\bibnamefont {Primas}},\ }\bibfield  {title} {\bibinfo {title} {Generalized perturbation theory in operator form},\ }\href {https://doi.org/10.1103/RevModPhys.35.710} {\bibfield  {journal} {\bibinfo  {journal} {Rev. Mod. Phys.}\ }\textbf {\bibinfo {volume} {35}},\ \bibinfo {pages} {710} (\bibinfo {year} {1963})}\BibitemShut {NoStop}%
\bibitem [{\citenamefont {Klein}(1974)}]{klein1974degenerate}%
  \BibitemOpen
  \bibfield  {author} {\bibinfo {author} {\bibfnamefont {D.}~\bibnamefont {Klein}},\ }\bibfield  {title} {\bibinfo {title} {Degenerate perturbation theory},\ }\href {https://doi.org/10.1063/1.1682018} {\bibfield  {journal} {\bibinfo  {journal} {The Journal of Chemical Physics}\ }\textbf {\bibinfo {volume} {61}},\ \bibinfo {pages} {786} (\bibinfo {year} {1974})}\BibitemShut {NoStop}%
\bibitem [{\citenamefont {Tong}\ and\ \citenamefont {Robicheaux}(2024)}]{PhysRevA.110.063701}%
  \BibitemOpen
  \bibfield  {author} {\bibinfo {author} {\bibfnamefont {W.}~\bibnamefont {Tong}}\ and\ \bibinfo {author} {\bibfnamefont {F.}~\bibnamefont {Robicheaux}},\ }\bibfield  {title} {\bibinfo {title} {Qualitatively altered driven {D}icke superradiance in extended systems due to infinitesimal perturbations},\ }\href {https://doi.org/10.1103/PhysRevA.110.063701} {\bibfield  {journal} {\bibinfo  {journal} {Phys. Rev. A}\ }\textbf {\bibinfo {volume} {110}},\ \bibinfo {pages} {063701} (\bibinfo {year} {2024})}\BibitemShut {NoStop}%
\bibitem [{\citenamefont {Li}\ \emph {et~al.}(2014)\citenamefont {Li}, \citenamefont {Petruccione},\ and\ \citenamefont {Koch}}]{li2014perturbative}%
  \BibitemOpen
  \bibfield  {author} {\bibinfo {author} {\bibfnamefont {A.~C.}\ \bibnamefont {Li}}, \bibinfo {author} {\bibfnamefont {F.}~\bibnamefont {Petruccione}},\ and\ \bibinfo {author} {\bibfnamefont {J.}~\bibnamefont {Koch}},\ }\bibfield  {title} {\bibinfo {title} {Perturbative approach to markovian open quantum systems},\ }\href {https://doi.org/10.1038/srep04887} {\bibfield  {journal} {\bibinfo  {journal} {Scientific reports}\ }\textbf {\bibinfo {volume} {4}},\ \bibinfo {pages} {4887} (\bibinfo {year} {2014})}\BibitemShut {NoStop}%
\bibitem [{\citenamefont {Huybrechts}\ and\ \citenamefont {Roscilde}(2024)}]{huybrechts2024quantum}%
  \BibitemOpen
  \bibfield  {author} {\bibinfo {author} {\bibfnamefont {D.}~\bibnamefont {Huybrechts}}\ and\ \bibinfo {author} {\bibfnamefont {T.}~\bibnamefont {Roscilde}},\ }\bibfield  {title} {\bibinfo {title} {Quantum correlations in the steady state of light-emitter ensembles from perturbation theory},\ }\bibfield  {journal} {\bibinfo  {journal} {arXiv preprint arXiv:2402.16824}\ }\href {https://doi.org/10.48550/arXiv.2402.16824} {10.48550/arXiv.2402.16824} (\bibinfo {year} {2024})\BibitemShut {NoStop}%
\bibitem [{\citenamefont {Thingna}\ and\ \citenamefont {Manzano}(2021)}]{thingna2021degenerated}%
  \BibitemOpen
  \bibfield  {author} {\bibinfo {author} {\bibfnamefont {J.}~\bibnamefont {Thingna}}\ and\ \bibinfo {author} {\bibfnamefont {D.}~\bibnamefont {Manzano}},\ }\bibfield  {title} {\bibinfo {title} {Degenerated {L}iouvillians and steady-state reduced density matrices},\ }\href {https://arxiv.org/pdf/2101.10236} {\bibfield  {journal} {\bibinfo  {journal} {Chaos: An Interdisciplinary Journal of Nonlinear Science}\ }\textbf {\bibinfo {volume} {31}} (\bibinfo {year} {2021})}\BibitemShut {NoStop}%
\bibitem [{\citenamefont {Albert}(2018)}]{albert2018lindbladians}%
  \BibitemOpen
  \bibfield  {author} {\bibinfo {author} {\bibfnamefont {V.~V.}\ \bibnamefont {Albert}},\ }\bibfield  {title} {\bibinfo {title} {{L}indbladians with multiple steady states: theory and applications},\ }\bibfield  {journal} {\bibinfo  {journal} {arXiv preprint arXiv:1802.00010}\ }\href {https://doi.org/10.48550/arXiv.1802.00010} {10.48550/arXiv.1802.00010} (\bibinfo {year} {2018})\BibitemShut {NoStop}%
\bibitem [{\citenamefont {G{\'o}mez}\ \emph {et~al.}(2018)\citenamefont {G{\'o}mez}, \citenamefont {Casta{\~n}o-Yepes},\ and\ \citenamefont {Thirumuruganandham}}]{gomez2018perturbation}%
  \BibitemOpen
  \bibfield  {author} {\bibinfo {author} {\bibfnamefont {E.~A.}\ \bibnamefont {G{\'o}mez}}, \bibinfo {author} {\bibfnamefont {J.~D.}\ \bibnamefont {Casta{\~n}o-Yepes}},\ and\ \bibinfo {author} {\bibfnamefont {S.~P.}\ \bibnamefont {Thirumuruganandham}},\ }\bibfield  {title} {\bibinfo {title} {Perturbation theory for open quantum systems at the steady state},\ }\href {https://doi.org/https://doi.org/10.1016/j.rinp.2018.06.038} {\bibfield  {journal} {\bibinfo  {journal} {Results in Physics}\ }\textbf {\bibinfo {volume} {10}},\ \bibinfo {pages} {353} (\bibinfo {year} {2018})}\BibitemShut {NoStop}%
\bibitem [{\citenamefont {Li}(2016)}]{li2016nonequilibrium}%
  \BibitemOpen
  \bibfield  {author} {\bibinfo {author} {\bibfnamefont {C.~Y.}\ \bibnamefont {Li}},\ }\emph {\bibinfo {title} {Nonequilibrium many-body physics with photons in circuit-{QED} lattices}},\ \href {https://bpb-us-e1.wpmucdn.com/sites.northwestern.edu/dist/2/1168/files/2022/01/thesis-v1.11.pdf} {Ph.D. thesis},\ \bibinfo  {school} {Northwestern University} (\bibinfo {year} {2016})\BibitemShut {NoStop}%
\bibitem [{\citenamefont {Krishna}\ \emph {et~al.}(2023)\citenamefont {Krishna}, \citenamefont {Solanki},\ and\ \citenamefont {Vinjanampathy}}]{krishna2023select}%
  \BibitemOpen
  \bibfield  {author} {\bibinfo {author} {\bibfnamefont {M.}~\bibnamefont {Krishna}}, \bibinfo {author} {\bibfnamefont {P.}~\bibnamefont {Solanki}},\ and\ \bibinfo {author} {\bibfnamefont {S.}~\bibnamefont {Vinjanampathy}},\ }\bibfield  {title} {\bibinfo {title} {Select topics in open quantum systems},\ }\href {https://doi.org/10.1007/s41745-022-00338-5} {\bibfield  {journal} {\bibinfo  {journal} {Journal of the Indian Institute of Science}\ }\textbf {\bibinfo {volume} {103}},\ \bibinfo {pages} {513} (\bibinfo {year} {2023})}\BibitemShut {NoStop}%
\bibitem [{\citenamefont {Iemini}\ \emph {et~al.}(2024)\citenamefont {Iemini}, \citenamefont {Chang},\ and\ \citenamefont {Marino}}]{PhysRevA.109.032204}%
  \BibitemOpen
  \bibfield  {author} {\bibinfo {author} {\bibfnamefont {F.}~\bibnamefont {Iemini}}, \bibinfo {author} {\bibfnamefont {D.}~\bibnamefont {Chang}},\ and\ \bibinfo {author} {\bibfnamefont {J.}~\bibnamefont {Marino}},\ }\bibfield  {title} {\bibinfo {title} {Dynamics of inhomogeneous spin ensembles with all-to-all interactions: Breaking permutational invariance},\ }\href {https://doi.org/10.1103/PhysRevA.109.032204} {\bibfield  {journal} {\bibinfo  {journal} {Phys. Rev. A}\ }\textbf {\bibinfo {volume} {109}},\ \bibinfo {pages} {032204} (\bibinfo {year} {2024})}\BibitemShut {NoStop}%
\bibitem [{\citenamefont {Ruostekoski}(2025)}]{ruostekoski2025superradiant}%
  \BibitemOpen
  \bibfield  {author} {\bibinfo {author} {\bibfnamefont {J.}~\bibnamefont {Ruostekoski}},\ }\bibfield  {title} {\bibinfo {title} {Superradiant phase transition in a large interacting driven atomic ensemble in free space},\ }\href {https://doi.org/10.1364/OPTICAQ.537927} {\bibfield  {journal} {\bibinfo  {journal} {Optica Quantum}\ }\textbf {\bibinfo {volume} {3}},\ \bibinfo {pages} {15} (\bibinfo {year} {2025})}\BibitemShut {NoStop}%
\bibitem [{\citenamefont {Ostermann}\ \emph {et~al.}(2024)\citenamefont {Ostermann}, \citenamefont {Rubies-Bigorda}, \citenamefont {Zhang},\ and\ \citenamefont {Yelin}}]{PhysRevResearch.6.023206}%
  \BibitemOpen
  \bibfield  {author} {\bibinfo {author} {\bibfnamefont {S.}~\bibnamefont {Ostermann}}, \bibinfo {author} {\bibfnamefont {O.}~\bibnamefont {Rubies-Bigorda}}, \bibinfo {author} {\bibfnamefont {V.}~\bibnamefont {Zhang}},\ and\ \bibinfo {author} {\bibfnamefont {S.~F.}\ \bibnamefont {Yelin}},\ }\bibfield  {title} {\bibinfo {title} {Breakdown of steady-state superradiance in extended driven atomic arrays},\ }\href {https://doi.org/10.1103/PhysRevResearch.6.023206} {\bibfield  {journal} {\bibinfo  {journal} {Phys. Rev. Res.}\ }\textbf {\bibinfo {volume} {6}},\ \bibinfo {pages} {023206} (\bibinfo {year} {2024})}\BibitemShut {NoStop}%
\bibitem [{\citenamefont {Goncalves}\ \emph {et~al.}(2025)\citenamefont {Goncalves}, \citenamefont {Bombieri}, \citenamefont {Ferioli}, \citenamefont {Pancaldi}, \citenamefont {Ferrier-Barbut}, \citenamefont {Browaeys}, \citenamefont {Shahmoon},\ and\ \citenamefont {Chang}}]{PRXQuantum.6.020303}%
  \BibitemOpen
  \bibfield  {author} {\bibinfo {author} {\bibfnamefont {D.}~\bibnamefont {Goncalves}}, \bibinfo {author} {\bibfnamefont {L.}~\bibnamefont {Bombieri}}, \bibinfo {author} {\bibfnamefont {G.}~\bibnamefont {Ferioli}}, \bibinfo {author} {\bibfnamefont {S.}~\bibnamefont {Pancaldi}}, \bibinfo {author} {\bibfnamefont {I.}~\bibnamefont {Ferrier-Barbut}}, \bibinfo {author} {\bibfnamefont {A.}~\bibnamefont {Browaeys}}, \bibinfo {author} {\bibfnamefont {E.}~\bibnamefont {Shahmoon}},\ and\ \bibinfo {author} {\bibfnamefont {D.}~\bibnamefont {Chang}},\ }\bibfield  {title} {\bibinfo {title} {Driven-dissipative phase separation in free-space atomic ensembles},\ }\href {https://doi.org/10.1103/PRXQuantum.6.020303} {\bibfield  {journal} {\bibinfo  {journal} {PRX Quantum}\ }\textbf {\bibinfo {volume} {6}},\ \bibinfo {pages} {020303} (\bibinfo {year} {2025})}\BibitemShut {NoStop}%
\bibitem [{\citenamefont {Li}\ and\ \citenamefont {Chesi}(2024)}]{PhysRevA.109.053702}%
  \BibitemOpen
  \bibfield  {author} {\bibinfo {author} {\bibfnamefont {J.}~\bibnamefont {Li}}\ and\ \bibinfo {author} {\bibfnamefont {S.}~\bibnamefont {Chesi}},\ }\bibfield  {title} {\bibinfo {title} {Routes to chaos in the balanced two-photon dicke model with qubit dissipation},\ }\href {https://doi.org/10.1103/PhysRevA.109.053702} {\bibfield  {journal} {\bibinfo  {journal} {Phys. Rev. A}\ }\textbf {\bibinfo {volume} {109}},\ \bibinfo {pages} {053702} (\bibinfo {year} {2024})}\BibitemShut {NoStop}%
\bibitem [{\citenamefont {Shah}\ \emph {et~al.}(2024)\citenamefont {Shah}, \citenamefont {Kirton}, \citenamefont {Felicetti},\ and\ \citenamefont {Alaeian}}]{shah2024dissipative}%
  \BibitemOpen
  \bibfield  {author} {\bibinfo {author} {\bibfnamefont {A.~J.}\ \bibnamefont {Shah}}, \bibinfo {author} {\bibfnamefont {P.}~\bibnamefont {Kirton}}, \bibinfo {author} {\bibfnamefont {S.}~\bibnamefont {Felicetti}},\ and\ \bibinfo {author} {\bibfnamefont {H.}~\bibnamefont {Alaeian}},\ }\bibfield  {title} {\bibinfo {title} {Dissipative phase transition in the two-photon dicke model},\ }\bibfield  {journal} {\bibinfo  {journal} {arXiv preprint arXiv:2412.14271}\ }\href {https://doi.org/10.48550/arXiv.2412.14271} {10.48550/arXiv.2412.14271} (\bibinfo {year} {2024})\BibitemShut {NoStop}%
\bibitem [{\citenamefont {Banerjee}\ \emph {et~al.}(2022)\citenamefont {Banerjee}, \citenamefont {Sharma},\ and\ \citenamefont {Bhattacherjee}}]{BANERJEE2022128287}%
  \BibitemOpen
  \bibfield  {author} {\bibinfo {author} {\bibfnamefont {P.}~\bibnamefont {Banerjee}}, \bibinfo {author} {\bibfnamefont {D.}~\bibnamefont {Sharma}},\ and\ \bibinfo {author} {\bibfnamefont {A.~B.}\ \bibnamefont {Bhattacherjee}},\ }\bibfield  {title} {\bibinfo {title} {Enhanced photon squeezing in two-photon dicke model},\ }\href {https://doi.org/https://doi.org/10.1016/j.physleta.2022.128287} {\bibfield  {journal} {\bibinfo  {journal} {Physics Letters A}\ }\textbf {\bibinfo {volume} {446}},\ \bibinfo {pages} {128287} (\bibinfo {year} {2022})}\BibitemShut {NoStop}%
\bibitem [{\citenamefont {Garbe}\ \emph {et~al.}(2017)\citenamefont {Garbe}, \citenamefont {Egusquiza}, \citenamefont {Solano}, \citenamefont {Ciuti}, \citenamefont {Coudreau}, \citenamefont {Milman},\ and\ \citenamefont {Felicetti}}]{PhysRevA.95.053854}%
  \BibitemOpen
  \bibfield  {author} {\bibinfo {author} {\bibfnamefont {L.}~\bibnamefont {Garbe}}, \bibinfo {author} {\bibfnamefont {I.~L.}\ \bibnamefont {Egusquiza}}, \bibinfo {author} {\bibfnamefont {E.}~\bibnamefont {Solano}}, \bibinfo {author} {\bibfnamefont {C.}~\bibnamefont {Ciuti}}, \bibinfo {author} {\bibfnamefont {T.}~\bibnamefont {Coudreau}}, \bibinfo {author} {\bibfnamefont {P.}~\bibnamefont {Milman}},\ and\ \bibinfo {author} {\bibfnamefont {S.}~\bibnamefont {Felicetti}},\ }\bibfield  {title} {\bibinfo {title} {Superradiant phase transition in the ultrastrong-coupling regime of the two-photon dicke model},\ }\href {https://doi.org/10.1103/PhysRevA.95.053854} {\bibfield  {journal} {\bibinfo  {journal} {Phys. Rev. A}\ }\textbf {\bibinfo {volume} {95}},\ \bibinfo {pages} {053854} (\bibinfo {year} {2017})}\BibitemShut {NoStop}%
\bibitem [{\citenamefont {Garbe}\ \emph {et~al.}(2020)\citenamefont {Garbe}, \citenamefont {Wade}, \citenamefont {Minganti}, \citenamefont {Shammah}, \citenamefont {Felicetti},\ and\ \citenamefont {Nori}}]{Garbe2020dissipation}%
  \BibitemOpen
  \bibfield  {author} {\bibinfo {author} {\bibfnamefont {L.}~\bibnamefont {Garbe}}, \bibinfo {author} {\bibfnamefont {P.}~\bibnamefont {Wade}}, \bibinfo {author} {\bibfnamefont {F.}~\bibnamefont {Minganti}}, \bibinfo {author} {\bibfnamefont {N.}~\bibnamefont {Shammah}}, \bibinfo {author} {\bibfnamefont {S.}~\bibnamefont {Felicetti}},\ and\ \bibinfo {author} {\bibfnamefont {F.}~\bibnamefont {Nori}},\ }\bibfield  {title} {\bibinfo {title} {Dissipation-induced bistability in the two-photon dicke model},\ }\href {https://doi.org/10.1038/s41598-020-69704-6} {\bibfield  {journal} {\bibinfo  {journal} {Scientific Reports}\ }\textbf {\bibinfo {volume} {10}},\ \bibinfo {pages} {13408} (\bibinfo {year} {2020})}\BibitemShut {NoStop}%
\bibitem [{\citenamefont {Crescente}\ \emph {et~al.}(2020)\citenamefont {Crescente}, \citenamefont {Carrega}, \citenamefont {Sassetti},\ and\ \citenamefont {Ferraro}}]{PhysRevB.102.245407}%
  \BibitemOpen
  \bibfield  {author} {\bibinfo {author} {\bibfnamefont {A.}~\bibnamefont {Crescente}}, \bibinfo {author} {\bibfnamefont {M.}~\bibnamefont {Carrega}}, \bibinfo {author} {\bibfnamefont {M.}~\bibnamefont {Sassetti}},\ and\ \bibinfo {author} {\bibfnamefont {D.}~\bibnamefont {Ferraro}},\ }\bibfield  {title} {\bibinfo {title} {Ultrafast charging in a two-photon dicke quantum battery},\ }\href {https://doi.org/10.1103/PhysRevB.102.245407} {\bibfield  {journal} {\bibinfo  {journal} {Phys. Rev. B}\ }\textbf {\bibinfo {volume} {102}},\ \bibinfo {pages} {245407} (\bibinfo {year} {2020})}\BibitemShut {NoStop}%
\bibitem [{\citenamefont {Chen}\ and\ \citenamefont {Zhang}(2018)}]{PhysRevA.97.053821}%
  \BibitemOpen
  \bibfield  {author} {\bibinfo {author} {\bibfnamefont {X.-Y.}\ \bibnamefont {Chen}}\ and\ \bibinfo {author} {\bibfnamefont {Y.-Y.}\ \bibnamefont {Zhang}},\ }\bibfield  {title} {\bibinfo {title} {Finite-size scaling analysis in the two-photon dicke model},\ }\href {https://doi.org/10.1103/PhysRevA.97.053821} {\bibfield  {journal} {\bibinfo  {journal} {Phys. Rev. A}\ }\textbf {\bibinfo {volume} {97}},\ \bibinfo {pages} {053821} (\bibinfo {year} {2018})}\BibitemShut {NoStop}%
\bibitem [{\citenamefont {Leppenen}\ and\ \citenamefont {Shahmoon}(2024)}]{leppenen2024quantumbistabilityinterplaycollective}%
  \BibitemOpen
  \bibfield  {author} {\bibinfo {author} {\bibfnamefont {N.}~\bibnamefont {Leppenen}}\ and\ \bibinfo {author} {\bibfnamefont {E.}~\bibnamefont {Shahmoon}},\ }\href {https://arxiv.org/abs/2404.02134} {\bibinfo {title} {Quantum bistability at the interplay between collective and individual decay}} (\bibinfo {year} {2024}),\ \Eprint {https://arxiv.org/abs/2404.02134} {arXiv:2404.02134 [quant-ph]} \BibitemShut {NoStop}%
\bibitem [{\citenamefont {Kirton}\ \emph {et~al.}(2019)\citenamefont {Kirton}, \citenamefont {Roses}, \citenamefont {Keeling},\ and\ \citenamefont {Dalla~Torre}}]{kirton2019introduction}%
  \BibitemOpen
  \bibfield  {author} {\bibinfo {author} {\bibfnamefont {P.}~\bibnamefont {Kirton}}, \bibinfo {author} {\bibfnamefont {M.~M.}\ \bibnamefont {Roses}}, \bibinfo {author} {\bibfnamefont {J.}~\bibnamefont {Keeling}},\ and\ \bibinfo {author} {\bibfnamefont {E.~G.}\ \bibnamefont {Dalla~Torre}},\ }\bibfield  {title} {\bibinfo {title} {Introduction to the {D}icke model: From equilibrium to nonequilibrium, and vice versa},\ }\href {https://doi.org/https://doi.org/10.1002/qute.201800043} {\bibfield  {journal} {\bibinfo  {journal} {Advanced Quantum Technologies}\ }\textbf {\bibinfo {volume} {2}},\ \bibinfo {pages} {1800043} (\bibinfo {year} {2019})}\BibitemShut {NoStop}%
\bibitem [{\citenamefont {EDMONDS}(1985)}]{edmonds1996angular}%
  \BibitemOpen
  \bibfield  {author} {\bibinfo {author} {\bibfnamefont {A.~R.}\ \bibnamefont {EDMONDS}},\ }\href {http://www.jstor.org/stable/j.ctt1cx3v9p} {\emph {\bibinfo {title} {Angular Momentum in Quantum Mechanics}}}\ (\bibinfo  {publisher} {Princeton University Press},\ \bibinfo {year} {1985})\BibitemShut {NoStop}%
\bibitem [{\citenamefont {Manzano}(2020)}]{manzano2020short}%
  \BibitemOpen
  \bibfield  {author} {\bibinfo {author} {\bibfnamefont {D.}~\bibnamefont {Manzano}},\ }\bibfield  {title} {\bibinfo {title} {A short introduction to the {L}indblad master equation},\ }\href {https://doi.org/10.1063/1.5115323} {\bibfield  {journal} {\bibinfo  {journal} {{AIP} Advances}\ }\textbf {\bibinfo {volume} {10}},\ \bibinfo {pages} {025106} (\bibinfo {year} {2020})}\BibitemShut {NoStop}%
\bibitem [{\citenamefont {Tong}\ \emph {et~al.}()\citenamefont {Tong}, \citenamefont {Alaeian},\ and\ \citenamefont {Robicheaux}}]{data_ODM_DPT}%
  \BibitemOpen
  \bibfield  {author} {\bibinfo {author} {\bibfnamefont {W.}~\bibnamefont {Tong}}, \bibinfo {author} {\bibfnamefont {H.}~\bibnamefont {Alaeian}},\ and\ \bibinfo {author} {\bibfnamefont {F.}~\bibnamefont {Robicheaux}},\ }\href@noop {} {\bibinfo {title} {Data for: Phase transitions in open dicke model: a degenerate perturbation theory approach, {P}urdue {U}niversity {R}esearch {R}epository (2025)}},\ \bibinfo {howpublished} {\url{https://doi.org/10.4231/MVHX-1V43}}\BibitemShut {NoStop}%
\bibitem [{\citenamefont {Wigner}(1932)}]{PhysRev.40.749}%
  \BibitemOpen
  \bibfield  {author} {\bibinfo {author} {\bibfnamefont {E.}~\bibnamefont {Wigner}},\ }\bibfield  {title} {\bibinfo {title} {On the quantum correction for thermodynamic equilibrium},\ }\href {https://doi.org/10.1103/PhysRev.40.749} {\bibfield  {journal} {\bibinfo  {journal} {Phys. Rev.}\ }\textbf {\bibinfo {volume} {40}},\ \bibinfo {pages} {749} (\bibinfo {year} {1932})}\BibitemShut {NoStop}%
\bibitem [{\citenamefont {Schleich}(2001)}]{doi:https://doi.org/10.1002/3527602976.ch3}%
  \BibitemOpen
  \bibfield  {author} {\bibinfo {author} {\bibfnamefont {W.~P.}\ \bibnamefont {Schleich}},\ }\bibinfo {title} {Wigner function},\ in\ \href {https://doi.org/https://doi.org/10.1002/3527602976.ch3} {\emph {\bibinfo {booktitle} {Quantum Optics in Phase Space}}}\ (\bibinfo  {publisher} {John Wiley \& Sons, Ltd},\ \bibinfo {year} {2001})\ Chap.~\bibinfo {chapter} {3}, pp.\ \bibinfo {pages} {67--98}\BibitemShut {NoStop}%
\bibitem [{\citenamefont {Santos}\ \emph {et~al.}(2022)\citenamefont {Santos}, \citenamefont {Cidrim}, \citenamefont {Villas-Boas}, \citenamefont {Kaiser},\ and\ \citenamefont {Bachelard}}]{PhysRevA.105.053715}%
  \BibitemOpen
  \bibfield  {author} {\bibinfo {author} {\bibfnamefont {A.~C.}\ \bibnamefont {Santos}}, \bibinfo {author} {\bibfnamefont {A.}~\bibnamefont {Cidrim}}, \bibinfo {author} {\bibfnamefont {C.~J.}\ \bibnamefont {Villas-Boas}}, \bibinfo {author} {\bibfnamefont {R.}~\bibnamefont {Kaiser}},\ and\ \bibinfo {author} {\bibfnamefont {R.}~\bibnamefont {Bachelard}},\ }\bibfield  {title} {\bibinfo {title} {Generating long-lived entangled states with free-space collective spontaneous emission},\ }\href {https://doi.org/10.1103/PhysRevA.105.053715} {\bibfield  {journal} {\bibinfo  {journal} {Phys. Rev. A}\ }\textbf {\bibinfo {volume} {105}},\ \bibinfo {pages} {053715} (\bibinfo {year} {2022})}\BibitemShut {NoStop}%
\bibitem [{\citenamefont {Santos}\ and\ \citenamefont {Bachelard}(2023)}]{PhysRevLett.130.053601}%
  \BibitemOpen
  \bibfield  {author} {\bibinfo {author} {\bibfnamefont {A.~C.}\ \bibnamefont {Santos}}\ and\ \bibinfo {author} {\bibfnamefont {R.}~\bibnamefont {Bachelard}},\ }\bibfield  {title} {\bibinfo {title} {Generation of maximally entangled long-lived states with giant atoms in a waveguide},\ }\href {https://doi.org/10.1103/PhysRevLett.130.053601} {\bibfield  {journal} {\bibinfo  {journal} {Phys. Rev. Lett.}\ }\textbf {\bibinfo {volume} {130}},\ \bibinfo {pages} {053601} (\bibinfo {year} {2023})}\BibitemShut {NoStop}%
\end{thebibliography}%

\end{document}